\PassOptionsToPackage{square,numbers}{natbib} 
\documentclass[10pt]{article}

\usepackage[utf8]{inputenc}
\usepackage[T1]{fontenc}
\usepackage[english]{babel}
\usepackage{graphicx}
\usepackage{amsmath,amssymb,amsbsy,amsthm,color,natbib}
\usepackage{multicol}
\usepackage{color} 
\usepackage{natbib}
\usepackage[linktoc=all,colorlinks=true,citecolor=blue,linkcolor=blue]{hyperref}
\usepackage[paperwidth=18cm,paperheight=29.7cm,hmargin=2cm,vmargin=2.5cm]{geometry}
\usepackage{authblk}
\usepackage{comment}
\usepackage{appendix}
\usepackage{bm}
\usepackage{booktabs}
\usepackage{tabularx}
\usepackage{comment}

\newtheorem{theorem}{Theorem}[section]
\newtheorem{corollary}[theorem]{Corollary}
\newtheorem{proposition}[theorem]{Proposition}

\theoremstyle{remark}
\newtheorem{remark}[theorem]{Remark}

\graphicspath{{Figures/}}

\title{Trajectory arclength reveals chaos}

\author[1,2]{Javier Jiménez-López\thanks{javier.jimenez.lopez@cern.ch}}
\author[2]{V. J. García-Garrido\thanks{vjose.garcia@uah.es (Corresponding author)}}
\affil[1]{Vrije Universiteit Brussel, Brussels, Belgium}
\affil[2]{Departamento de F\'isica y Matem\'aticas, Facultad de Ciencias, Universidad de Alcal\'a, 28805 Alcal\'a de Henares, Madrid, Spain.}

\begin{document}

\maketitle


\begin{abstract}

In this paper, we investigate the time-averaged phase space arclength of a trajectory, expressed through a normalized Lagrangian descriptor, as a trajectory-based diagnostic for distinguishing regular and chaotic dynamics. The proposed indicator, $\Omega$, is computed by integrating the equations of motion together with a single scalar arclength-accumulation equation and therefore does not require variational equations, neighboring trajectories, or spatial derivatives of a Lagrangian descriptor field. Using Birkhoff's ergodic theorem together with estimates for quasiperiodic motion and correlation-based arguments for chaotic ergodic dynamics, we show that regular and chaotic trajectories exhibit different finite-time behavior despite the convergence of $\Omega$ in both regimes. For quasiperiodic trajectories with Diophantine frequency vectors, deviations from the limiting average decay as $\mathcal{O}(\tau^{-1})$, whereas sufficiently mixing chaotic dynamics exhibit root-mean-square fluctuations of order $\tau^{-1/2}$ under suitable correlation assumptions. This difference in finite-time convergence and fluctuation scaling provides the theoretical basis for using $\Omega$ to discriminate between the two dynamical regimes. We validate the proposed approach in the Hénon--Heiles and Fermi--Pasta--Ulam systems, comparing the resulting classification of initial conditions with SALI as a reference chaos indicator. The results support $\Omega$ as a complementary single-trajectory diagnostic for chaos detection and as a practical source of scalar features for the analysis of trajectory ensembles, particularly in settings involving large numbers of trajectories or high-dimensional systems.

\end{abstract}

\noindent \textbf{keywords:} Phase space, Chaos indicators, Lagrangian descriptors, Hamiltonian dynamics, Dynamical systems.


\section{Introduction} 
\label{sec:Introduction}

Identifying the dynamical regime associated with a numerically computed orbit is a fundamental problem in nonlinear dynamics. In practice, trajectories are observed over finite intervals, and behavior that appears regular at moderate times need not remain so over longer integrations. A useful diagnostic must therefore do more than provide an asymptotic distinction between regularity and chaos: it should extract reliable information from finite-time data while remaining feasible for large collections of initial conditions and systems with many degrees of freedom. Such questions arise in applications ranging from transport in celestial mechanics \cite{Daquin2016,Daquin2019} and chemical reaction dynamics \cite{doi:10.1142/14372} to nonlinear beam motion in particle accelerators \cite{papaphilippou2014,montanari2025}.

Several established diagnostics determine the character of an orbit by monitoring the evolution of infinitesimal perturbations. Lyapunov exponents quantify exponential instability \cite{Wolf1985,Skokos2010a}, while finite-time indicators such as FLI, SALI, RLI, and GALI reveal different aspects of the dynamics of deviation vectors \cite{Froeschle1997a,Froeschle1997b,Sandor2004,Skokos2001,Skokos2003,Skokos2004,Skokos2008,moges2025}. The multiparticle formulation of GALI replaces deviation vectors with nearby trajectories \cite{MANYMANDA2025}. Although these techniques are well established, evaluating tangent dynamics or multiple trajectories may complicate their use in some models. Complementary information can be extracted from individual orbits, for example through frequency analysis \cite{Laskar1993}, wavelet-based time--frequency representations \cite{VelaArevalo2001,Chandre2003}, and the convergence of Birkhoff averages \cite{Levanjic2010,Levanjic2015,Das2016}. The latter approach is particularly relevant here because it shifts attention from the instantaneous separation of trajectories to the statistical properties of an observable sampled along a single orbit.

Time averages provide a natural bridge between orbit geometry and ergodic theory. For an integrable observable and an invariant probability measure, Birkhoff's ergodic theorem establishes almost-everywhere convergence of the average along trajectories \cite{5400911,10.1063/1.166399}. Convergence alone, however, does not specify how quickly a finite-time estimate approaches its limiting value. This distinction is important for numerical classification: the magnitude, temporal variability, and decay of finite-time corrections may contain more useful information than the limiting average itself. Moreover, these properties depend on the dynamical regime and, in chaotic regions, on the decay of temporal correlations.

Lagrangian descriptors (LDs) offer a convenient setting in which to investigate this idea. LDs are scalar quantities accumulated along trajectories and have been widely used to study the geometry of phase space \cite{Mancho2013,Lopesino2017}. Their use as the basis of chaos indicators has also been explored in several recent studies \cite{Hille22,zimper23,Daquin2022,caliman2025,jimenezlopez2025c}. For an autonomous flow $\dot{\mathbf{x}}=\mathbf{f}(\mathbf{x})$, let $\mathcal{L}_{f}(\mathbf{x}_0,\tau)$ denote the forward LD obtained from the arclength of the orbit starting at $\mathbf{x}_0$. We consider its time-normalized form,
\begin{equation}
\Omega(\mathbf{x}_0,\tau)
=\frac{\mathcal{L}_{f}(\mathbf{x}_0,\tau)}{\tau}
=\frac{1}{\tau}\int_{0}^{\tau}\bigl\|\mathbf{f}(\mathbf{x}(t;\mathbf{x}_0))\bigr\|\,\mathrm{d}t.
\label{eq:omega_introduction}
\end{equation}
Evaluating $\Omega$ requires only the reference trajectory and the accumulation of one scalar quantity; neither variational equations nor neighboring orbits are needed.

The normalized LD, $\Omega$, was previously introduced as an auxiliary quantity in our investigation of the LD difference between neighboring trajectories, $\Delta\mathcal{L}$ \cite{jimenezlopez2025c}. In that work, the proposed diagnostic was based on comparing the arclengths of two nearby orbits. The question examined here is different: \emph{can the finite-time evolution of the average arclength itself distinguish regular from chaotic motion?} In this work we investigate $\Omega$ as an independent observable and develop a theoretical framework to characterize the convergence and fluctuation properties that form the basis for distinguishing between different dynamical regimes in Hamiltonian systems.

Our analysis separates two mechanisms. For periodic motion, and for sufficiently smooth quasiperiodic dynamics under appropriate nonresonance conditions, Fourier arguments lead to finite-time corrections to $\Omega$ of order $\tau^{-1}$. By contrast, for chaotic trajectories exhibiting sufficiently rapid decay of correlations, fluctuations of the time average are expected to have a characteristic magnitude of order $\tau^{-1/2}$. These statements require different hypotheses: Birkhoff convergence alone does not imply either rate, and slow correlation decay may produce departures from the latter scaling. Making these assumptions explicit helps delimit the circumstances in which the temporal behavior of $\Omega$ provides a useful chaos diagnostic.

We examine this approach numerically in two Hamiltonian settings with different dimensionality and dynamical complexity: the Fermi--Pasta--Ulam (FPU) lattice and the H\'enon--Heiles system. Classifications obtained from $\Omega$ are compared with those produced by SALI (the Smaller Alignment Index), which serves as a reference diagnostic, and we investigate cases in which the two methods disagree. These examples test whether the finite-time behavior of a single trajectory-based observable can provide a practical alternative to diagnostics that explicitly track orbital instability.

The paper is outlined as follows. Section~\ref{sec:Methodology} introduces the LD formulation and develops the estimates for the finite-time behavior of $\Omega$ on regular and chaotic trajectories. Section~\ref{sec:Results} presents the numerical experiments for the H\'enon--Heiles and FPU systems and compares the results with SALI. Section~\ref{sec:Conclusions} discusses the implications and limitations of the approach. Appendix~\ref{sec:ground-truth} describes the reference classification based on SALI, while Appendix~\ref{sec:miss} examines the cases for which the classifications differ.

\section{Methodology} 
\label{sec:Methodology}

In this section we present the mathematical framework underlying the chaos indicator $\Omega$ based on Lagrangian descriptors. We begin in Sec.~\ref{sec:LDs} by reviewing the theoretical foundations of Lagrangian descriptors for continuous Hamiltonian systems, with particular emphasis on their arclength-based definition. We then turn to Sec.~\ref{sec:theory}, where we introduce the time-averaged Lagrangian descriptor, $\Omega$, derive its asymptotic behavior for both regular (Sec.~\ref{sec:theory_reg}) and chaotic (Sec.~\ref{sec:theory_chao}) motion, and describe in Sec.~\ref{sec:theory_class} how to use its properties to classify an ensemble of initial conditions.


\subsection{Lagrangian descriptors} \label{sec:LDs}

Originally developed as a mathematical framework to analyze transport and mixing processes in geophysical flows \cite{madrid2009,mendoza10}, Lagrangian descriptors have emerged as a powerful, trajectory-based scalar diagnostic tool for revealing phase space structures in general dynamical systems \cite{Garcia2022a}. In this work we will focus on applying this method in the context of continuous Hamiltonian systems, although this approach can be used similarly for discrete systems (maps) \cite{Lopesino2015}.

Consider a Hamiltonian system with $N$ degrees-of-freedom, $\mathbf{q} = (q_1, \, ..., \, q_N) \in \mathbb{R}^{N}$, and conjugate momenta $\mathbf{p} = (p_1, \, ..., \, p_N) \in \mathbb{R}^{N}$, which is defined by a scalar function $H(\mathbf{x})$ that is conserved along trajectories, and $\mathbf{x} = (\mathbf{q}, \mathbf{p}) \in \mathbb{R}^{2N}$ represents the state vector in the phase space of the system. In this context, Hamilton's equations of motion are given by the following set of ordinary differential equations:
\begin{equation} \label{eq:eq_motion}
    \dot{\mathbf{x}} = J \, \nabla H (\mathbf{x}) = \mathbf{f}(\mathbf{x}) \, ,
\end{equation}
where:
\begin{equation} \label{eq:Poisson_matrix}
    J = \begin{pmatrix}
        \mathbb{O}_N  & \mathbb{I}_N \\[.2cm] -\mathbb{I}_N & \mathbb{O}_N
    \end{pmatrix} \, ,
\end{equation}
is known as the Poisson matrix. Here, $\mathbb{O}_N$ and $\mathbb{I}_N$ denote the zero and the identity matrices of order $N$, respectively.

A Lagrangian descriptor is a scalar function, $\mathcal{L}(\mathbf{x}_0, t_0, \tau)$, associated with an initial condition $\mathbf{x}_0$ at time $t_0$. It is defined from the accumulation along the phase space trajectory of a non-negative scalar quantity, $\mathcal{F}(\mathbf{x}, t)$, over the time interval $[t_0-\tau,, t_0+\tau]$. The total LD value is obtained by adding the forward- and backward-time contributions:
\begin{equation}
    \mathcal{L}(\mathbf{x}_0, t_0, \tau) = \mathcal{L}_f(\mathbf{x}_0, t_0, \tau) + \mathcal{L}_b(\mathbf{x}_0, t_0, \tau) \, ,
\end{equation}
where the forward $(\mathcal{L}_f)$ and backward $(\mathcal{L}_b)$ components are given by:
\begin{equation}
    \mathcal{L}_f(\mathbf{x}_0,t_0,\tau) = \int_{t_0}^{t_0+\tau} \mathcal{F}(\mathbf{x}(t;\mathbf{x}_0),t) \, dt \quad,\quad
    \mathcal{L}_b(\mathbf{x}_0,t_0,\tau) = \int_{t_0-\tau}^{t_0} \mathcal{F}(\mathbf{x}(t;\mathbf{x}_0),t) \, dt \;.
    \label{ld_fw_bw}
\end{equation}

The integrand $\mathcal{F}$ used to define LDs typically quantifies some geometric or physical property of the trajectory. In this work, we adopt the arclength formulation, in which $\mathcal{F}$ is given by the Euclidean norm of the vector field:
\[
\mathcal{F}(\mathbf{x}, t) = \|\dot{\mathbf{x}}\| = \|\nabla H(\mathbf{x}, t)\|.
\]
With this choice, the LD represents the phase space arclength of the trajectory generated by the initial condition $\mathbf{x}_0$ over the time interval $[t_0 - \tau, t_0 + \tau]$. Note that for discrete systems, the integral definition is replaced by a finite sum over iterations \cite{Lopesino2015}. 

A fundamental property of the scalar field provided by LDs is that it can accurately reveal invariant phase space structures—such as equilibria, stable and unstable manifolds, invariant tori, and periodic orbits—that govern the dynamical evolution of trajectories \cite{Mancho2013,Lopesino2017}. Computationally, LDs are straightforward to implement, since the integrals in Eq.\eqref{ld_fw_bw} can be rewritten as an initial value problem using the Fundamental Theorem of Calculus. For example, the forward component of the Lagrangian descriptor satisfies:
\begin{equation}
    \begin{cases}
        \dfrac{d \mathcal{L}_f}{d \tau} = \mathcal{F}(\mathbf{x}(t_0 + \tau;\mathbf{x}_0),t_0 + \tau) \\[.3cm]
         \mathcal{L}_f(\mathbf{x}_0,t_0,0) = 0
    \end{cases}
\end{equation}
and thus it can be integrated simultaneously with the equations of motion in Eq.\eqref{eq:eq_motion}.

\subsection{Time-averaged arclength} \label{sec:theory}
In this subsection we introduce the time-averaged arclength of a trajectory starting at the initial condition $\mathbf{x}_0$ at time $t_0$, which is given by:
\begin{equation}
    \Omega = \dfrac{\mathcal{L}_f(\mathbf{x}_0,t_0,\tau)}{\tau} \;.
    \label{eq:omega_def}
\end{equation}
We investigate whether this quantity contains finite-time information that can be used to distinguish regular and chaotic trajectories. Unlike a conventional LD phase space portrait, which is constructed from an ensemble of initial conditions, $\Omega$ is a scalar time series obtained from a single forward trajectory. This framework avoids spatial differentiation of an LD field and comparison with neighboring initial conditions ~\cite{Hille22, Daquin2022, jimenez2024}. Next we will present the asymptotic and convergence properties of $\Omega$ both for regular periodic and quasiperiodic motion and also for chaotic orbits. Note that the definition of $\Omega$ in Eq.\eqref{eq:omega_def} only uses forward integration, since backward integration would reveal the same dynamical nature of the trajectory.

\subsubsection{Time-averaged arclength for regular trajectories}
\label{sec:theory_reg}

We first establish the asymptotic behavior of the time-averaged arclength for quasiperiodic trajectories evolving on a smooth invariant torus.

\begin{theorem}
\label{thm:regular_omega}
Let:
\begin{equation}
\mathcal{T}=K(\mathbb{T}^d)\subset\mathbb{R}^{2N} \quad,\quad
    \mathbb{T}^d=(\mathbb{R}/2\pi\mathbb{Z})^d \;,
\end{equation}
be a compact invariant torus, where $d\geq 1$ and
$K:\mathbb{T}^d\rightarrow\mathbb{R}^{2N}$ is a $C^1$ embedding such that $DK(\boldsymbol{\theta})$ is injective for every $\boldsymbol{\theta}\in\mathbb{T}^d$. Suppose that the dynamics on
$\mathcal{T}$ is conjugate to a rigid rotation with frequency vector
$\boldsymbol{\omega}\in\mathbb{R}^d$ \cite{HARO2021105859}, namely:
\begin{equation}
    \mathbf{f}(K(\boldsymbol{\theta}))
    =
    DK(\boldsymbol{\theta})\,\boldsymbol{\omega}\;,
\end{equation}
so that:
\begin{equation}
    \boldsymbol{\theta}(t)
    =
    \boldsymbol{\theta}_0+\boldsymbol{\omega}t
    \quad (\mathrm{mod}\ 2\pi) \;.
\end{equation}

Define:
\begin{equation}
    g(\boldsymbol{\theta})
    =
    \|\mathbf{f}(K(\boldsymbol{\theta}))\|
    =
    \|DK(\boldsymbol{\theta})\,\boldsymbol{\omega}\| \;,
\end{equation}
and assume that $\boldsymbol{\omega}$ satisfies the Diophantine condition:
\begin{equation}
    |\boldsymbol{k}\cdot\boldsymbol{\omega}|
    \geq
    \xi |\boldsymbol{k}|^{-\nu} \;\;,\;\;
    \boldsymbol{k}\in\mathbb{Z}^d\setminus\{\boldsymbol{0}\},
\end{equation}
for some $\xi>0$ and $\nu\geq0$, and that
$g\in C^r(\mathbb{T}^d)$ for an integer $r>\nu+d$ \cite{de2001tutorial}.

Then the time-averaged arclength:
\begin{equation}
\label{eq:Omega}
    \Omega(\tau)
    =
    \frac{\mathcal{L}_f(\mathbf{x}_0,t_0,\tau)}{\tau}
    =
    \frac{1}{\tau}
    \int_0^\tau
    g(\boldsymbol{\theta}_0+\boldsymbol{\omega}t)\,
    \mathrm{d}t \;,
\end{equation}
satisfies:
\begin{equation}
\label{eq:regular_asymptotic}
    \Omega(\tau)
    =
    \Omega_\infty+\mathcal{O}(\tau^{-1}) \;,
\end{equation}
as $\tau \to +\infty$ uniformly with respect to the initial phase
$\boldsymbol{\theta}_0$, where:
\begin{equation}
\label{eq:regular_limit}
    \Omega_\infty
    =
    \frac{1}{(2\pi)^d}
    \int_{\mathbb{T}^d}
    g(\boldsymbol{\theta})\,\mathrm{d}\boldsymbol{\theta}
    >0 \;.
\end{equation}
\end{theorem}

\begin{proof}
Since $K$ is an embedding, $DK(\boldsymbol{\theta})$ is injective at every point of $\mathbb{T}^d$. Moreover, the Diophantine condition implies
$\boldsymbol{\omega}\neq\boldsymbol{0}$. Hence:
\begin{equation}
    DK(\boldsymbol{\theta})\,\boldsymbol{\omega}
    \neq \boldsymbol{0} \;,
\end{equation}
for every $\boldsymbol{\theta}\in\mathbb{T}^d$, and therefore:
\begin{equation}
    g(\boldsymbol{\theta})>0
    \;\;,\;\;
    \text{for all }
    \boldsymbol{\theta}\in\mathbb{T}^d \;.
\end{equation}

Since $g$ is continuous and $\mathbb{T}^d$ is compact, $g$ attains a strictly positive minimum. It follows immediately that its angular average satisfies:
\begin{equation}
    \Omega_\infty
    =
    \frac{1}{(2\pi)^d}
    \int_{\mathbb{T}^d}
    g(\boldsymbol{\theta})\,\mathrm{d}\boldsymbol{\theta}
    >0 \;.
\end{equation}

To estimate the difference between the time average $\Omega(\tau)$ and the angular average $\Omega_\infty$, we introduce the Fourier coefficients of $g$:
\begin{equation}
    \widehat{g}_{\boldsymbol{k}}
    =
    \frac{1}{(2\pi)^d}
    \int_{\mathbb{T}^d}
    g(\boldsymbol{\theta})
    e^{-i\boldsymbol{k}\cdot\boldsymbol{\theta}}\,
    \mathrm{d}\boldsymbol{\theta} \quad,\quad
    \boldsymbol{k}\in\mathbb{Z}^d \;,
\end{equation}
where the zero Fourier mode is precisely $\widehat{g}_{\boldsymbol{0}}=\Omega_\infty $. For $\boldsymbol{k}\neq\boldsymbol{0}$, choose an index $j$ such that:
\begin{equation}
    |k_j|
    =
    \max_{1\leq \ell\leq d}|k_\ell| \;.
\end{equation}

Integrating by parts with respect to $\theta_j$ gives:
\begin{equation}
    \widehat{g}_{\boldsymbol{k}}
    =
    \frac{1}{ik_j}
    \frac{1}{(2\pi)^d}
    \int_{\mathbb{T}^d}
    \partial_{\theta_j}g(\boldsymbol{\theta})
    e^{-i\boldsymbol{k}\cdot\boldsymbol{\theta}}\,
    \mathrm{d}\boldsymbol{\theta} \;,
\end{equation}
where the boundary terms vanish because $g$ is $2\pi$-periodic in each angular coordinate. Repeating the integration by parts $r$ times yields:
\begin{equation}
    \widehat{g}_{\boldsymbol{k}}
    =
    \frac{1}{(ik_j)^r}
    \frac{1}{(2\pi)^d}
    \int_{\mathbb{T}^d}
    \partial_{\theta_j}^r g(\boldsymbol{\theta})
    e^{-i\boldsymbol{k}\cdot\boldsymbol{\theta}}\,
    \mathrm{d}\boldsymbol{\theta} \;.
\end{equation}

Consequently:
\begin{equation}
    |\widehat{g}_{\boldsymbol{k}}|
    \leq
    \frac{1}{|k_j|^r}
    \frac{1}{(2\pi)^d}
    \int_{\mathbb{T}^d}
    |\partial_{\theta_j}^r g(\boldsymbol{\theta})|\,
    \mathrm{d}\boldsymbol{\theta}
    \leq
    \frac{\|\partial_{\theta_j}^r g\|_\infty}{|k_j|^r}
    \leq
    C_g \|\boldsymbol{k}\|^{-r} \;,
\label{eq:regular_coeffcients_bound}
\end{equation}
where we have used that:
\begin{equation}
    |k_j|
    \geq
    \frac{\|\boldsymbol{k}\|}{\sqrt{d}}
    >0 \;,
\end{equation}
and:
\begin{equation}
    C_g
    =
    d^{r/2}
    \max_{1\leq j\leq d}
    \|\partial_{\theta_j}^r g\|_\infty \;,
\end{equation}
is independent of $\boldsymbol{k}$. Since $r>\nu+d\geq d$, we have $r>d$, and therefore:
\begin{equation}
    \sum_{\boldsymbol{k}\in\mathbb{Z}^d}
    |\widehat{g}_{\boldsymbol{k}}|
    <\infty \;.
\end{equation}

Thus, the Fourier series of $g$ converges absolutely and uniformly, and may be integrated term by term along the trajectory:
\begin{equation}
    g(\boldsymbol{\theta}_0+\boldsymbol{\omega}t)
    =
    \sum_{\boldsymbol{k}\in\mathbb{Z}^d}
    \widehat{g}_{\boldsymbol{k}}
    e^{i\boldsymbol{k}\cdot\boldsymbol{\theta}_0}
    e^{it\boldsymbol{k}\cdot\boldsymbol{\omega}} \;.
\end{equation}

For each nonzero Fourier mode:
\begin{equation}
    \int_0^\tau
    e^{it\boldsymbol{k}\cdot\boldsymbol{\omega}}\,
    \mathrm{d}t
    =
    \frac{
        e^{i\tau\boldsymbol{k}\cdot\boldsymbol{\omega}}-1
    }{
        i\boldsymbol{k}\cdot\boldsymbol{\omega}
    } \;.
\end{equation}

Substituting the Fourier representation into
Eq.\eqref{eq:Omega} and separating the zero mode gives:
\begin{equation}
\label{eq:diff}
    \Omega(\tau)-\Omega_\infty
    =
    \frac{1}{\tau}
    \sum_{\boldsymbol{k}\neq\boldsymbol{0}}
    \widehat{g}_{\boldsymbol{k}}
    e^{i\boldsymbol{k}\cdot\boldsymbol{\theta}_0}
    \frac{
        e^{i\tau\boldsymbol{k}\cdot\boldsymbol{\omega}}-1
    }{
        i\boldsymbol{k}\cdot\boldsymbol{\omega}
    } \;.
\end{equation}

It remains to control the small divisors $\boldsymbol{k}\cdot\boldsymbol{\omega}$. By the Diophantine condition
and Eq.\eqref{eq:regular_coeffcients_bound}:
\begin{equation}
    S    =
    \sum_{\boldsymbol{k}\neq\boldsymbol{0}}
    \frac{
        |\widehat{g}_{\boldsymbol{k}}|
    }{
        |\boldsymbol{k}\cdot\boldsymbol{\omega}|
    }\leq
    \frac{C_g}{\xi}
    \sum_{\boldsymbol{k}\neq\boldsymbol{0}}
    |\boldsymbol{k}|^{-(r-\nu)}
    <\infty \;,
\end{equation}
since $r-\nu>d$. Taking absolute values in
Eq.\eqref{eq:diff} and using
$|e^{i\alpha}-1|\leq2$ and
$|e^{i\boldsymbol{k}\cdot\boldsymbol{\theta}_0}|=1$ gives:
\begin{equation}
    |\Omega(\tau)-\Omega_\infty|
    \leq
    \frac{2}{\tau}
    \sum_{\boldsymbol{k}\neq\boldsymbol{0}}
    \frac{
        |\widehat{g}_{\boldsymbol{k}}|
    }{
        |\boldsymbol{k}\cdot\boldsymbol{\omega}|
    }
    =
    \frac{2S}{\tau} \;.
\end{equation}

Since the bound is independent of $\boldsymbol{\theta}_0$, the convergence is uniform in the initial phase. Hence:
\begin{equation}
    \Omega(\tau)
    =
    \Omega_\infty+\mathcal{O}(\tau^{-1}) \;,
\end{equation}
as $\tau \to +\infty$, which proves the result.
\end{proof}

The preceding result applies in particular to periodic trajectories. In this case, the limiting value admits a simple geometric interpretation.

\begin{corollary}
\label{cor:periodic_omega}
Let $\mathbf{x}(t)$ be a non-constant periodic trajectory of period $T>0$, so that:
\begin{equation}
    \mathbf{x}(t+T)=\mathbf{x}(t) \;.
\end{equation}
and suppose that:
\begin{equation}
    L_T
    =
    \int_0^T
    \|\mathbf{f}(\mathbf{x}(t))\|\,
    \mathrm{d}t \;,
\end{equation}
denotes the arclength traveled during one period. Then:
\begin{equation}
    \Omega(\tau)
    =
    \frac{L_T}{T}
    +
    \mathcal{O}(\tau^{-1}) \quad,\quad
    \tau\rightarrow\infty \;.
\end{equation}

In particular:
\begin{equation}
    \Omega_\infty=\frac{L_T}{T}>0 \;.
\end{equation}
\end{corollary}

\begin{proof}

A periodic orbit can be parametrized by a single angular coordinate, which we may choose as:
\begin{equation}
    \theta(t)=\frac{2\pi}{T}t \;.
\end{equation}

Therefore, its angular average is:
\begin{equation}
    \Omega_\infty
    =
    \dfrac{1}{2\pi}
    \int_0^{2\pi}
    g(\theta)\,\mathrm{d}\theta
    \nonumber\\
    =
    \dfrac{1}{T}
    \int_0^T
    \|\mathbf{f}(\mathbf{x}(t))\|\,
    \mathrm{d}t
    =
    \frac{L_T}{T} \;.
\end{equation}

Since the trajectory is non-constant, $L_T>0$, and hence
$\Omega_\infty>0$. To obtain the convergence rate directly, write:
\begin{equation}
    \tau=nT+s \;\;,\;\;
    n=\left\lfloor\frac{\tau}{T}\right\rfloor \;\;,\;\;
    0\leq s<T \;.
\end{equation}

The trajectory completes $n$ full periods followed by a remaining interval of duration $s$. By periodicity:
\begin{equation}
    \Omega(\tau)
    =
    \frac{
        nL_T+
        \displaystyle\int_0^s
        \|\mathbf{f}(\mathbf{x}(t))\|\,\mathrm{d}t
    }{\tau} \;.
\end{equation}
Using $nT=\tau-s$ and $L_T=T\Omega_\infty$, we obtain:
\begin{equation}
    \Omega(\tau)-\Omega_\infty
    =
    \dfrac{1}{\tau}
    \left[
        \int_0^s
        \|\mathbf{f}(\mathbf{x}(t))\|\,\mathrm{d}t
        -
        s\Omega_\infty
    \right]=
    \dfrac{1}{\tau}
    \int_0^s
    \left(
        \|\mathbf{f}(\mathbf{x}(t))\|
        -
        \Omega_\infty
    \right)\mathrm{d}t \;.
\end{equation}

Therefore:
\begin{equation}
    |\Omega(\tau)-\Omega_\infty|
    \leq
    \frac{1}{\tau}
    \int_0^s
    \left|
        \|\mathbf{f}(\mathbf{x}(t))\|
        -
        \Omega_\infty
    \right|
    \mathrm{d}t \;.
\end{equation}

Since the integrand is continuous and periodic, there exists
$M>0$ such that:
\begin{equation}
    \left|
        \|\mathbf{f}(\mathbf{x}(t))\|
        -
        \Omega_\infty
    \right|
    \leq M \;,
\end{equation}
for all $t$. Since $s<T$, it follows that:
\begin{equation}
    |\Omega(\tau)-\Omega_\infty|
    \leq
    \frac{Ms}{\tau}
    \leq
    \frac{MT}{\tau} \;,
\end{equation}
and hence:
\begin{equation}
    \Omega(\tau)
    =
    \Omega_\infty+\mathcal{O}(\tau^{-1}) \;.
\end{equation}
\end{proof}

\begin{remark}
\label{rem:weighted_birkhoff}

The convergence of time averages for quasiperiodic trajectories can be substantially accelerated by replacing the uniform time average with a smoothly weighted Birkhoff average~\cite{Duignan2023,SanderMeiss2020}. Under suitable regularity assumptions, such weighted averages can achieve
superpolynomial convergence for quasiperiodic trajectories with Diophantine rotation vectors. In the present work, however, we retain the standard unweighted time average because the finite-time convergence and fluctuation properties of $\Omega(\tau)$ themselves provide the information used to distinguish regular from chaotic motion.
\end{remark}

\subsubsection{Time-averaged arclength in ergodic chaotic regions}
\label{sec:theory_chao}

In what follows, we consider trajectories evolving in an invariant region that is ergodic with respect to the microcanonical measure. Ergodicity alone is not sufficient to characterize chaotic dynamics, since quasiperiodic motion on an invariant torus may also be ergodic with respect to its natural invariant measure. Our interest here is specifically in chaotic ergodic regions exhibiting sufficiently fast decay of temporal correlations. This additional dynamical property is introduced below through the assumption that the autocorrelation of the phase space speed is absolutely integrable.

Under this assumption, the finite-time behavior of $\Omega(\tau)$ differs from that obtained for quasiperiodic motion. Rather than the deterministic $\mathcal{O}(\tau^{-1})$ bound derived above, the deviations of $\Omega(\tau)$ from its limiting value exhibit a slower statistical decay, with a root-mean-square magnitude of order $\tau^{-1/2}$. Such fluctuations are characteristic of the mixing chaotic regime considered here~\cite{doi:10.1142/S0218127417300014,NAIK2019104907,MONTES2021105860}. The following theorem makes this statement precise by relating the finite-time fluctuations of $\Omega(\tau)$ to the autocorrelation function of the phase space speed.

\begin{theorem}
\label{thm:chaotic_omega}
Let $\Phi^t$ denote the Hamiltonian flow generated by $H$ and consider $\Sigma_E$ to be a compact component of the energy surface $H^{-1}(E)$ such that:
\begin{equation}
      \nabla H(\mathbf{x})\neq\mathbf{0}
    \quad,\quad
    \text{for all }\mathbf{x}\in\Sigma_E \;.
\end{equation}

Suppose also that $C\subseteq\Sigma_E$ is a compact invariant set of positive microcanonical measure, and that the Hamiltonian flow restricted to $C$ is ergodic with respect to the normalized microcanonical measure:
\begin{equation}
\label{eq:microcanonical_measure}
    \mathrm{d}\mu(\mathbf{x})
    =
    \dfrac{1}{Z_C}
    \dfrac{\mathrm{d}\sigma(\mathbf{x})}
         {\|\nabla H(\mathbf{x})\|} \quad,
    \qquad
    Z_C
    =
    \int_C
    \frac{\mathrm{d}\sigma(\mathbf{x})}
         {\|\nabla H(\mathbf{x})\|} \;,
\end{equation}
where $\mathrm{d}\sigma$ denotes the induced surface measure on the constant energy surface \cite{PhysRevLett.78.772}. If we define:
\begin{equation}
\label{eq:Omega_chaotic}
    \Omega(\tau;\mathbf{x}_0)
    =
    \frac{1}{\tau}
    \int_0^\tau
    \|\mathbf{f}(\Phi^t\mathbf{x}_0)\|\,
    \mathrm{d}t \;,
\end{equation}
then, for $\mu$-almost every initial condition
$\mathbf{x}_0\in C$ we have that:
\begin{equation}
\label{eq:chaotic_birkhoff}
    \lim_{\tau\to\infty}
    \Omega(\tau;\mathbf{x}_0)
    =
    \Omega_\infty \;,
\end{equation}
where:
\begin{equation}
    \Omega_\infty
    =
    \int_C
    \|\mathbf{f}(\mathbf{x})\|\,
    \mathrm{d}\mu(\mathbf{x})
    >0 \;.
\end{equation}

Furthermore, take:
\begin{equation}
\label{eq:h_chaotic}
    h(\mathbf{x})
    =
    \|\mathbf{f}(\mathbf{x})\|-\Omega_\infty \;,
\end{equation}
and define its stationary autocorrelation function:
\begin{equation}
\label{eq:alpha_chaotic}
    \alpha(s) = \int_C h(\mathbf{x}) \, h(\Phi^s\mathbf{x})\,
    \mathrm{d}\mu(\mathbf{x}) \;.
\end{equation}

If the autocorrelation function is absolutely integrable, that is:
\begin{equation}
\label{eq:integrable_corr}
    I = \int_0^\infty |\alpha(s)|\,\mathrm{d}s  <\infty \;,
\end{equation}
then:
\begin{equation}
\label{eq:chaotic_variance_bound}
    \operatorname{Var}_{\mu}[\Omega(\tau)] \leq \dfrac{2I}{\tau} \;,
\end{equation}
and consequently:
\begin{equation}
\label{eq:chaotic_op}
    \Omega(\tau) = \Omega_\infty + \mathcal{O}_p(\tau^{-1/2}) \;.
\end{equation}

Moreover:
\begin{equation}
\label{eq:limit}
    \lim_{\tau\to\infty}
    \tau\,
    \operatorname{Var}_{\mu}[\Omega(\tau)]
    =
    2\int_0^\infty
    \alpha(s)\,\mathrm{d}s \;.
\end{equation}

Hence, if we set $\displaystyle{\sigma_\Omega^2 = 2\int_0^\infty
    \alpha(s)\,\mathrm{d}s}$, and assume that it is strictly positive, then:
\begin{equation}
    \operatorname{Var}_{\mu}[\Omega(\tau)]
    =
    \dfrac{\sigma_\Omega^2}{\tau}
    +
    o(\tau^{-1}) \quad,\quad \operatorname{Std}_{\mu}[\Omega(\tau)]
    \sim
    \sigma_\Omega\,\tau^{-1/2} \;.
\end{equation}
\end{theorem}

\begin{proof}
Since $\Sigma_E$ is compact and $\mathbf{f}$ is continuous, $\|\mathbf{f}(\mathbf{x})\|$ is bounded on $\Sigma_E$, and therefore it is integrable with respect to $\mu$. Birkhoff's ergodic theorem
\cite{Birkhoff1931} implies that, for $\mu$-almost every $\mathbf{x}_0\in C$:
\begin{equation}
    \lim_{\tau\to\infty}
    \frac{1}{\tau}
    \int_0^\tau
    \|\mathbf{f}(\Phi^t\mathbf{x}_0)\|\,
    \mathrm{d}t
    =
    \int_C
    \|\mathbf{f}(\mathbf{x})\|\,
    \mathrm{d}\mu(\mathbf{x}),
\end{equation}
where ergodicity ensures that the limiting time average is independent
of the initial condition for $\mu$-almost every
$\mathbf{x}_0\in C$. This establishes
Eq.\eqref{eq:chaotic_birkhoff}.

For a Hamiltonian system in canonical coordinates, $\mathbf{f}(\mathbf{x}) = J\nabla H(\mathbf{x})$, where the symplectic matrix $J$ is invertible. Hence the assumption $\nabla H(\mathbf{x})\neq\mathbf{0}$ on $\Sigma_E$ implies $\mathbf{f}(\mathbf{x})\neq\mathbf{0}$ for all $\mathbf{x}\in\Sigma_E$. Since $\|\mathbf{f}(\mathbf{x})\|$ is continuous on the compact set
$C$, it attains a strictly positive minimum there. Consequently:
\begin{equation}
    \Omega_\infty
    =
    \int_C
    \|\mathbf{f}(\mathbf{x})\|\,
    \mathrm{d}\mu(\mathbf{x})
    >0.
\end{equation}

We now study the finite-time fluctuations about this limiting value. Calculating the time-average of Eq.\eqref{eq:h_chaotic} we get:
\begin{equation}
\label{eq:chaotic_error}
    \Omega(\tau;\mathbf{x}_0)-\Omega_\infty
    =
    \frac{1}{\tau}
    \int_0^\tau
    h(\Phi^t\mathbf{x}_0)\,
    \mathrm{d}t.
\end{equation}

Here and below, probabilistic quantities refer only to the sampling of
the initial condition $\mathbf{x}_0$ according to $\mu$; the
Hamiltonian evolution itself remains deterministic. Since $\mu$ is invariant under the flow:
\begin{equation}
    \mathbb{E}_{\mu}
    \left[
        h(\Phi^t\mathbf{x}_0)
    \right]
    =
    \int_C h(\mathbf{x})\,\mathrm{d}\mu(\mathbf{x})
    =
    0 \;.
\end{equation}

Therefore:
\begin{equation}
    \mathbb{E}_{\mu}
    \left[
        \Omega(\tau)-\Omega_\infty
    \right]
    =
    0 \;,
\end{equation}
and the mean-square error coincides with the variance. The invariance of $\mu$ also implies stationarity. For $t\geq u$,
a change of variables $\mathbf{y}=\Phi^u\mathbf{x}$ gives:
\begin{align}
    \mathbb{E}_{\mu}
    \left[
        h(\Phi^t\mathbf{x}_0)
        h(\Phi^u\mathbf{x}_0)
    \right]
    &=
    \int_C
    h(\Phi^t\mathbf{x})
    h(\Phi^u\mathbf{x})\,
    \mathrm{d}\mu(\mathbf{x})
    \nonumber\\[.2cm]
    &=
    \int_C
    h(\Phi^{t-u}\mathbf{y})
    h(\mathbf{y})\,
    \mathrm{d}\mu(\mathbf{y})=
    \alpha(t-u) \;.
\end{align}

Hence:
\begin{align}
    \operatorname{Var}_{\mu}[\Omega(\tau)]
    &=
    \frac{1}{\tau^2}
    \int_0^\tau
    \int_0^\tau
    \mathbb{E}_{\mu}
    \left[
        h(\Phi^t\mathbf{x}_0)
        h(\Phi^u\mathbf{x}_0)
    \right]
    \,\mathrm{d}u\,\mathrm{d}t.
\end{align}

Using the symmetry of the covariance kernel in $t$ and $u$, this can be written as \cite{10.1561/0900000004}:
\begin{align}
    \operatorname{Var}_{\mu}[\Omega(\tau)]
    &=
    \frac{2}{\tau^2}
    \int_0^\tau
    \int_0^t
    \alpha(t-u)\,
    \mathrm{d}u\,\mathrm{d}t
 =
    \frac{2}{\tau^2}
    \int_0^\tau
    (\tau-s)\alpha(s)\,
    \mathrm{d}s
    \nonumber\\[.2cm]
    &=
    \frac{2}{\tau}
    \int_0^\tau
    \left(1-\frac{s}{\tau}\right)
    \alpha(s)\,
    \mathrm{d}s.
\label{eq:variance_correlation}
\end{align}

If the autocorrelation is absolutely integrable, then:
\begin{align}
    \operatorname{Var}_{\mu}[\Omega(\tau)]
    \leq
    \frac{2}{\tau}
    \int_0^\tau
    \left(1-\frac{s}{\tau}\right)
    |\alpha(s)|\,\mathrm{d}s\leq
    \frac{2}{\tau}
    \int_0^\infty
    |\alpha(s)|\,\mathrm{d}s
    =
    \frac{2I}{\tau} \;,
\end{align}
which proves Eq.\eqref{eq:chaotic_variance_bound}. In particular,
the root-mean-square fluctuation satisfies:
\begin{equation}
    \left(
        \mathbb{E}_{\mu}
        \left[
            |\Omega(\tau)-\Omega_\infty|^2
        \right]
    \right)^{1/2}
    =
    \mathcal{O}(\tau^{-1/2}) \;.
\end{equation}

The corresponding probabilistic bound follows from Chebyshev's
inequality. Indeed, for any $M>0$ we have that:
\begin{align}
    \mu
    \left(
        |\Omega(\tau)-\Omega_\infty|
        >
        \frac{M}{\sqrt{\tau}}
    \right)
    \leq
    \frac{
        \operatorname{Var}_{\mu}[\Omega(\tau)]
    }{
        M^2/\tau
    }
\leq
    \frac{2I}{M^2} \;.
\end{align}

Thus, for every $\varepsilon\in(0,1)$, choosing:
\begin{equation}
    M_\varepsilon
    \geq
    \sqrt{\dfrac{2I}{\varepsilon}} \;,
\end{equation}
gives:
\begin{equation}
    \mu
    \left(
        |\Omega(\tau)-\Omega_\infty|
        \leq
        \frac{M_\varepsilon}{\sqrt{\tau}}
    \right)
    \geq
    1-\varepsilon \;,
\end{equation}
for all $\tau>0$. Therefore:
\begin{equation}
    \Omega(\tau)
    =
    \Omega_\infty
    +
    \mathcal{O}_p(\tau^{-1/2}) \;,
\end{equation}
where the subscript $p$ denotes boundedness in probability with respect to initial conditions distributed according to $\mu$ \cite{Vaart_1998}. Finally, multiplying Eq.\eqref{eq:variance_correlation} by $\tau$
gives:
\begin{equation}
    \tau\,
    \operatorname{Var}_{\mu}[\Omega(\tau)]
    =
    2
    \int_0^\infty
    \mathbf{1}_{[0,\tau]}(s)
    \left(1-\frac{s}{\tau}\right)
    \alpha(s)\,
    \mathrm{d}s \;.
\end{equation}

For every fixed $s\geq0$,
\begin{equation}
    \mathbf{1}_{[0,\tau]}(s)
    \left(1-\frac{s}{\tau}\right)
    \longrightarrow 1
    \qquad
    \text{as }\tau\to\infty \;,
\end{equation}
while its absolute value is bounded by $1$. Since $|\alpha|$ is integrable, the dominated convergence theorem gives:
\begin{equation}
    \lim_{\tau\to\infty}
    \tau\,
    \operatorname{Var}_{\mu}[\Omega(\tau)]
    =
    2\int_0^\infty
    \alpha(s)\,\mathrm{d}s \;.
\end{equation}

If the latter quantity is strictly positive, defining:
\begin{equation}
    \sigma_\Omega^2
    =
    2\int_0^\infty
    \alpha(s)\,\mathrm{d}s \;,
\end{equation}
yields:
\begin{equation}
    \operatorname{Var}_{\mu}[\Omega(\tau)]
    =
    \frac{\sigma_\Omega^2}{\tau}
    +
    o(\tau^{-1}) \;,
\end{equation}
and hence:
\begin{equation}
    \operatorname{Std}_{\mu}[\Omega(\tau)]
    \sim
    \sigma_\Omega\,\tau^{-1/2} \;.
\end{equation}
\end{proof}

\begin{remark}
\label{rem:chaotic_correlations}
The absolute integrability condition:
\begin{equation}
       \int_0^\infty |\alpha(s)|\,\mathrm{d}s<\infty \;,
\end{equation}
is not required for the almost-sure convergence of
$\Omega(\tau)$, which follows from ergodicity through Birkhoff's ergodic theorem. Its role here is to provide a quantitative estimate of the finite-time fluctuations.

If the autocorrelation is not absolutely integrable, no universal convergence rate follows from ergodicity alone. In particular, the failure of absolute integrability does not by itself imply a specific rate slower than $\tau^{-1/2}$; the asymptotic behavior then depends on
the detailed long-time correlation structure of the flow
\cite{10.1561/0900000004}.
\end{remark}

\subsubsection{Finite-time classification with the \texorpdfstring{$\Omega$}{Omega} indicator}
\label{sec:theory_class}

The preceding results show that the time-averaged arclength $\Omega(\tau)$ converges to a finite positive limit for both quasiperiodic and ergodic chaotic motion under the corresponding assumptions. Therefore, the distinction between the two dynamical regimes cannot be based on convergence versus nonconvergence. Instead, the relevant information is contained in the finite-time fluctuations about the limiting value. The following proposition makes this difference explicit by comparing their asymptotic scaling in the regular and chaotic cases, thereby providing the theoretical basis for using $\Omega$ as a finite-time indicator of dynamical character.

\begin{proposition}
\label{prop:regular_chaotic_comparison}

Under the assumptions of Theorems~\ref{thm:regular_omega} and \ref{thm:chaotic_omega}, the time-averaged arclength $\Omega(\tau)$ converges to a finite positive limiting value $\Omega_\infty$ for both quasiperiodic and ergodic chaotic motion. However, the finite-time fluctuations about this limit exhibit different asymptotic scalings.

For a quasiperiodic trajectory on a Diophantine invariant torus, there exists a constant $C>0$, independent of the initial phase, such that:
\begin{equation}
\label{eq:regular_comparison_bound}
    |\Omega(\tau)-\Omega_\infty|
    \leq
    \frac{C}{\tau} \;.
\end{equation}
Consequently:
\begin{equation}
\label{eq:regular_rescaled}
    \sqrt{\tau}\,
    |\Omega(\tau)-\Omega_\infty|
    \longrightarrow 0
    \qquad
    \text{as }\tau\to\infty \;.
\end{equation}

On the other hand, for trajectories with initial conditions distributed according to the invariant measure of an ergodic chaotic component with absolutely integrable autocorrelation and strictly positive asymptotic variance:
\begin{equation}
    \sigma_\Omega^2
    =
    2\int_0^\infty \alpha(s)\,\mathrm{d}s
    >0 \;,
\end{equation}
the mean-square fluctuation satisfies:
\begin{equation}
\label{eq:chaotic_comparison_variance}
    \mathbb{E}_{\mu}
    \left[
        |\Omega(\tau)-\Omega_\infty|^2
    \right]
    =
    \frac{\sigma_\Omega^2}{\tau}
    +
    o(\tau^{-1}) \;,
\end{equation}
and hence:
\begin{equation}
\label{eq:chaotic_comparison_rms}
    \left(
        \mathbb{E}_{\mu}
        \left[
            |\Omega(\tau)-\Omega_\infty|^2
        \right]
    \right)^{1/2}
    \sim
    \sigma_\Omega\,\tau^{-1/2} \;.
\end{equation}

Equivalently:
\begin{equation}
\label{eq:chaotic_rescaled}
    \left(
        \mathbb{E}_{\mu}
        \left[
            \left|
                \sqrt{\tau}\,
                (\Omega(\tau)-\Omega_\infty)
            \right|^2
        \right]
    \right)^{1/2}
    \longrightarrow
    \sigma_\Omega>0.
\end{equation}

Thus, although $\Omega(\tau)$ converges in both dynamical regimes, the finite-time deviations exhibit fundamentally different asymptotic behavior: for quasiperiodic motion they are deterministically bounded by $\mathcal{O}(\tau^{-1})$, whereas for chaotic motion their root-mean-square magnitude scales as $\tau^{-1/2}$ under the stated correlation assumptions.
\end{proposition}

\begin{proof}
For quasiperiodic motion, Theorem~\ref{thm:regular_omega} gives the uniform estimate:
\begin{equation}
    |\Omega(\tau)-\Omega_\infty|
    \leq
    \dfrac{C}{\tau} \;,
\end{equation}
for some constant $C>0$. Multiplying by $\sqrt{\tau}$ yields:
\begin{equation}
    \sqrt{\tau}\,
    |\Omega(\tau)-\Omega_\infty|
    \leq
    \frac{C}{\sqrt{\tau}} \;,
\end{equation}
and therefore:
\begin{equation}
    \sqrt{\tau}\,
    |\Omega(\tau)-\Omega_\infty|
    \longrightarrow 0 \;.
\end{equation}

For chaotic motion, Theorem~\ref{thm:chaotic_omega} gives:
\begin{equation}
    \operatorname{Var}_{\mu}[\Omega(\tau)]
    =
    \frac{\sigma_\Omega^2}{\tau}
    +
    o(\tau^{-1}) \;,
\end{equation}
provided that the autocorrelation is absolutely integrable and $\sigma_\Omega^2>0$. Since invariance of $\mu$ implies:
\begin{equation}
    \mathbb{E}_{\mu}[\Omega(\tau)]
    =
    \Omega_\infty \;,
\end{equation}
the variance is equal to the mean-square deviation from
$\Omega_\infty$. Hence:
\begin{equation}
    \mathbb{E}_{\mu}
    \left[
        |\Omega(\tau)-\Omega_\infty|^2
    \right]
    =
    \frac{\sigma_\Omega^2}{\tau}
    +
    o(\tau^{-1}) \;.
\end{equation}

Taking square roots gives:
\begin{equation}
    \left(
        \mathbb{E}_{\mu}
        \left[
            |\Omega(\tau)-\Omega_\infty|^2
        \right]
    \right)^{1/2}
    \sim
    \sigma_\Omega\,\tau^{-1/2} \;.
\end{equation}

Multiplying by $\sqrt{\tau}$ then yields:
\begin{equation}
    \left(
        \mathbb{E}_{\mu}
        \left[
            \left|
                \sqrt{\tau}\,
                (\Omega(\tau)-\Omega_\infty)
            \right|^2
        \right]
    \right)^{1/2}
    \longrightarrow
    \sigma_\Omega \;,
\end{equation}
which establishes the different finite-time fluctuation scalings in the two regimes.
\end{proof}

\begin{remark}
\label{rem:finite_time_discrimination}
Proposition~\ref{prop:regular_chaotic_comparison} does not distinguish regular and chaotic dynamics through convergence versus nonconvergence: under the stated assumptions, $\Omega(\tau)$ converges in both cases.
Rather, the distinction lies in the organization and magnitude of the finite-time fluctuations about the limiting value. For quasiperiodic motion these fluctuations are deterministically bounded by $\mathcal{O}(\tau^{-1})$, whereas for sufficiently mixing chaotic motion their root-mean-square magnitude is asymptotically of order
$\tau^{-1/2}$.

The different finite-time scaling of $\Omega(\tau)$ provides a mechanism for distinguishing regular and chaotic trajectories. Quasiperiodic trajectories exhibit deviations from their limiting value that decay as $\mathcal{O}(\tau^{-1})$, whereas sufficiently mixing chaotic dynamics exhibit root-mean-square fluctuations of order $\tau^{-1/2}$. Consequently, chaotic trajectories are expected to retain larger finite-time fluctuations as the integration time increases. This scaling difference supports the use of $\Omega$ as a practical chaos indicator, although at a fixed finite integration time the distributions associated with the two dynamical regimes may overlap, so that the theoretical result does not provide a universal trajectory-by-trajectory classification threshold.
\end{remark}

To characterize the convergence rate of $\Omega$ from the computational perspective, we use a block-average fluctuation analysis closely related to the blocking techniques commonly used in the study of correlated time series \cite{Flyvbjerg1989}. Let $L(t)$ denote the the cumulative quantity defining the finite-time average such that:
\begin{equation}
    \Omega (t) = \dfrac{L(t)}{t} \, .
\end{equation}

After discarding an initial transient, the trajectory is partitioned into non-overlapping intervals of size $\Delta$. The average associated with the $k-$th block is then reconstructed directly from the stored $\Omega$ time series as:
\begin{equation}
    \Omega^{(\Delta)}_k = \dfrac{L(t_k + \Delta) - L(t_k)}{\Delta} = \dfrac{(t_k + \Delta)\Omega(t_k + \Delta) - t_k\Omega(t_k)}{\Delta} \, ,
\end{equation}
so that for each block size $\Delta$ we compute the sample standard deviation:
\begin{equation}
    \sigma_{\mathrm{block}}(\Delta) = \sqrt{\dfrac{1}{N_{\Delta} - 1} \, \sum_{i=1}^{N_{\Delta}} \left( \Omega^{(\Delta)}_k - \overline{\Omega}^{(\Delta)} \right)^2} \, ,
\end{equation}
where $N_{\Delta}$ is the number of complete blocks of size $\Delta$ that fit inside the part of the trajectory under analysis and $\overline{\Omega}^{(\Delta)}$ denotes the mean of all the block averages for the same block size. The scaling of $\sigma_{\mathrm{block}} (\Delta)$ provides a distinction between periodic/quasiperiodic motion and chaotic dynamics. In the first case, the oscillatory contributions decay inversely with time as proved in Sec.~\ref{sec:theory_reg}, leading asymptotically to:
\begin{equation}
    \sigma_{\mathrm{block}}(\Delta) \sim \Delta^{-1} \, .
\end{equation}

On the other hand, for sufficiently mixing chaotic dynamics with an integrable correlation function, the central limit theorem implies \cite{Melbourne2005, SanderMeiss2020}:
\begin{equation}
    \mathrm{Var}\left[ \Omega^{(\Delta)}_k \right] \sim \Delta^{-1} \implies \sigma_{\mathrm{block}}(\Delta) \sim \Delta^{-1/2} \, ,
\end{equation}
following our result from Sec.~\ref{sec:theory_chao}. We therefore fit a function of the form $\sigma_{\mathrm{block}}(\Delta) = A \Delta^{-\alpha}$ expecting a binomial distribution with one peak concentrated around $\alpha = 1$ (corresponding to regular behavior) and another one near $\alpha = 1/2$, linked to chaotic motion.

In addition to the block scaling analysis, the convergence rate can also be inferred from the residual fluctuations of $\Omega(t)$ around its asymptotic value. To do so, we will use the final value of the time series as a numerical estimator of $\Omega_{\infty}$, such that over a fixed time interval $[t_a, t_b]$ we define the root-mean-square deviation as:
\begin{equation}
    \sigma_{\Omega} = \sqrt{\dfrac{1}{M} \sum_{i = 1}^{M} \left( \Omega (t_i) - \Omega_{\infty} \right)^2} \, ,
\end{equation}
where $\left\lbrace t_i \right\rbrace_{i=0}^M$ are the sampled times within the interval $[t_a, t_b]$. In order to remove the dependence on the value of the observable, we define the dimensionless normalized fluctuation amplitude $\sigma_{\Omega}/\Omega_{\infty}$. This quantity is analogous to the coefficient of variation, although the fluctuations are evaluated relative to the estimated asymptotic values instead of the sample mean. However, we will denote it as CV in the rest of the paper for simplicity.

In the continuous time limit, the numerator can be written as:
\begin{equation}
    \sigma^2_{\Omega} \simeq \dfrac{1}{t_b - t_a} \int_{t_a}^{t_b} \left[ \Omega(t) - \Omega_{\infty}\right] ^2 \, \mathrm{d}t \, ,
\end{equation}
so that for regular motion we have:
\begin{equation}
    \dfrac{\sigma_{\Omega}}{\Omega_{\infty}} \sim \dfrac{1}{\Omega_{\infty} \sqrt{t_a t_b}} \, ,
\end{equation}
which directly follows from Theorem~\ref{thm:regular_omega}. Similarly, in the case of sufficiently mixing chaotic dynamics, it follows from Theorem~\ref{thm:chaotic_omega} that:
\begin{equation}
    \dfrac{\sigma_{\Omega}}{\Omega_{\infty}} \sim \dfrac{1}{\Omega_{\infty}} \sqrt{\dfrac{\log(t_b / t_a)}{t_b - t_a}} \, .
\end{equation}
It is then clear that the quantity $\sigma_{\Omega}/\Omega_{\infty}$ will considerably differ for regular and chaotic motion, allowing us to use it for the task of classifying initial conditions. In the case of considering an ensemble of initial conditions, these two methods will produce, as already mentioned, bimodal distributions, such that the classification itself consists in finding the value of $\alpha$ or of the CV that separates both populations in the best way possible (with the least possible overlap). 

As we have obtained the true nature of the trajectories under analysis with the SALI indicator (see App.~\ref{sec:ground-truth}), it is then possible to optimize the threshold selection for each ensemble of initial conditions. This approach aims to test the robustness of both methodologies for the study of initial conditions ensembles that may contain both regular and chaotic trajectories. Then, the threshold is determined by the maximization of the Matthews correlation coefficient (MCC), defined as \cite{Matthews1975, Chicco2020}:
\begin{equation} \label{eq:MCC}
    \mathrm{MCC} = \dfrac{\mathrm{TP} \cdot \mathrm{TN} - \mathrm{FP} \cdot \mathrm{FN}}{\sqrt{(\mathrm{TP + FP})(\mathrm{TP + FN})(\mathrm{TN + FP})(\mathrm{TN + FN})}} \, 
\end{equation}
where TP is a chaotic orbit correctly classified as chaotic, TN a regular orbit correctly classified as regular, FP a regular orbit incorrectly classified as chaotic and FN a chaotic orbit incorrectly classified as regular. By definition, $\mathrm{MCC} \in [-1,1]$ where a value of $1$ indicates perfect classification. With this approach we aim to test the robustness of the calculated threshold (obtained through the maximization of the MCC) among the different ensembles of initial conditions, each one corresponding to a different energy, considered for the H\'enon-Heiles system in the next section.


\section{Results} 
\label{sec:Results}

In this section, we present the results obtained by applying the time-averaged Lagrangian descriptor, $\Omega$, to characterize regular and chaotic trajectories in two well-known Hamiltonian systems: the H\'enon--Heiles \cite{henon1964applicability} and Fermi--Pasta--Ulam (FPU) \cite{fermi1955} systems. As described in Sec.~\ref{sec:theory_class}, trajectories are classified as regular or chaotic from the finite-time behavior of the $\Omega$ time series using block-scaling analysis and residual fluctuations (encoded in the CV). We assess the performance of these classification procedures under different numerical settings by comparing their results with those obtained using SALI as a reference chaos indicator. 

\subsection{The H\'enon-Heiles System}
\label{subsec:HH}

To validate our theoretical predictions for the behavior of the $\Omega$ indicator, we analyze first the H\'enon-Heiles system, which is a two DoF Hamiltonian system given by \cite{henon1964applicability}:
\begin{equation}
   H(x,y,p_x,p_y) = \frac{p_x^2}{2} + \frac{p_y^2}{2} + \frac{1}{2}(x^2+y^2) + x^2y - \frac{1}{3}y^3 \, , 
    \label{eq:Ham_HH}
\end{equation} 
widely used in the field to test the performance of chaos indicators and mathematical tools for the analysis of dynamical systems. The initial conditions used in the different ensembles where randomly sampled on the energetically accessible part of the Poncar\'e section defined by $x = 0$ and $p_x > 0$. The trajectories considered were integrated for a total time $\tau = 10^{5}$ with the Dormand-Prince $8(9)$ ODE solver \cite{prince1981} implemented in the PYTHON package SciPy \cite{Virtanen:2019joe}, ensuring a conservation of energy within $10^{-9}$, achieved setting the relative and absolute tolerances to $10^{-10}$.

\begin{figure}[htbp]
    \centering
    A) \includegraphics[scale = 0.325]{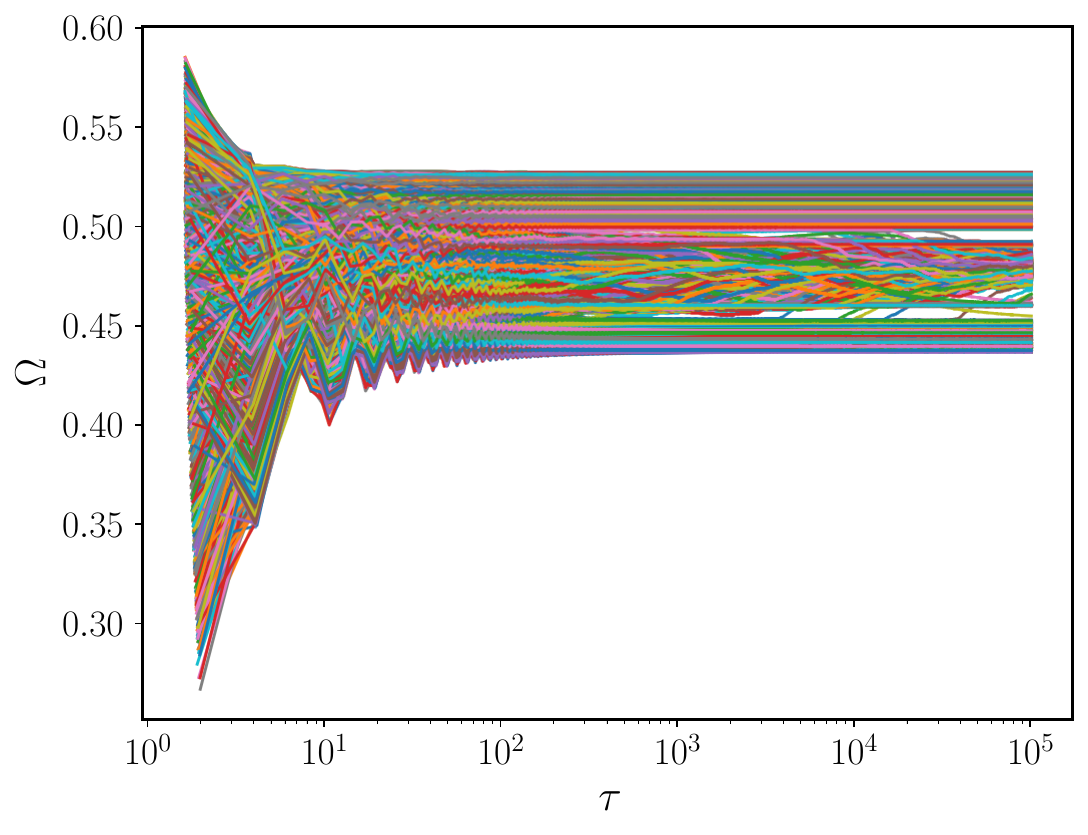}
    B) \includegraphics[scale = 0.3175]{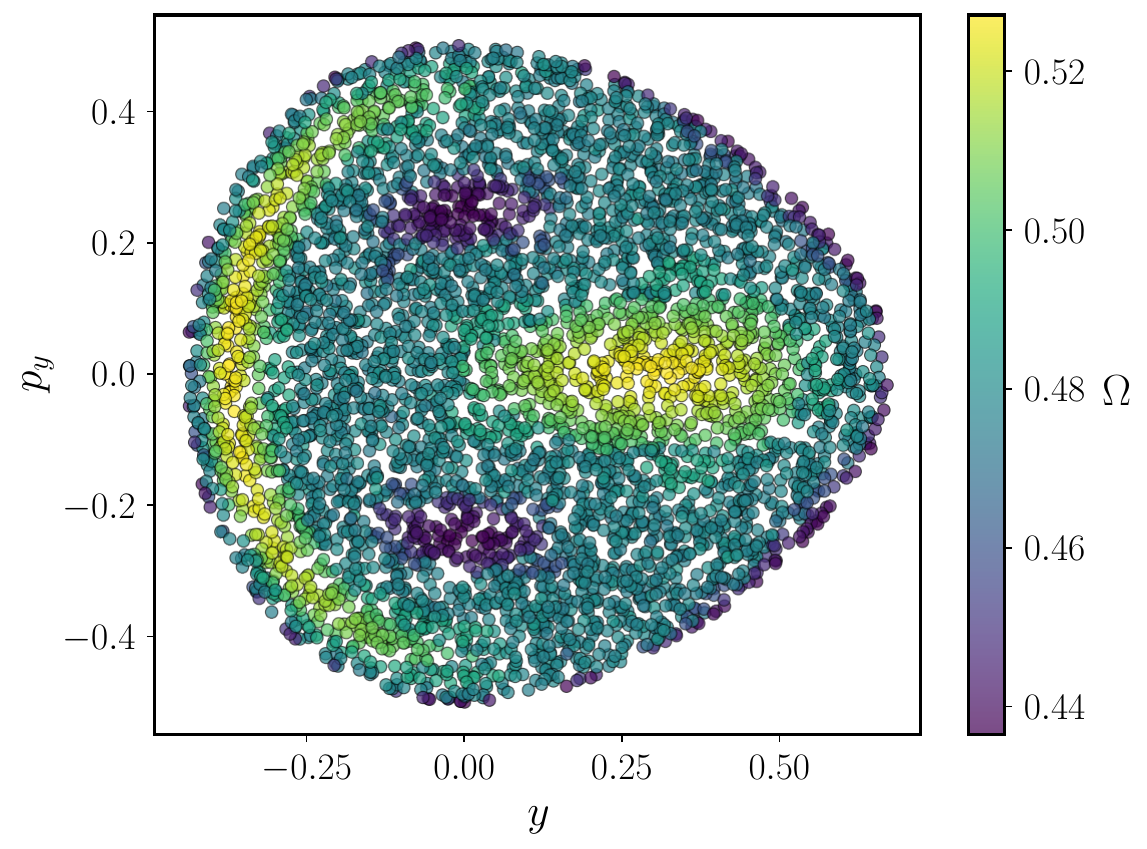} 
    \caption{Illustration of Birkhoff's Ergodic Partition Theorem using $\Omega$ for the H\'enon-Heiles system at energy $H = 1/8$. A) Time evolution of $\Omega$ for an ensemble of $5000$ trajectories. B) Value of $\Omega$ for the same ensemble of initial conditions depicted in the Poincar\'e section given by $x = 0$ and $p_x > 0$ (shown in Fig.~\ref{fig:Poincare_sec_HH}). Panel A) shows that the chaotic initial conditions (according to the criteria given for $\Omega$) are confined between two regimes of regular trajectories. This is consistent with Birkhoff's Ergodic Partition Theorem for systems with two DoF. Furthermore, the value of $\Omega$ effectively identifies the different structures present in the system's phase space, as it can be seen in panel B).}
    \label{fig:Birkhoff}
\end{figure}

\begin{figure}[htbp]
    \centering
    \includegraphics[scale = 0.45]{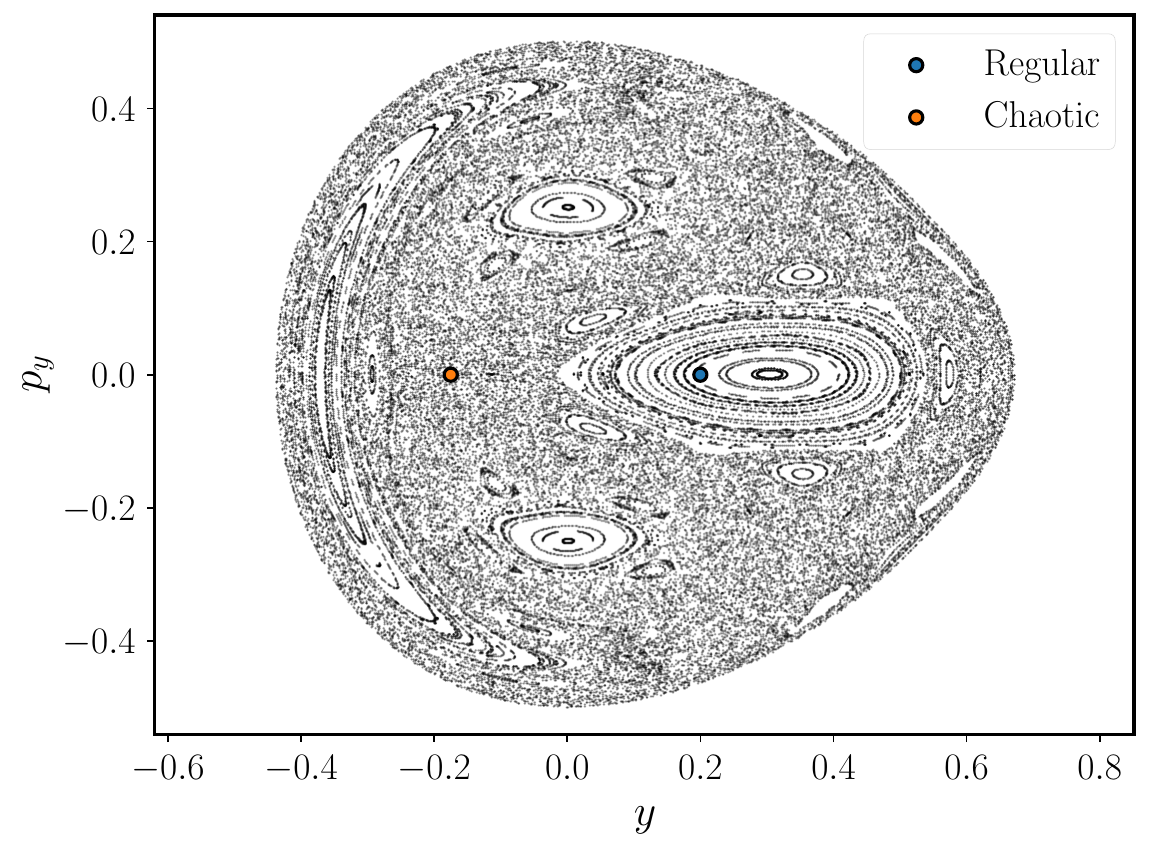}
    \caption{Poincar\'e section for the H\'enon-Heiles system defined by $x = 0$ and $p_x > 0$ calculated at energy $H = 1/8$. A regular (blue) and a chaotic (orange) initial condition are superimposed, illustrating the mixed phase space structure that is present this system.}
    \label{fig:Poincare_sec_HH}
\end{figure}

Fig.~\ref{fig:Birkhoff} A) shows the time evolution of $\Omega$ for an ensemble of $5000$, randomly sampled, initial conditions, while in Fig.~\ref{fig:Birkhoff} B) we display the corresponding values on the Poincar\'e section (defined by $x = 0$ and $p_x > 0$). In it, we see that the band-like organization present in the time evolution of $\Omega$ directly maps into the different phase space structures. A comparison with the corresponding Poincar\'e section, shown in Fig.~\ref{fig:Poincare_sec_HH}, confirms that the time averaged arclenght retains useful information about the underlying phase space organization. This is consistent with the common limiting average expected within an ergodic component, although a single scalar average might not uniquely identify every invariant component. Therefore, the classification uses the finite time fluctuations of $\Omega$ rather than the limiting value its time evolution achieves.

\begin{figure}[htbp]
    \centering
    A) \includegraphics[scale = 0.52]{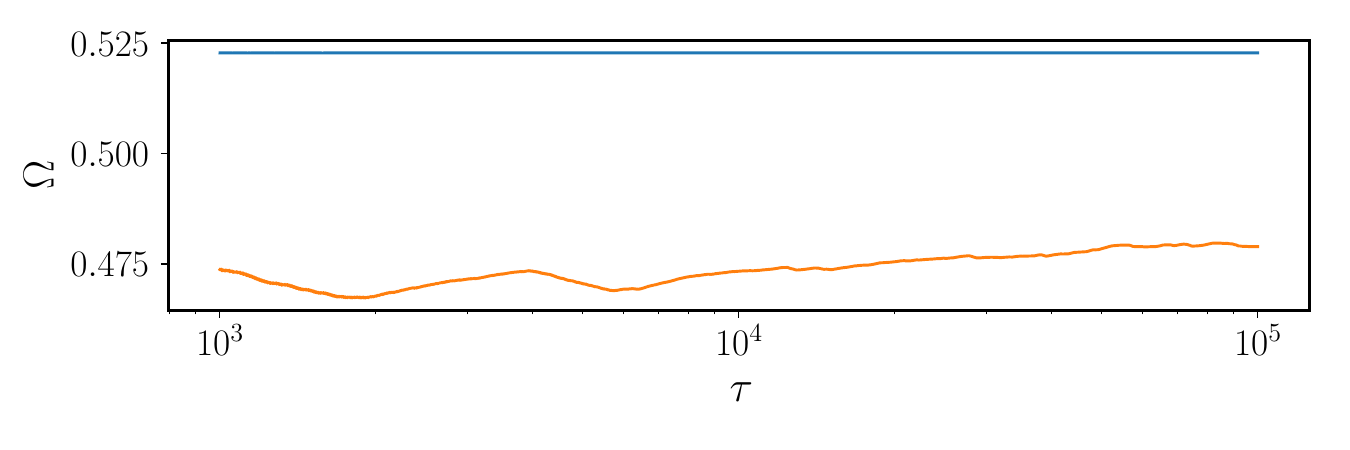} \\
    B) \includegraphics[scale = 0.3]{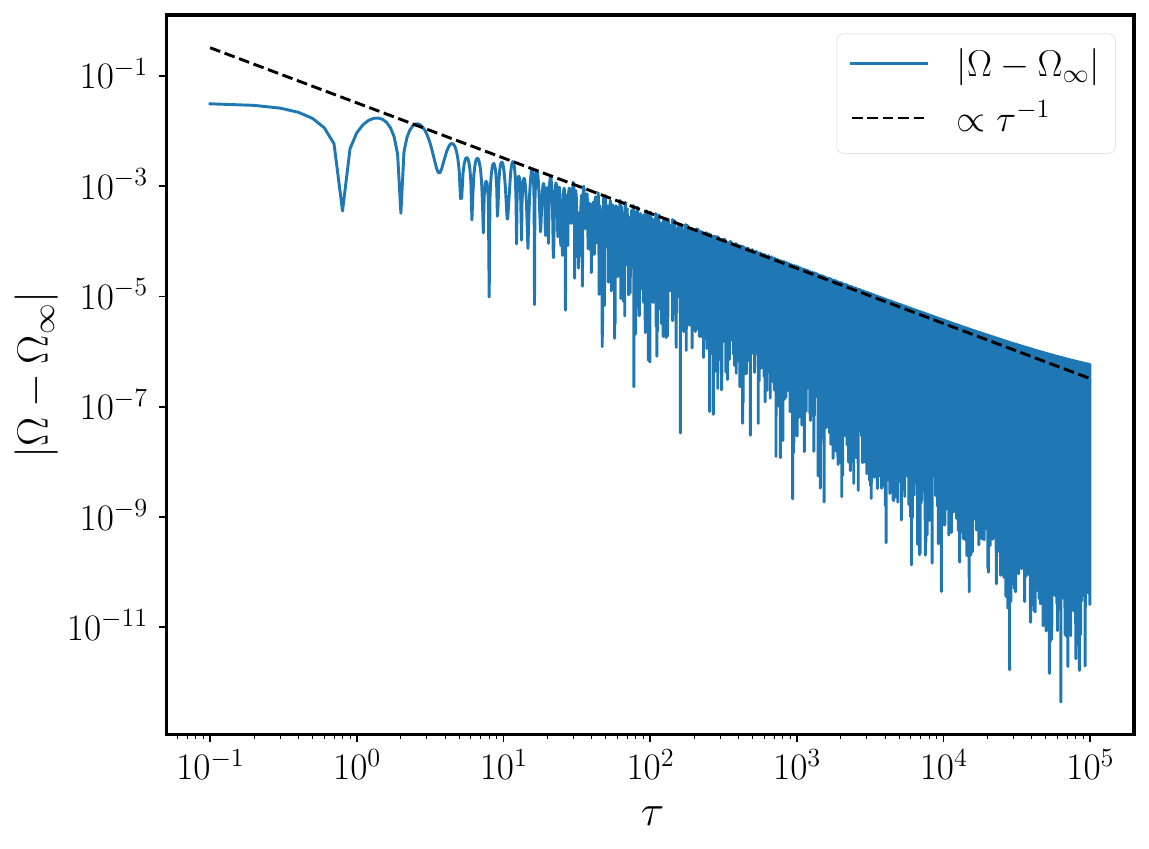}
    C) \includegraphics[scale = 0.3]{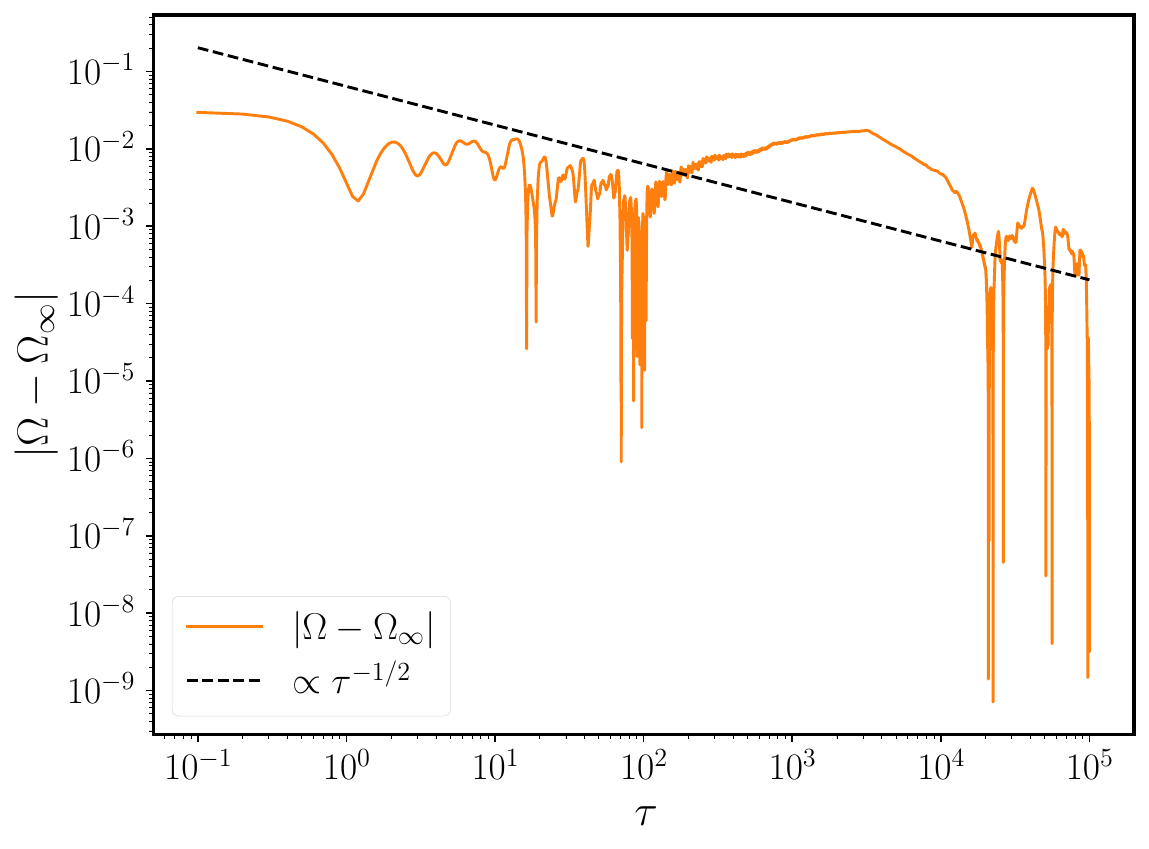}
    \caption{Time evolution of $\Omega$ in the H\'enon-Heiles system. A) Evolution of the time-averaged Lagrangian descriptor $\Omega$ defined in Eq.~\eqref{eq:omega_def}, for a regular (blue) and a chaotic (orange) initial condition in the H\'enon-Heiles system. B) Evolution of $|\Omega - \Omega_{\infty}|$ for a regular trajectory, in perfect agreement with the prediction of Theorem~\ref{thm:regular_omega}. C) Analogous results for a chaotic initial condition, were in this case there is a significant deviation from the $\tau^{-1/2}$ convergence rate derived in Theorem~\ref{thm:chaotic_omega} due to the particularities of the H\'enon-Heiles system. This deviation can be explained by referring to the phenomenon of dynamical sickness, which is present in this trajectory, and illustrated in Fig.~\ref{fig:sticky_1}. Note that the initial conditions are the ones displayed in Fig.~\ref{fig:Poincare_sec_HH}.}
    \label{fig:Omega_evol}
\end{figure}

We now consider two representative initial conditions generated on the Poincar\'e section defined by $x = 0$ and $p_x > 0$ at a total energy $H = 1/8$: one corresponding to a regular trajectory (blue) and the other one associated to a chaotic orbit (orange). These initial conditions are superimposed on the section shown in Fig.~\ref{fig:Poincare_sec_HH}. In Fig.~\ref{fig:Omega_evol} A) we display the time evolution of the time-averaged LD, $\Omega$, for both trajectories. For the regular trajectory (blue), $\Omega$ exhibits an almost flat evolution in time, which is consistent with the decaying of its fluctuations, shown in Fig.~\ref{fig:Omega_evol} B), where we observe the predicted $\tau^{-1}$ behavior stated in Theorem~\ref{thm:regular_omega}. On the other hand, the chaotic initial condition (orange) shows a very different behavior. In Fig.~\ref{fig:Omega_evol} A), the time evolution of $\Omega$ exhibit considerable fluctuations, whose time evolution is presented in Fig.~\ref{fig:Omega_evol} C). The prediction from Theorem~\ref{thm:chaotic_omega} is that the fluctuations for a chaotic initial condition, under the assumptions stated in it, should be $\tau^{-1/2}$. However, this is not what is observed for this particular initial condition. 

The reason behind this discrepancy between the prediction from the theorem and the observation arises from dynamical stickiness \cite{PhysRevLett.55.2741}, a phenomenon typical of systems with mixed phase space, in which regular and chaotic regions coexist. Stickiness is a rich and complex dynamical phenomenon with many well-documented consequences, including memory effects and anomalous transport \cite{Zaslavsky2002}, a characteristic bimodal distribution of finite-time Lyapunov exponents \cite{Szezech2005, daSilva2015}, alignment of covariant Lyapunov vectors near regular boundaries \cite{Ginelli2007}, and algebraic decay of time correlations \cite{Chirikov1999, Manchein2019}. During its exploration of phase space, a chaotic trajectory may pass near a regular region and exhibit quasi-regular motion for finite intervals of time. This intermittent dynamics produces long-term temporal correlations along the trajectory, distinguishing it from a purely random or uniformly hyperbolic process \cite{PhysRevLett.55.2741,10.1143/PTPS.98.36}. 

\begin{figure}[htbp]
    \centering
    A) \includegraphics[scale = 0.48]{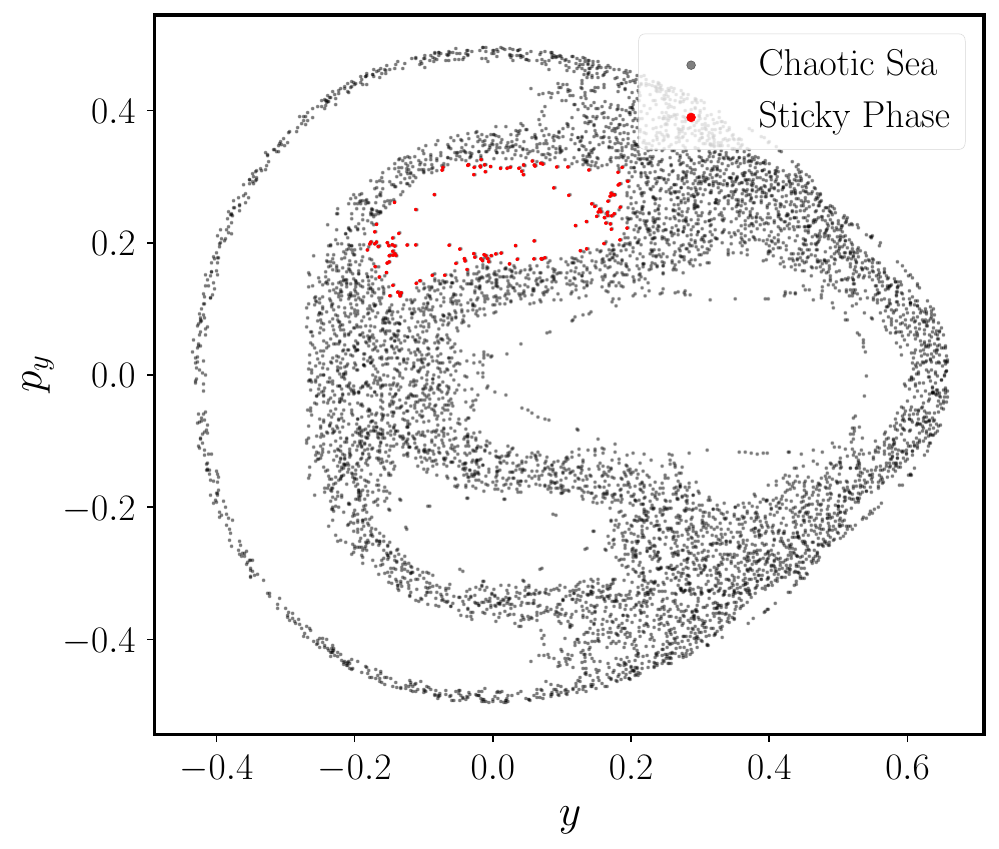} \\
    B) \includegraphics[scale = 0.33]{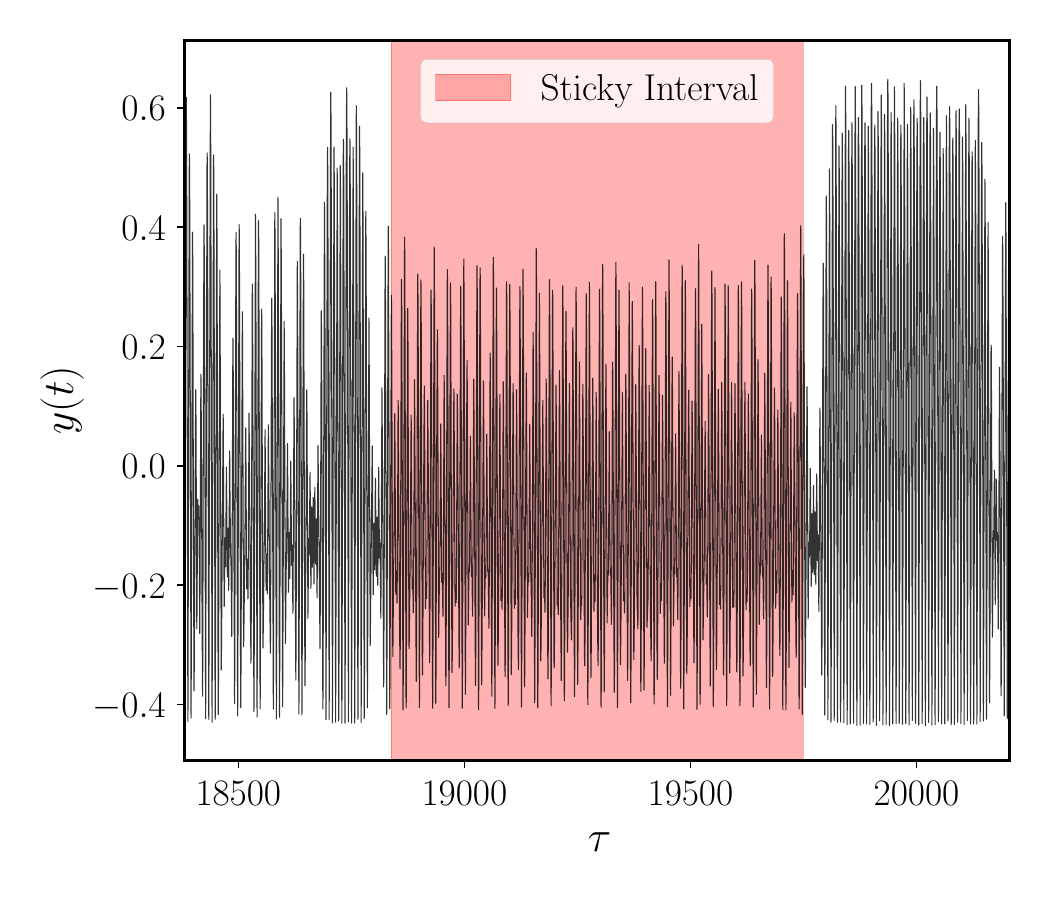} 
    C) \includegraphics[scale = 0.33]{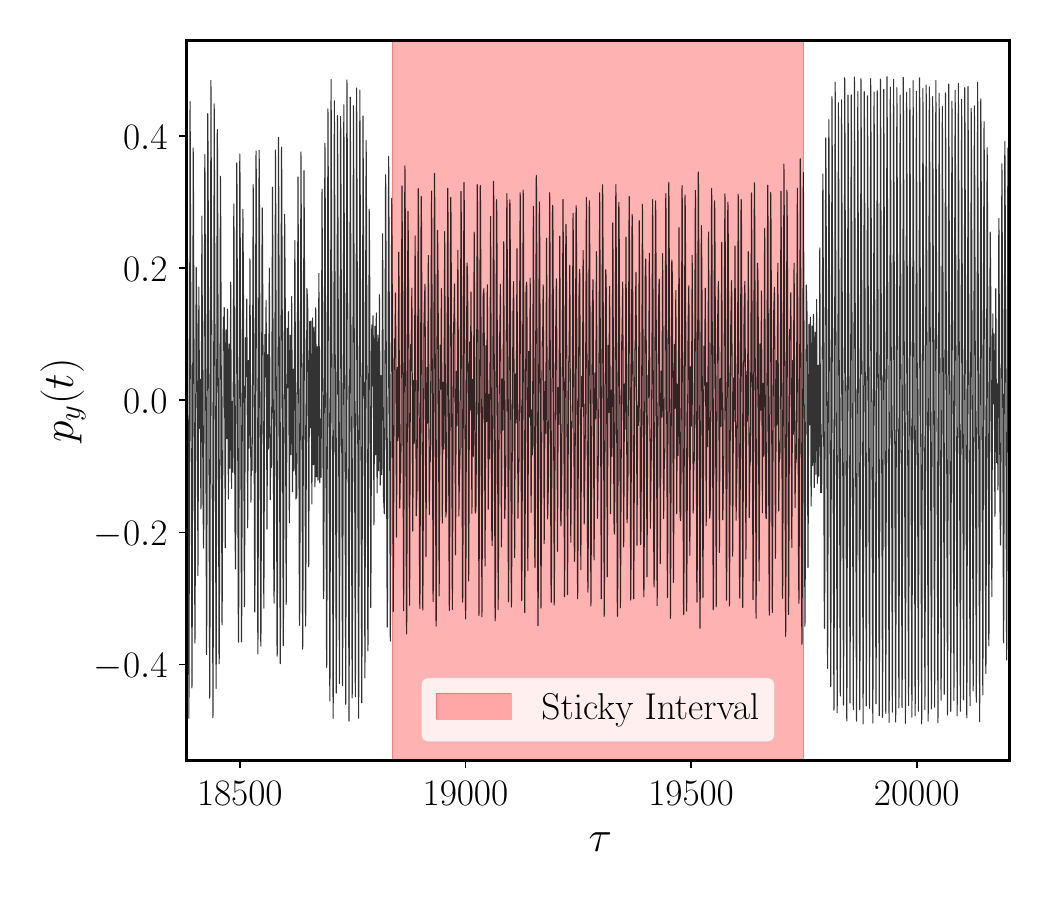}
    \caption{Dynamical stickiness in the H\'enon-Heiles system. A) Intersections of a chaotic trajectory with the Poincar\'e section defined by $x=0$ and $p_x>0$. Gray points correspond to the chaotic sea, while red points indicate the sticky regime, in which the trajectory lingers near the boundary of a regular island. B) and C) Time series of $y(t)$ and $p_y(t)$, respectively. The pink shaded region marks the sticky interval, which can be compared with the result in Fig.~\ref{fig:Omega_evol} C) to understand why the fluctuations of $\Omega$ do not decay as $\tau^{-1/2}$. Note that the initial condition is the one displayed in Fig.~\ref{fig:Poincare_sec_HH} (orange).}
    \label{fig:sticky_1}
\end{figure}

In this particular case, looking at the Poincar\'e section in Fig.~\ref{fig:Poincare_sec_HH}, it is clear that the H\'enon-Heiles system has a highly mixed phase space, and thus, dynamical stickiness is expected to be present in it. In fact, it is, and for the chaotic initial condition we are considering under analysis, Fig.~\ref{fig:sticky_1} depicts one sticky interval that the trajectory undergoes during its evolution. As the correlation in this case becomes algebraic, it can be easily derived from Theorem~\ref{thm:chaotic_omega} that the decay of the fluctuations in such scenario will follow the trend $\tau^{-\gamma / 2}$, where $\gamma \in (0,1)$, implying a slower decay than the $\tau^{-1/2}$ trend derived for the case of integrable correlations. This, however, is not problematic, because in this case the convergence can be slower than $\tau^{-1/2}$ but not faster, meaning that, in principle and for $\tau \rightarrow \infty$, there will still be a clear separation between the regular and chaotic trajectories in terms of the convergence rate of $\Omega$. 

As the purpose of deriving efficient chaos indicators is to classify large ensembles of initial conditions, we will consider four sets of $25000$, randomly sampled, initial conditions for the H\'enon-Heiles system on the Poincar\'e section defined by $x = 0$ and $p_x > 0$. Each set corresponds to a different energy value, with the values considered being $H = 1/15$, $1/12$, $1/10$ and $1/8$. Each initial condition is integrated for $\tau = 10^5$ time units using the Dormand-Prince $8(9)$ ODE solver \cite{prince1981} implemented in the PYTHON package SciPy \cite{Virtanen:2019joe}, which as previously mentioned, provides a good enough energy conservation for our analysis.

\begin{figure}[htbp]
    \centering
    A) \includegraphics[scale = 0.3]{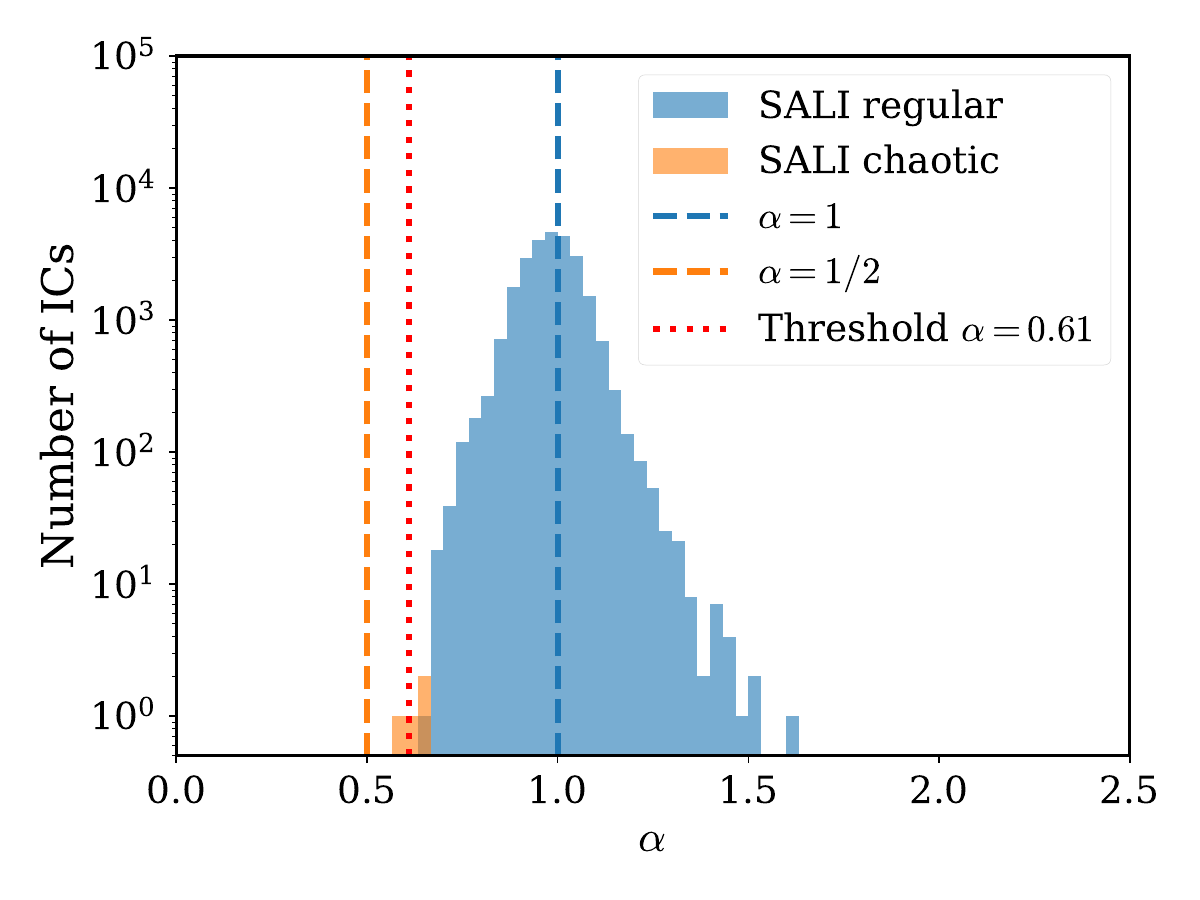} 
    B) \includegraphics[scale = 0.3]{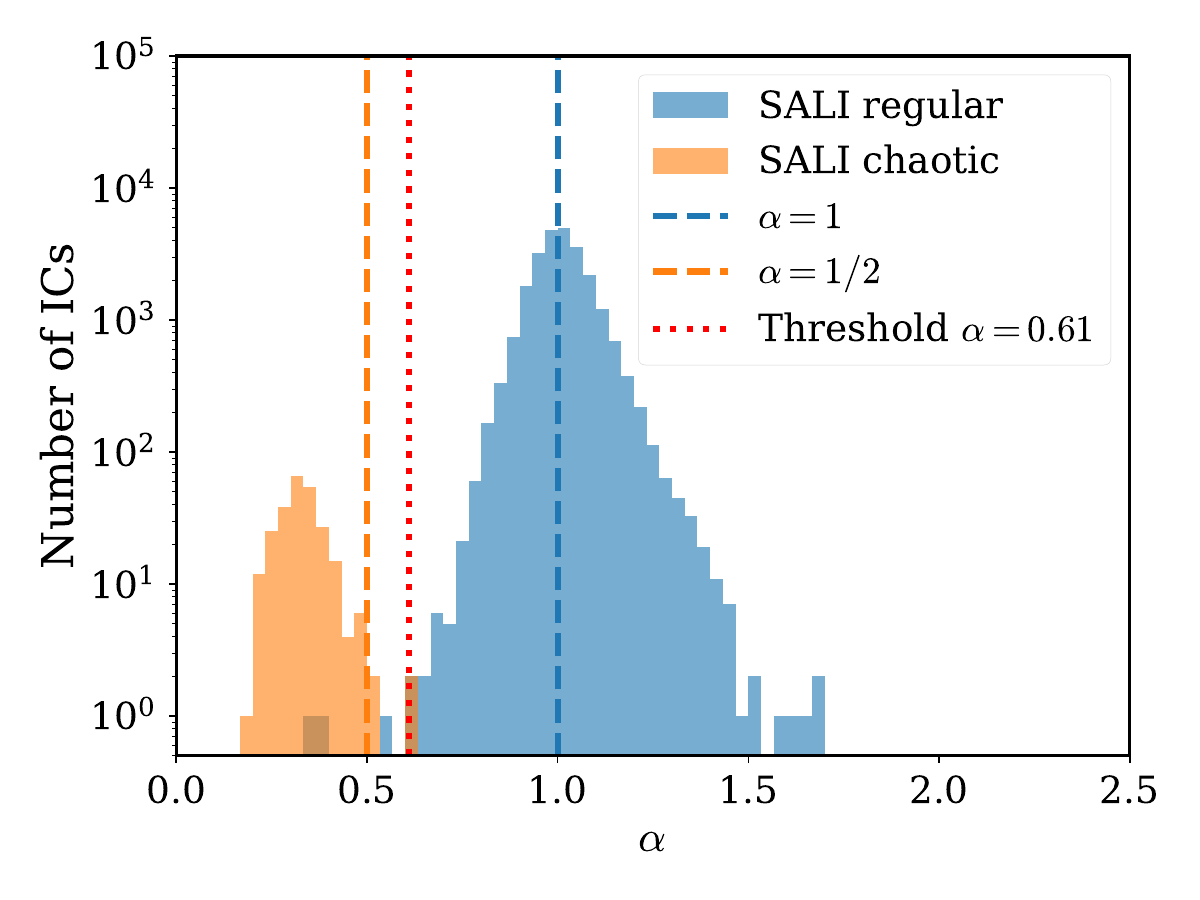} \\
    C) \includegraphics[scale = 0.3]{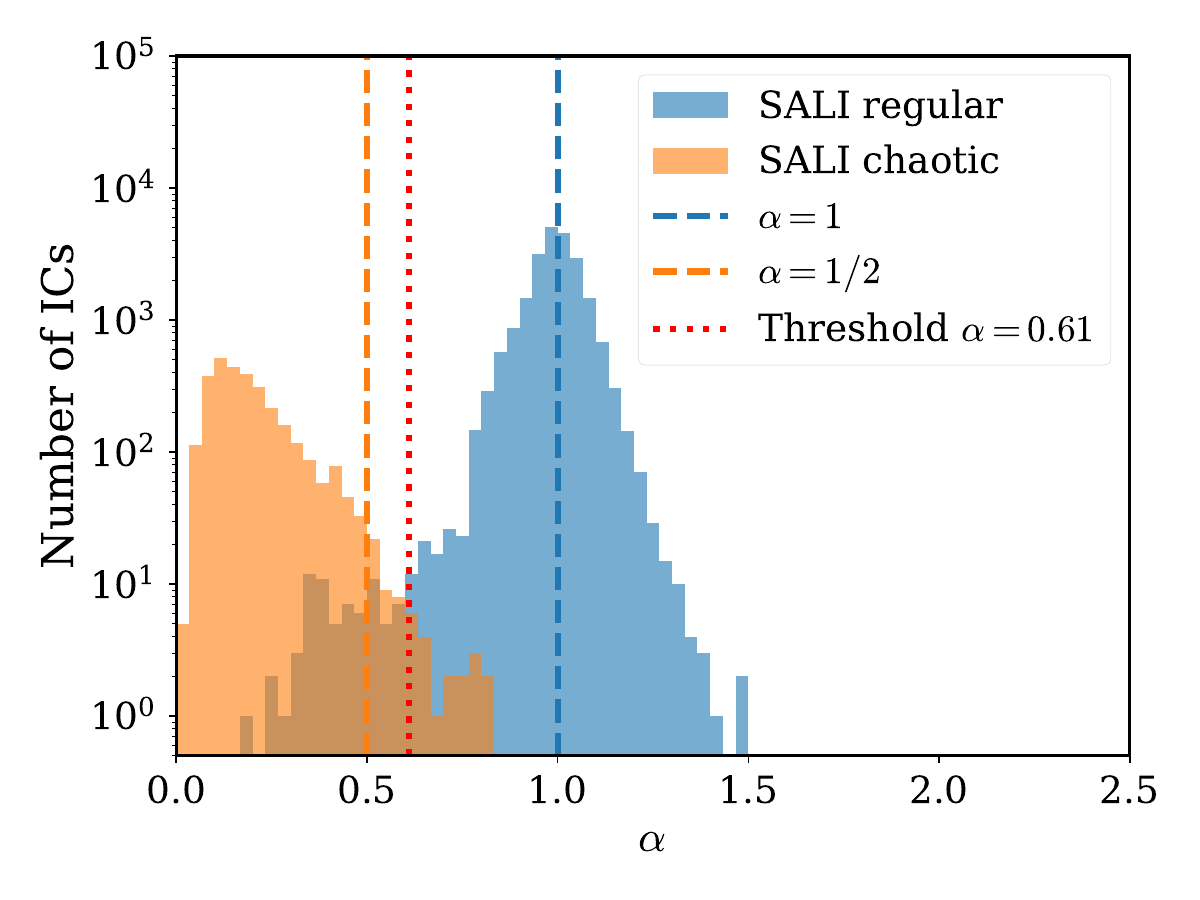} 
    D) \includegraphics[scale = 0.3]{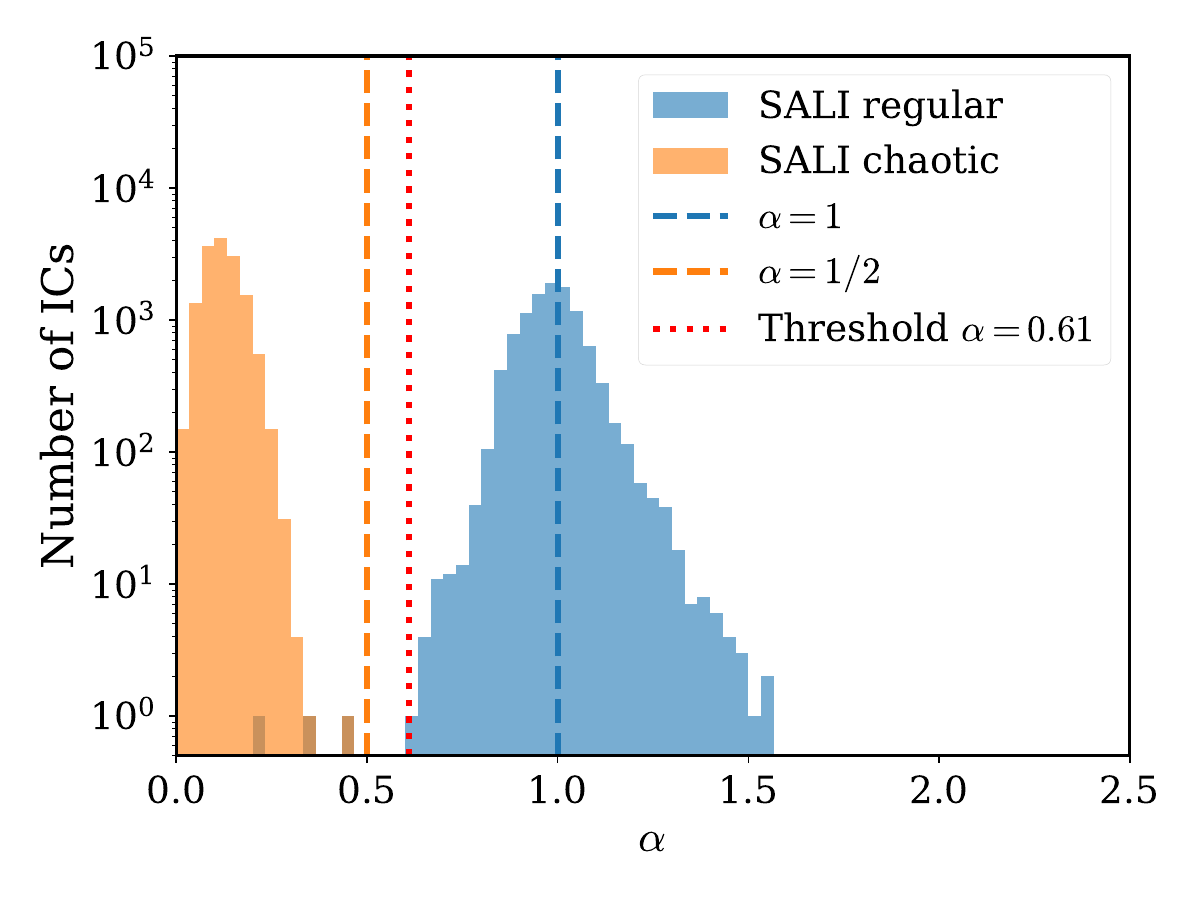}
    \caption{Histograms of the block-fluctuation exponent $\alpha$ for different energies in the H\'enon-Heiles system. These figures display the distributions of $\alpha$ for $25000$ initial conditions, randomly sampled on the Poincar\'e section defined by $x = 0$ and $p_x > 0$, for the energies $H = 1/15$ (A), $H = 1/12$ (B), $H = 1/10$ (C) and $H = 1/8$ (D). The threshold is determined by the maximization of the MCC as described at the end of Sec.~\ref{sec:theory_class}. The blue and orange vertical lines respectively indicate the theoretical prediction for regular and chaotic populations. A detailed analysis of the incorrectly classified initial conditions is given in App.~\ref{sec:miss}.}
    \label{fig:alpha_hists}
\end{figure}

\begin{figure}[htbp]
    \centering
    A) \includegraphics[scale = 0.3]{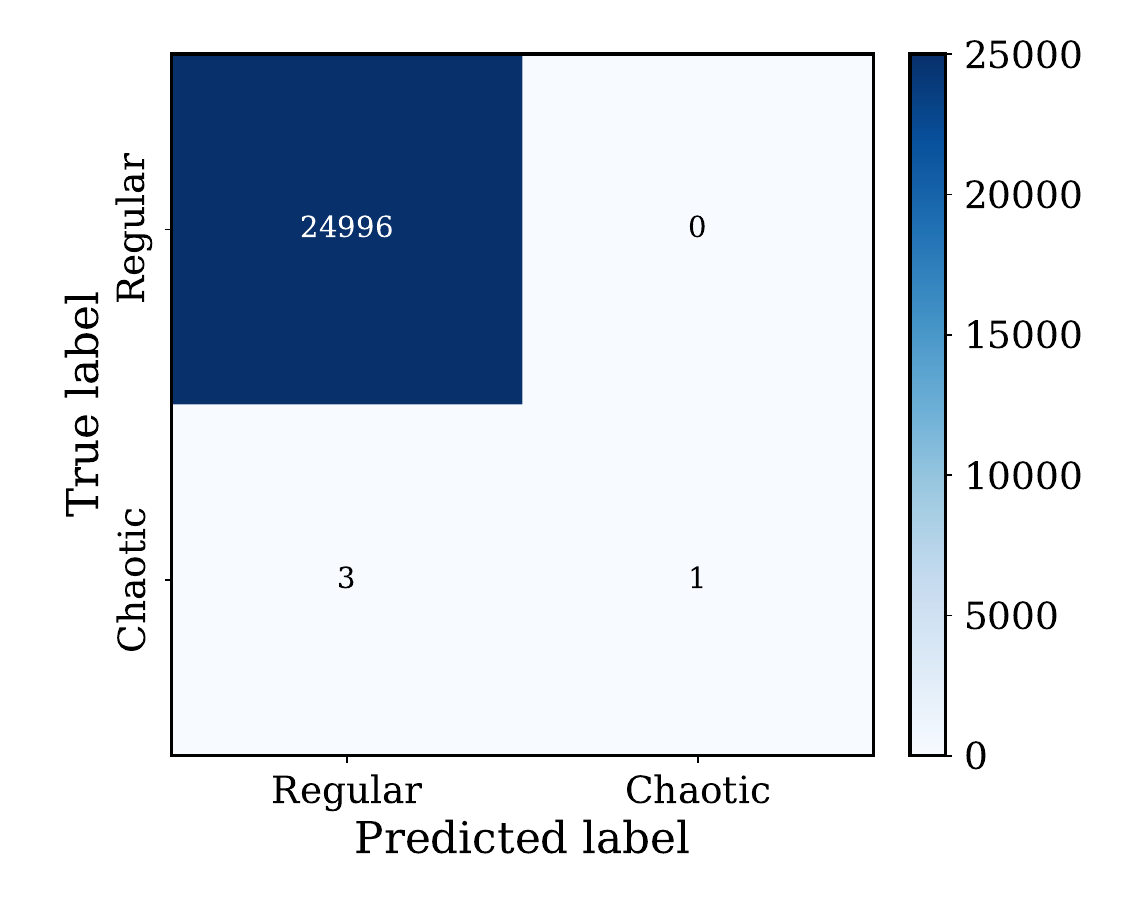} 
    B) \includegraphics[scale = 0.3]{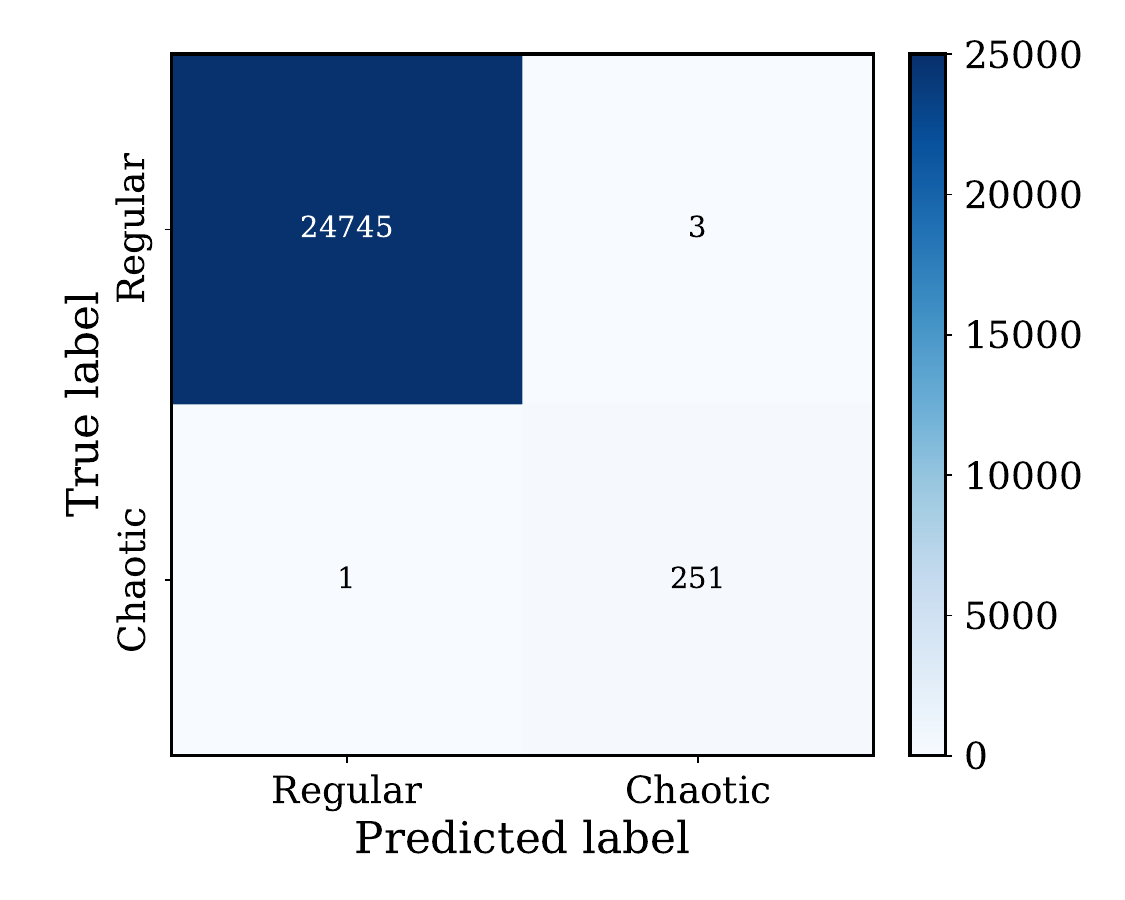} \\
    C) \includegraphics[scale = 0.3]{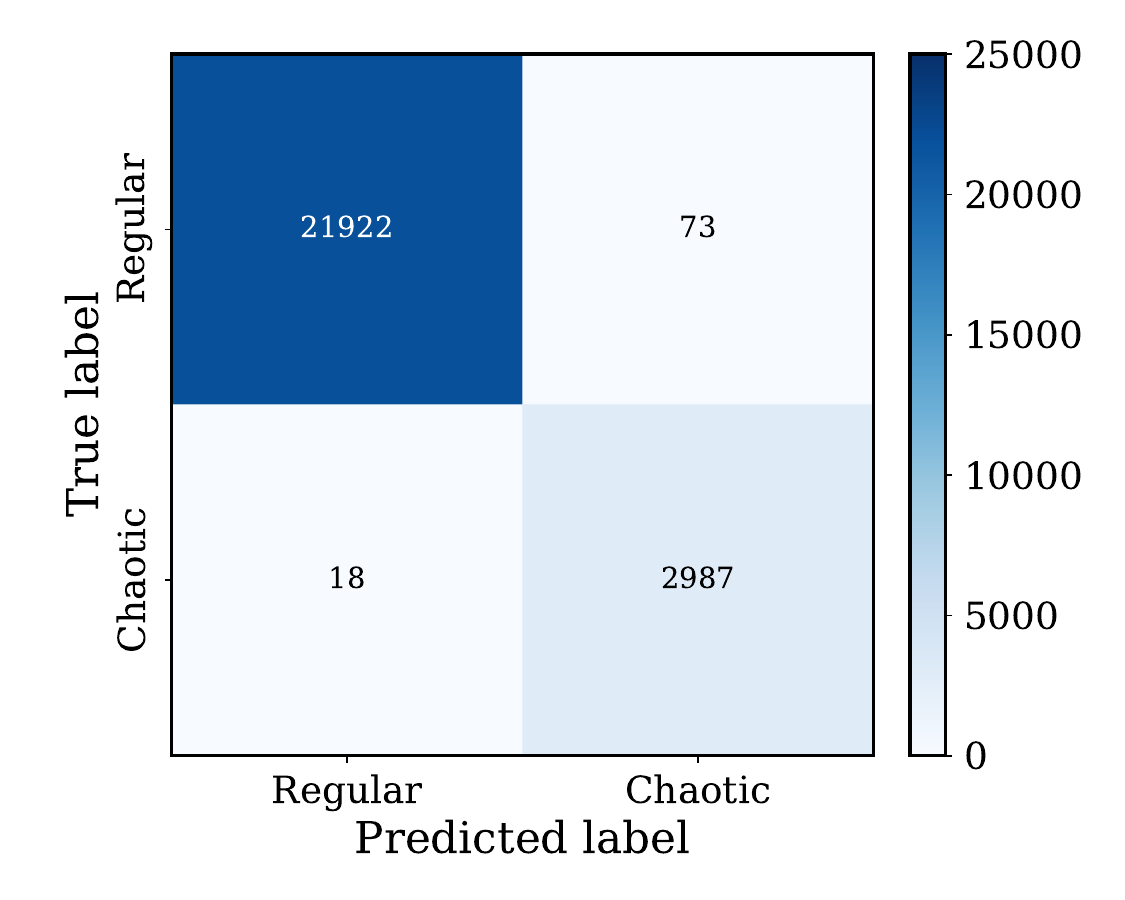} 
    D) \includegraphics[scale = 0.3]{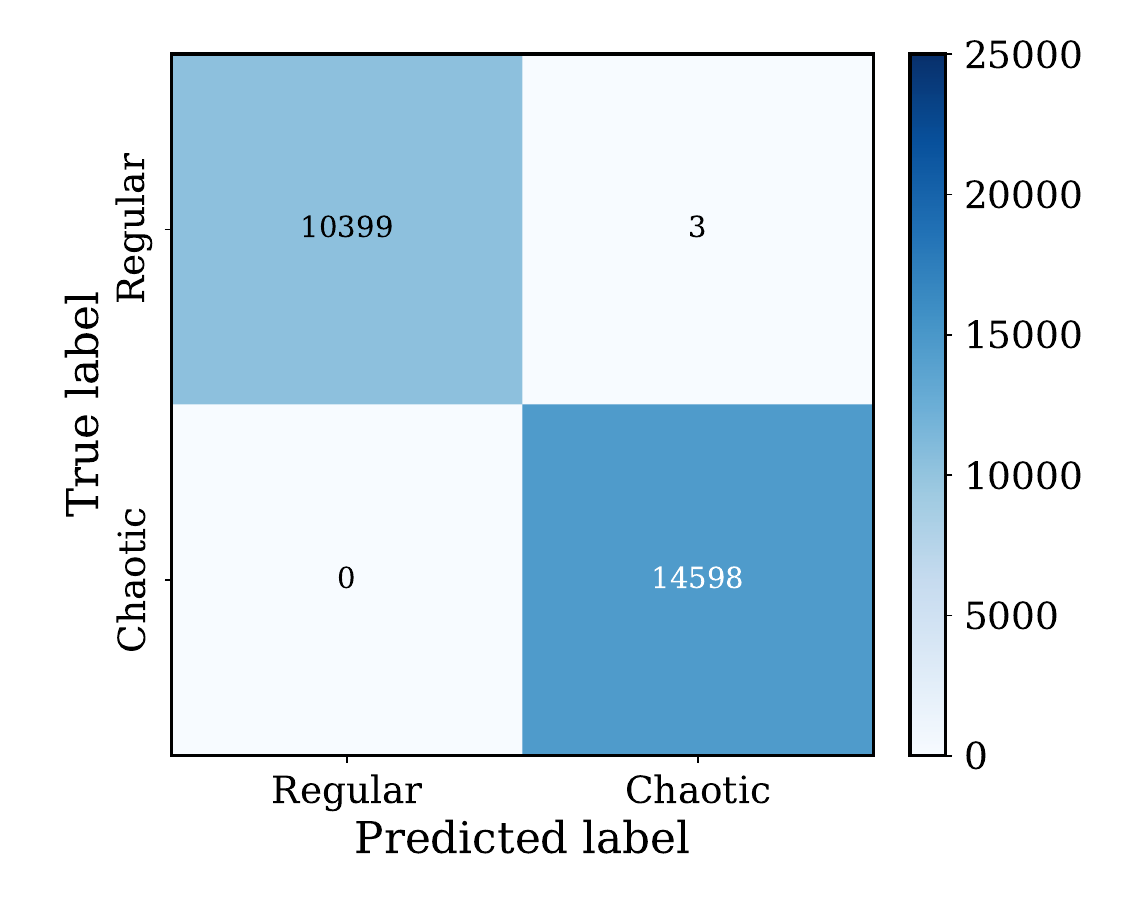}
    \caption{Confusion matrices for the classification conducted with the block-fluctuation exponent $\alpha$ for different energies in the H\'enon-Heiles system, correspnding to the histograms in Fig.~\ref{fig:alpha_hists}. A total of $25000$ initial conditions, randomly sampled on the Poincar\'e section defined by $x = 0$ and $p_x > 0$, were considered for each energy value. The different energies are: $H = 1/15$ (A), $H = 1/12$ (B), $H = 1/10$ (C) and $H = 1/8$ (D). A detailed analysis of the incorrectly classified initial conditions is given in App.~\ref{sec:miss}.}
    \label{fig:alpha_cm}
\end{figure}

\begin{figure}[htbp]
    \centering
    A) \includegraphics[scale = 0.3]{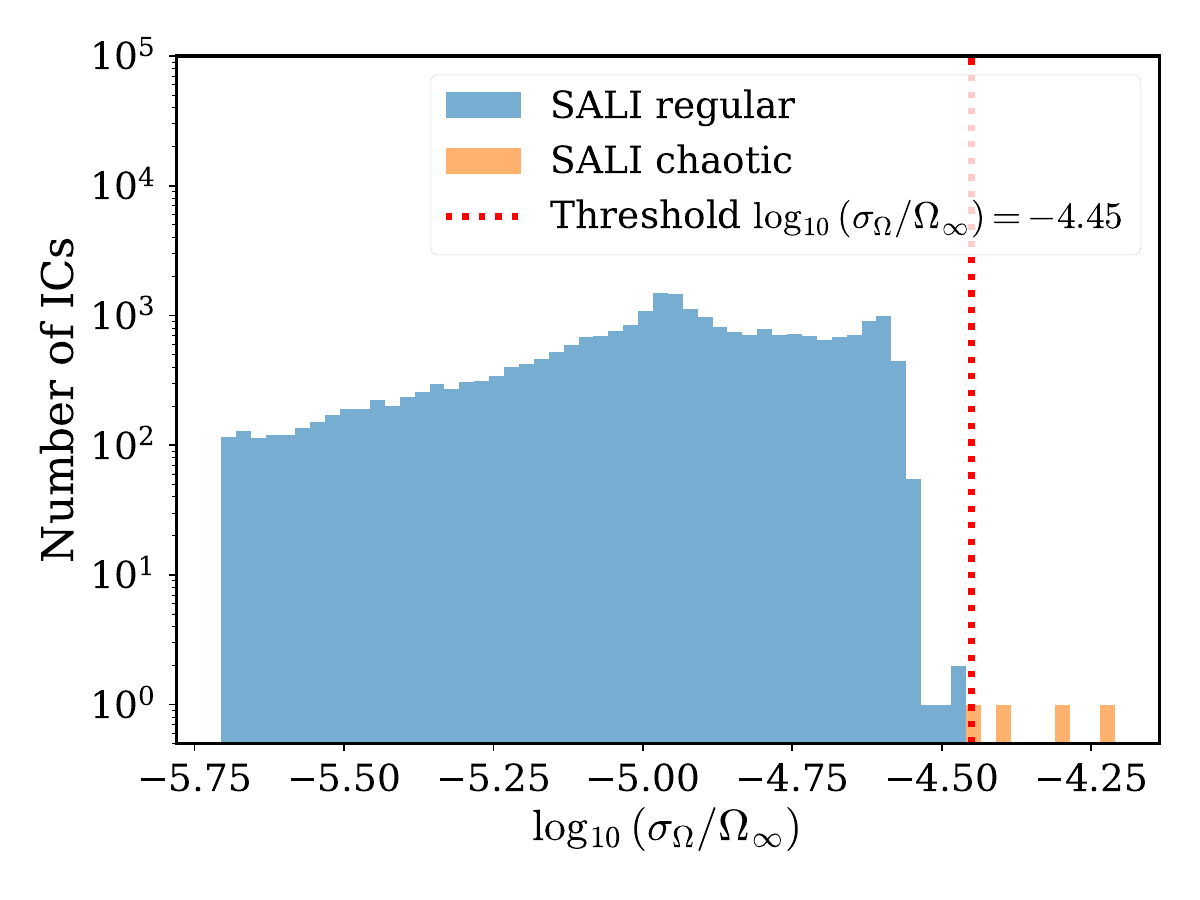} 
    B) \includegraphics[scale = 0.3]{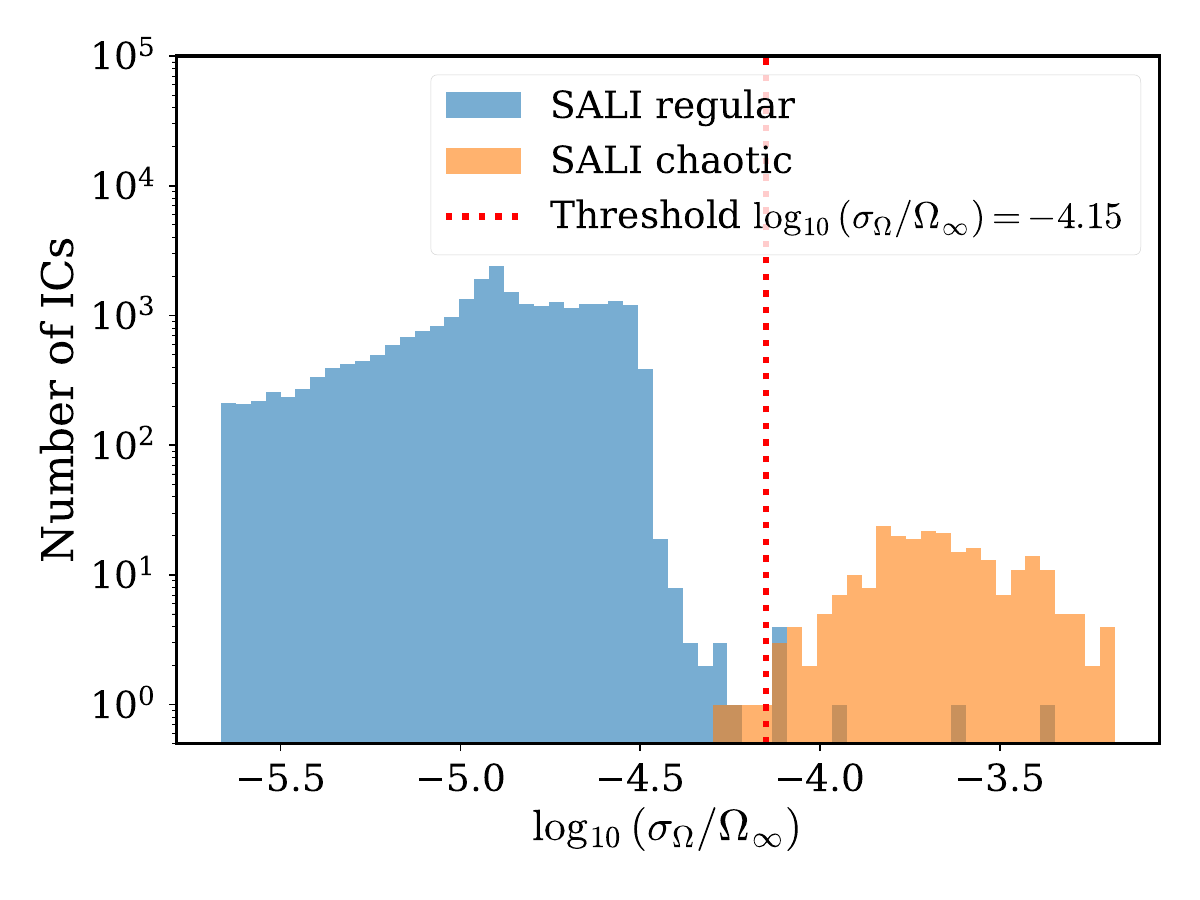} \\
    C) \includegraphics[scale = 0.3]{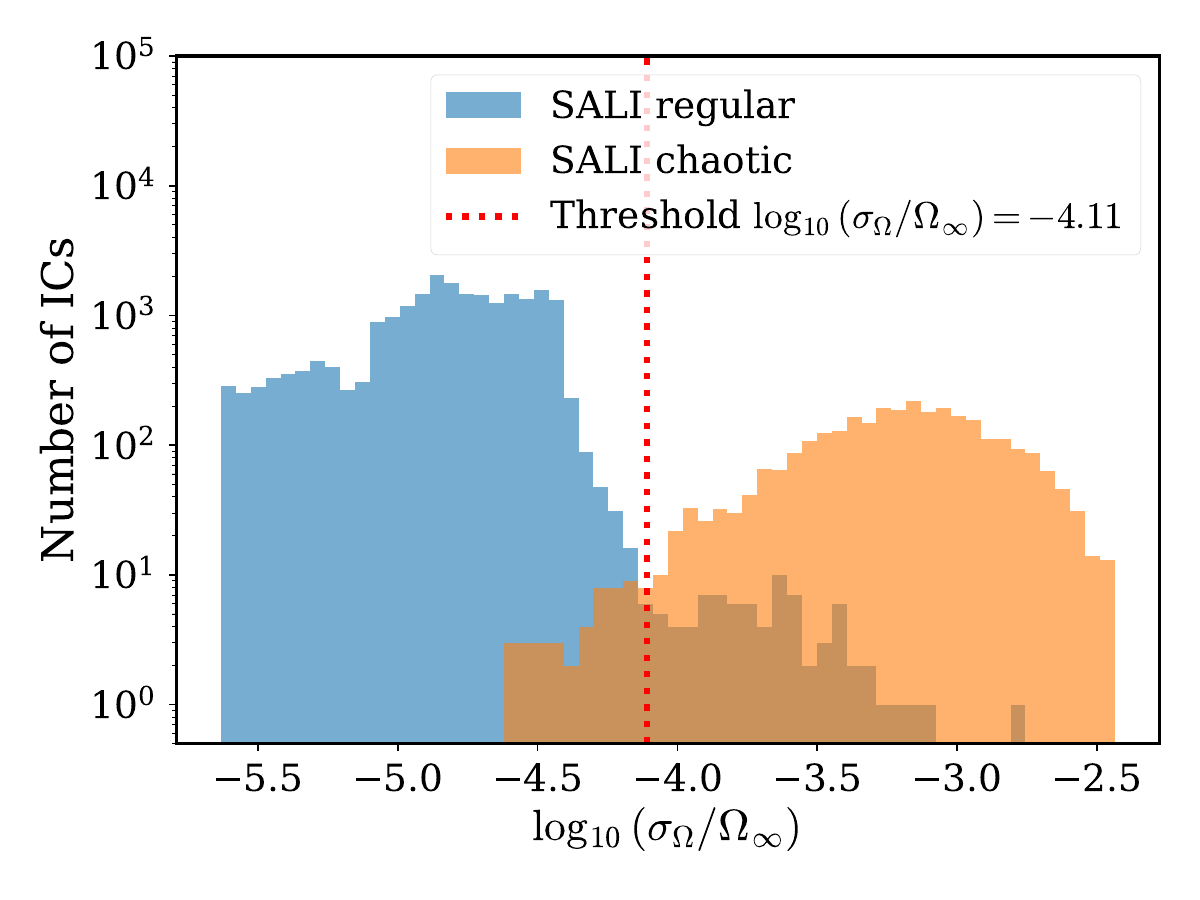} 
    D) \includegraphics[scale = 0.3]{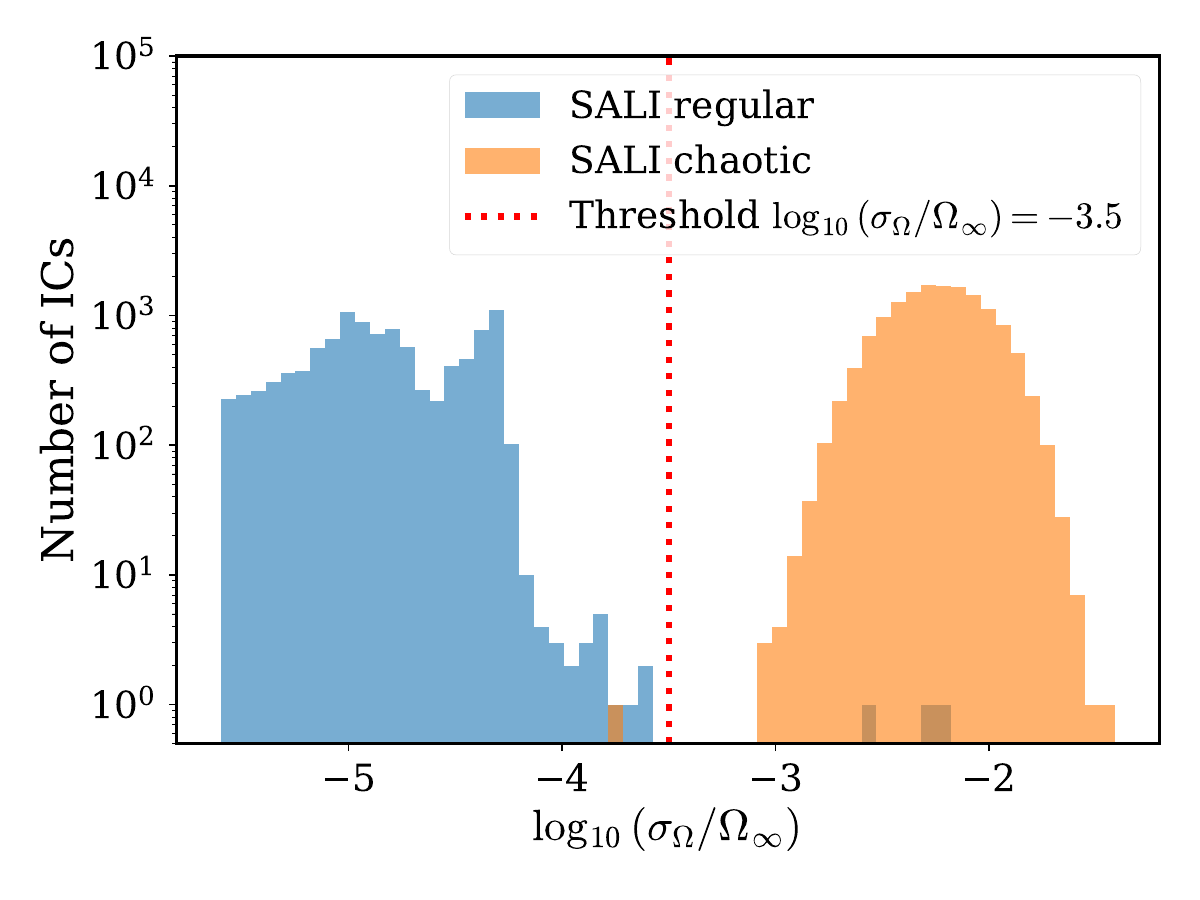}
    \caption{Histograms of normalized fluctuation amplitude $\mathrm{CV} = \sigma_{\Omega} / \Omega_{\infty}$ for different energies in the H\'enon-Heiles system. These figures display the distributions of the CV for $25000$ initial conditions, randomly sampled on the Poincar\'e section defined by $x = 0$ and $p_x > 0$, for the energies $H = 1/15$ (A), $H = 1/12$ (B), $H = 1/10$ (C) and $H = 1/8$ (D). Contrary to the histograms for the block-fluctuation exponent show on Fig.~\ref{fig:alpha_hists}, in this case, the threshold determined by the minimization of the MCC can not be used across the different energies, and thus, and independent study of each histogram has to be done in order to determine each threshold. A detailed analysis of the incorrectly classified initial conditions is given in App.~\ref{sec:miss}.}
    \label{fig:CV_hists}
\end{figure}

\begin{figure}[htbp]
    \centering
    A) \includegraphics[scale = 0.3]{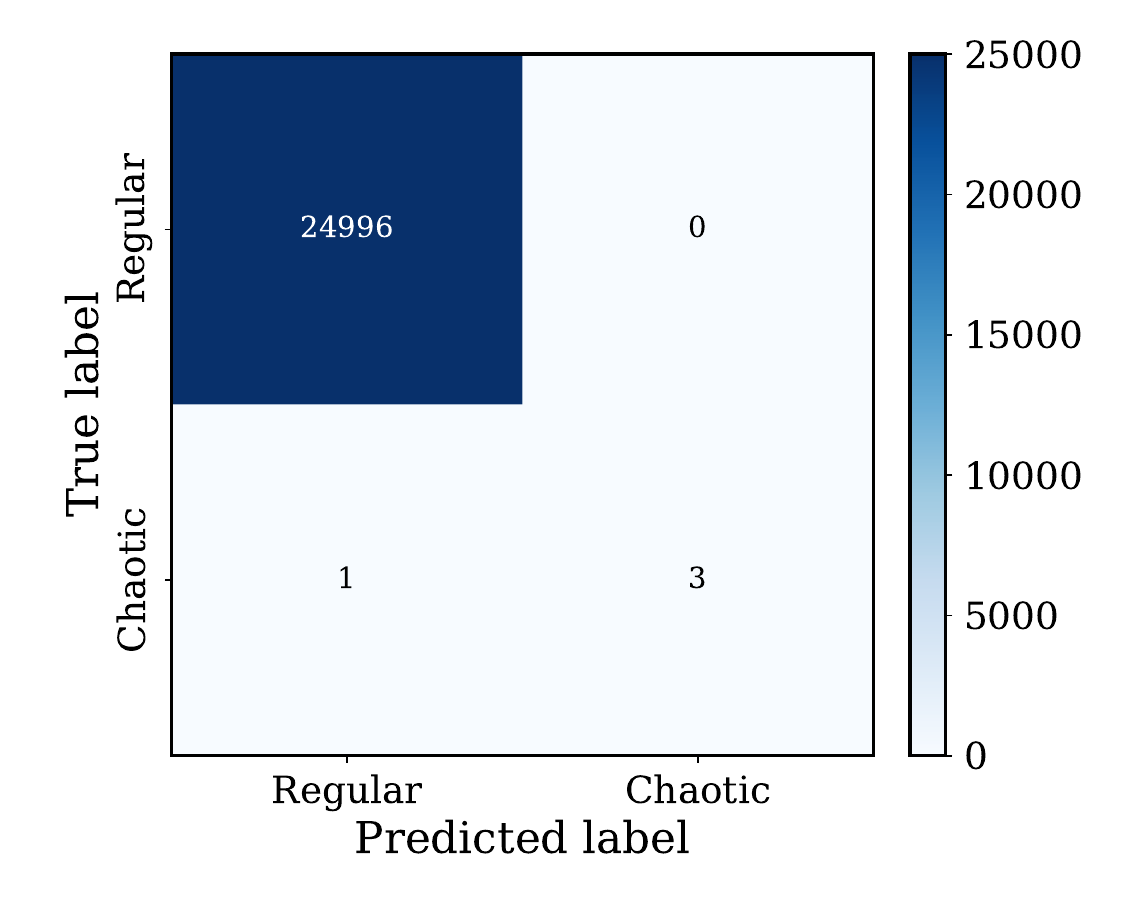} 
    B) \includegraphics[scale = 0.3]{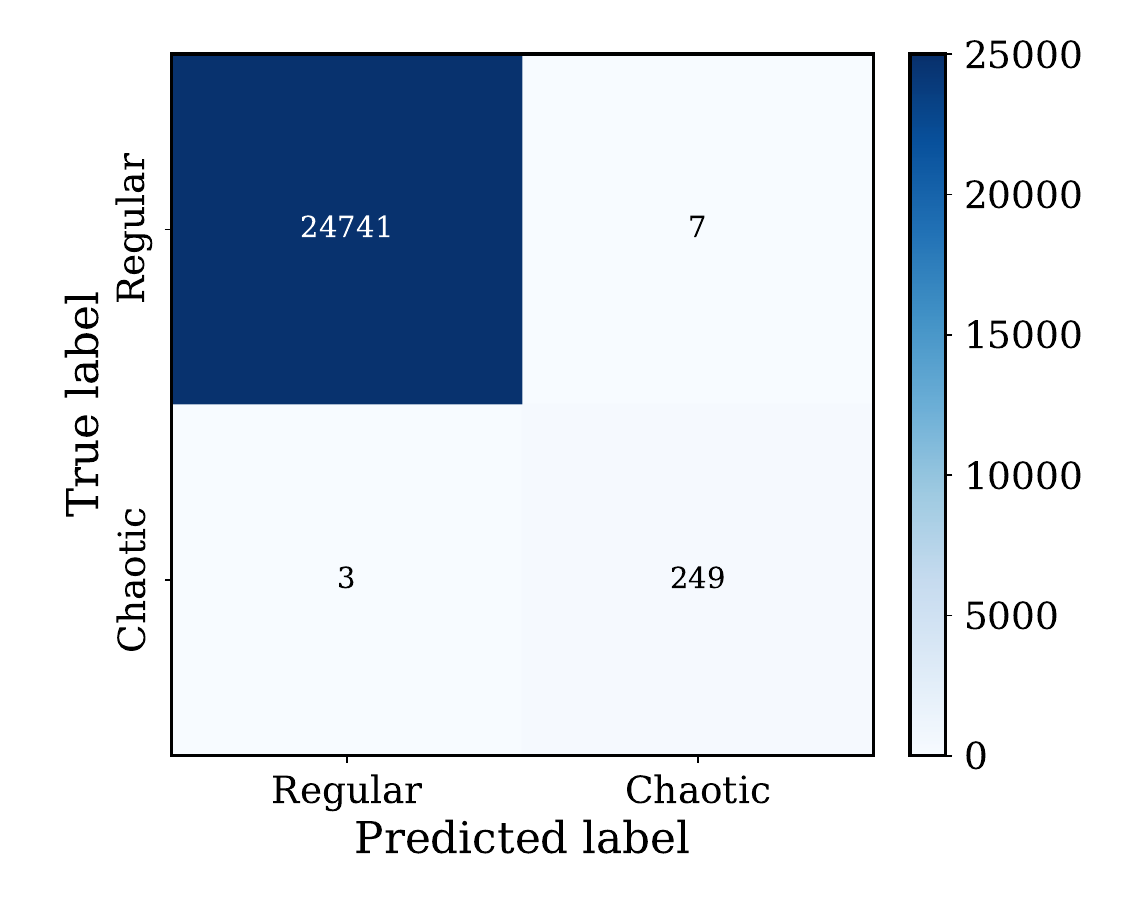} \\
    C) \includegraphics[scale = 0.3]{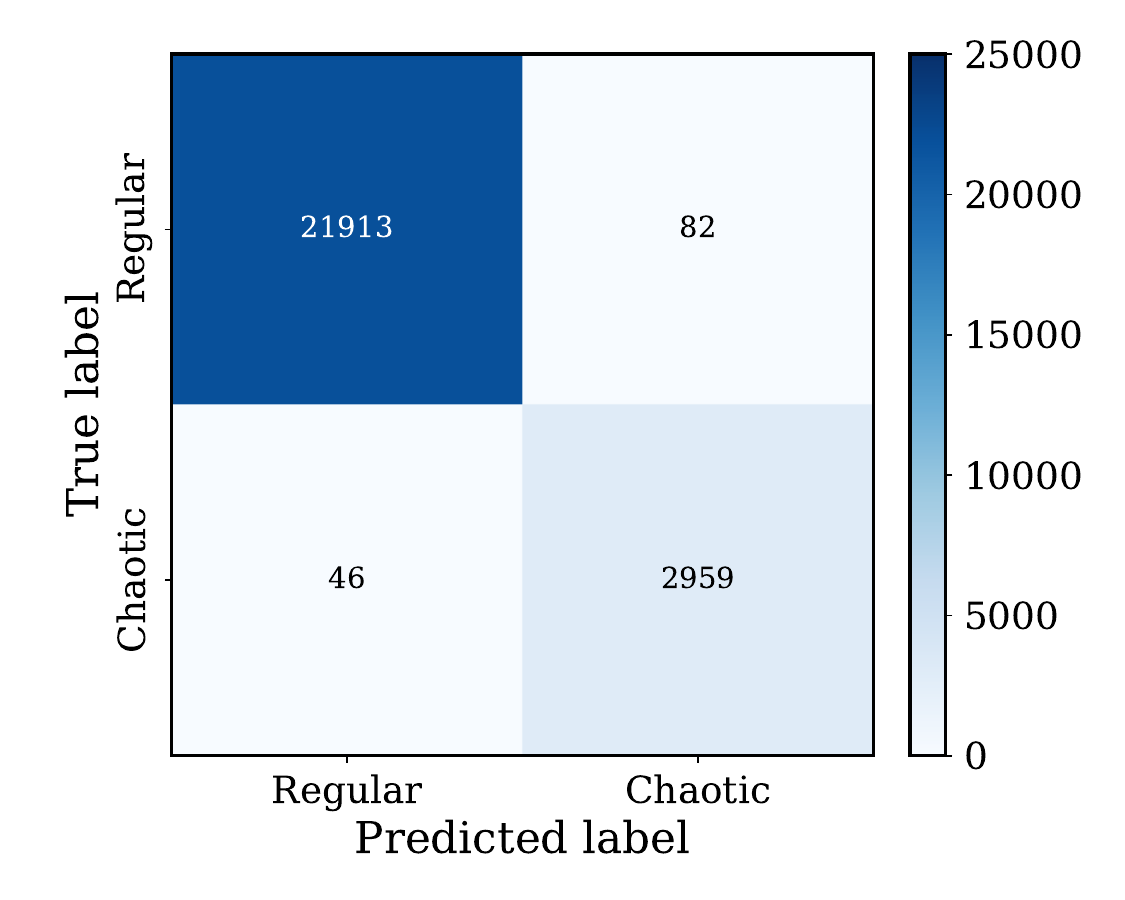} 
    D) \includegraphics[scale = 0.3]{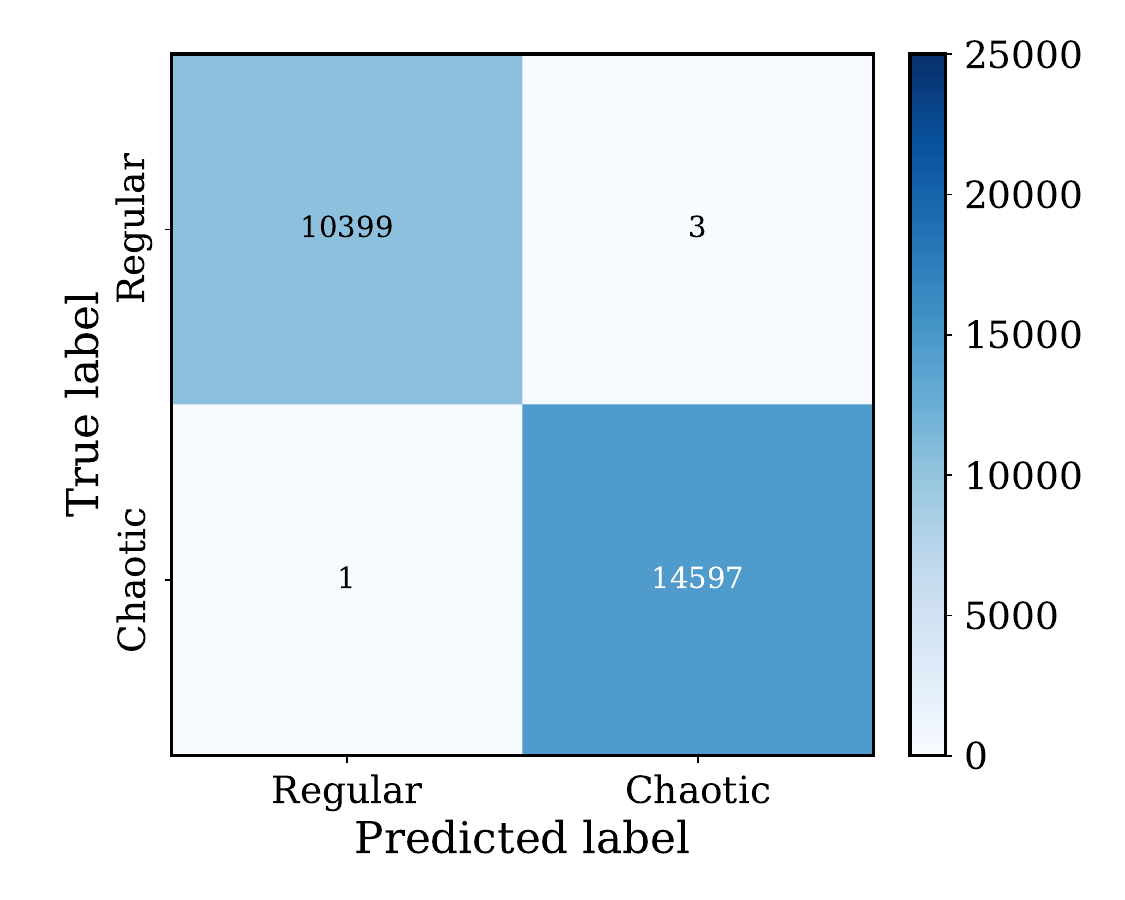}
    \caption{Confusion matrices for the classification conducted with the normalized fluctuation amplitude $\mathrm{CV} = \sigma_{\Omega} / \Omega_{\infty}$ for different energies in the H\'enon-Heiles system, corresponding to the histograms in Fig.~\ref{fig:CV_hists}. A total of $25000$ initial conditions, randomly sampled on the Poincar\'e section defined by $x = 0$ and $p_x > 0$, were considered for each energy value. The different energies are: $H = 1/15$ (A), $H = 1/12$ (B), $H = 1/10$ (C) and $H = 1/8$ (D). A detailed analysis of the incorrectly classified initial conditions is given in App.~\ref{sec:miss}.}
    \label{fig:CV_cm}
\end{figure}

In order to test the performance of the two methods for ensemble classification based on the convergence of $\Omega$, we will examine the histograms obtained for the block-fluctuation exponent (Fig.~\ref{fig:alpha_hists}) and the normalized fluctuation amplitude (Fig.~\ref{fig:CV_hists}) as well as the confusion matrices, given in Fig.\ref{fig:alpha_cm} and Fig.~\ref{fig:CV_cm}, respectively, as they provide a clear picture of the classification results. As a third metric, we consider the accuracy for each ensemble and for the two methods defined as:
\begin{equation} \label{eq:acc}
    \mathrm{Accuracy} = \dfrac{\mathrm{TP} + \mathrm{TN}}{\mathrm{TP} + \mathrm{FP} + \mathrm{FN} + \mathrm{TN}} \, ,
\end{equation}
where the different contributions to this formula are the same ones as in Eq.\eqref{eq:MCC}.

The block-fluctuation exponent distributions in Fig.~\ref{fig:alpha_hists} show a regular population concentrated near $\alpha = 1$, and a chaotic population that predominantly lies towards lower values of the exponent. These chaotic distributions are generally not centered at $\alpha = 1/2$, which is consistent with the dynamical phenomena that occur in systems with mixed a phase space. Nevertheless, the separation between the regular and chaotic populations is sufficient for the common threshold $\alpha = 0.61$, obtained by maximizing the MCC in Eq.\eqref{eq:MCC}, to yield accuracies between $99.636 \, \%$ and $99.988 \, \%$ across all four ensembles (Table.~\ref{tab:accuracy}).

The distributions of the normalized fluctuation amplitude in Fig.~\ref{fig:CV_hists} provide a complementary distinction, with chaotic trajectories exhibiting larger value, as argued in Sec.~\ref{sec:theory_class}. With the energy dependent thresholds specified in each of the histograms, this method achieves accuracies between $99.488 \, \%$ and $99.996 \, \%$. The strongest overlap between the regular and chaotic populations for both features occurs at $H = 1/10$, where the block-fluctuation exponent and CV produce $91$ and $121$ misclassifications, respectively, out of $25000$ total trajectories in the ensemble. Thus, even in the ensemble with the largest observed overlap, the fraction of disagreements with the ground truth classification obtained with SALI remains below $0.52 \, \%$ for both methods. The misclassifications for each energy and ensemble can be seen in the confusion matrices in Fig.~\ref{fig:alpha_cm} and Fig.~\ref{fig:CV_cm}, for the block-fluctuation exponent and the CV, respectively.

The results obtained for the ensemble with energy $H = 1/8$ are particularly informative because both dynamical classes are well represented: SALI identified $10402$ regular trajectories and $14598$ chaotic ones. The block-fluctuation exponent correctly identifies all chaotic trajectories while producing only three misclassifications, yielding an accuracy of $99.988 \, \%$. For the same ensemble, the CV classifier produces one chaotic misclassification and three regular ones, for a total accuracy of $99.984 \, \%$. The high overall scores in this ensemble therefore reflect the successful identification of both classes.

\begin{table}[htbp]
\centering
\caption{Agreement of $\Omega$ with the SALI classification for the H\'enon-Heiles system ensembles, measured wit the accuracy metric as defined in Eq.\eqref{eq:acc}, for both classification methodologies proposed in Sec.~\ref{sec:theory_class}. The histograms for both methods are given in Fig.\ref{fig:alpha_hists} (see Fig.~\ref{fig:alpha_cm} for the corresponding confusion matrices) and Fig.~\ref{fig:CV_hists} (see Fig.~\ref{fig:CV_cm} for the corresponding confusion matrices), respectively.}
\label{tab:accuracy}
\vspace{0.5cm}
\begin{tabular}{lccc}
\toprule
Energy &
Accuracy $(\alpha)$ &
Accuracy (CV)  \\
\midrule
$1/15$        & 0.99988 & 0.99996  \\
$1/12$        & 0.99984 & 0.99960  \\
$1/10$        & 0.99636 & 0.99488 \\
$1/8$        & 0.99988 & 0.99984  \\
\bottomrule
\end{tabular}
\end{table}

At lower energies the system becomes more regular, and therefore, the interpretation is affected by the strong class imbalance. In the case of $H = 1/12$, $\alpha$ identifies $251$ of the $252$ chaotic trajectories, compared with the $249$ for the CV, while producing only three and seven false positives, respectively. At $H = 1/15$ only four trajectories are identified as chaotic by SALI. Both methods correctly identify all regular trajectories, whereas the block-fluctuation exponent identifies one of the chaotic trajectories as regular and the CV identifies three. Consequently, the near-unity overall accuracies at these energies reflect the dominant regular population, making the confusion matrices essential to understand the performance of the classification. Across the four ensembles, the block-fluctuation exponent $\alpha$ yields a fewer misclassifications the three highest energies, while the CV performs better at the lowest one.

In App.~\ref{sec:miss} we examine the remaining misclassifications in phase space and through their $\Omega$ time series. Fig.~\ref{fig:mis_P_sec_alpha}, \ref{fig:mis_P_sec_cv} and \ref{fig:Omega_mis_convergence} show that these trajectories include cases with irregular transients and slowly changing residuals. The integrations extended to $10^6$ and $10^7$ in Fig.~\ref{fig:Omega_longer_time} further illustrate the sensitivity of their convergence to the observation interval. These examples help to delimit the finite time classification problem; they do not by themselves establish an improved ensemble accuracy at longer integration times. Nevertheless, the small number of disagreements, particularly at $H = 1/8$, supports the usage of either feature as a practical diagnostic within the sampled ensembles.

\subsection{The Fermi--Pasta--Ulam System}

\begin{figure}[htbp]
    \centering
    A) \includegraphics[scale = 0.53]{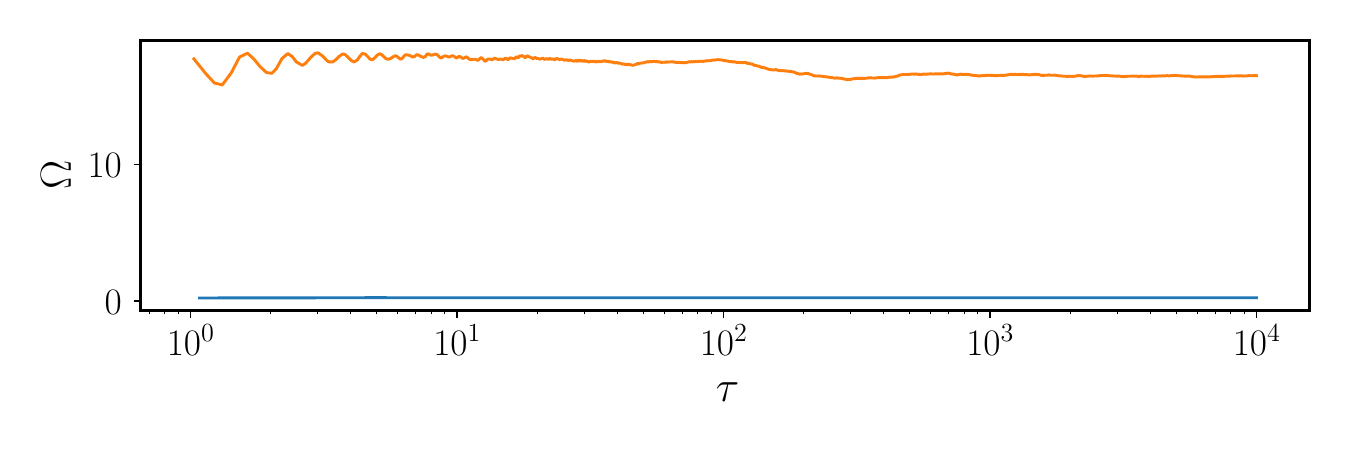} \\
    B) \includegraphics[scale = 0.3]{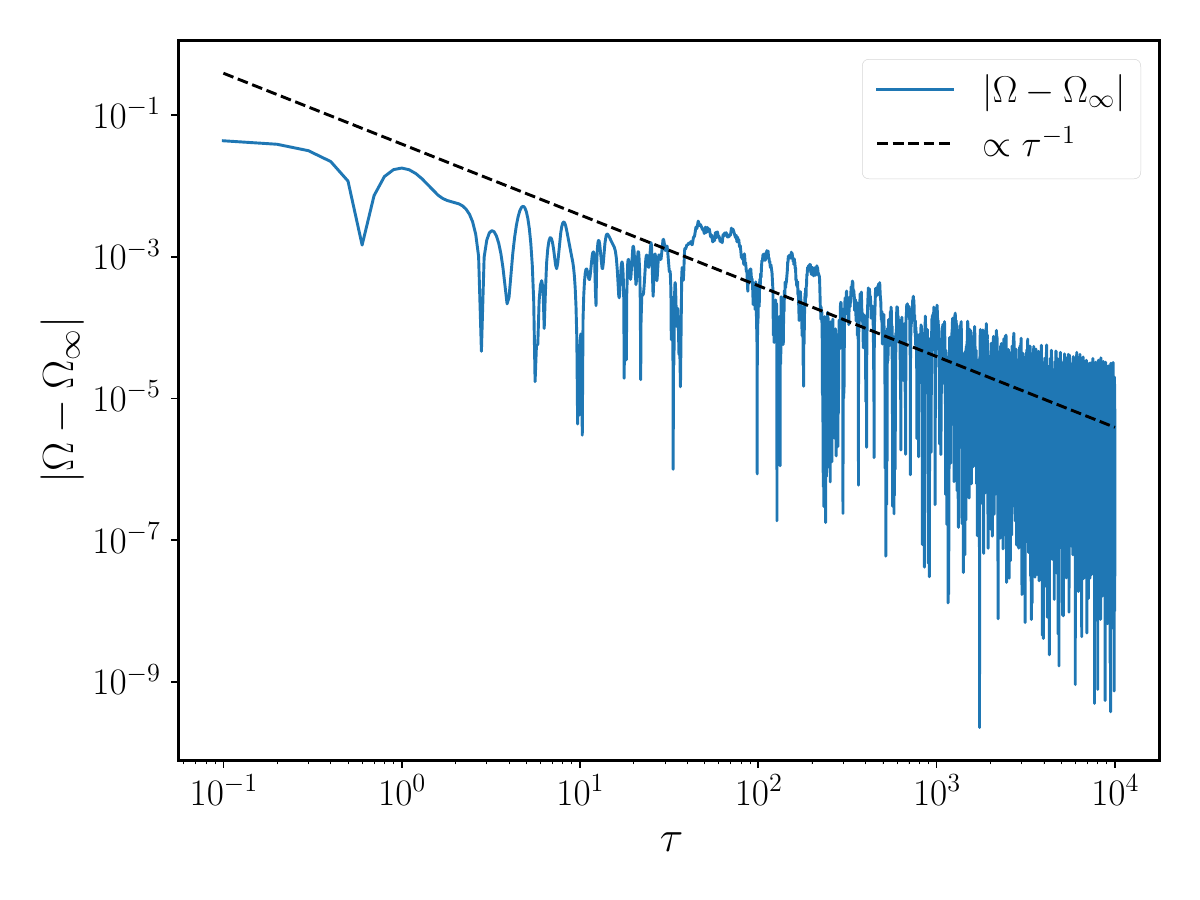} 
    C) \includegraphics[scale = 0.3]{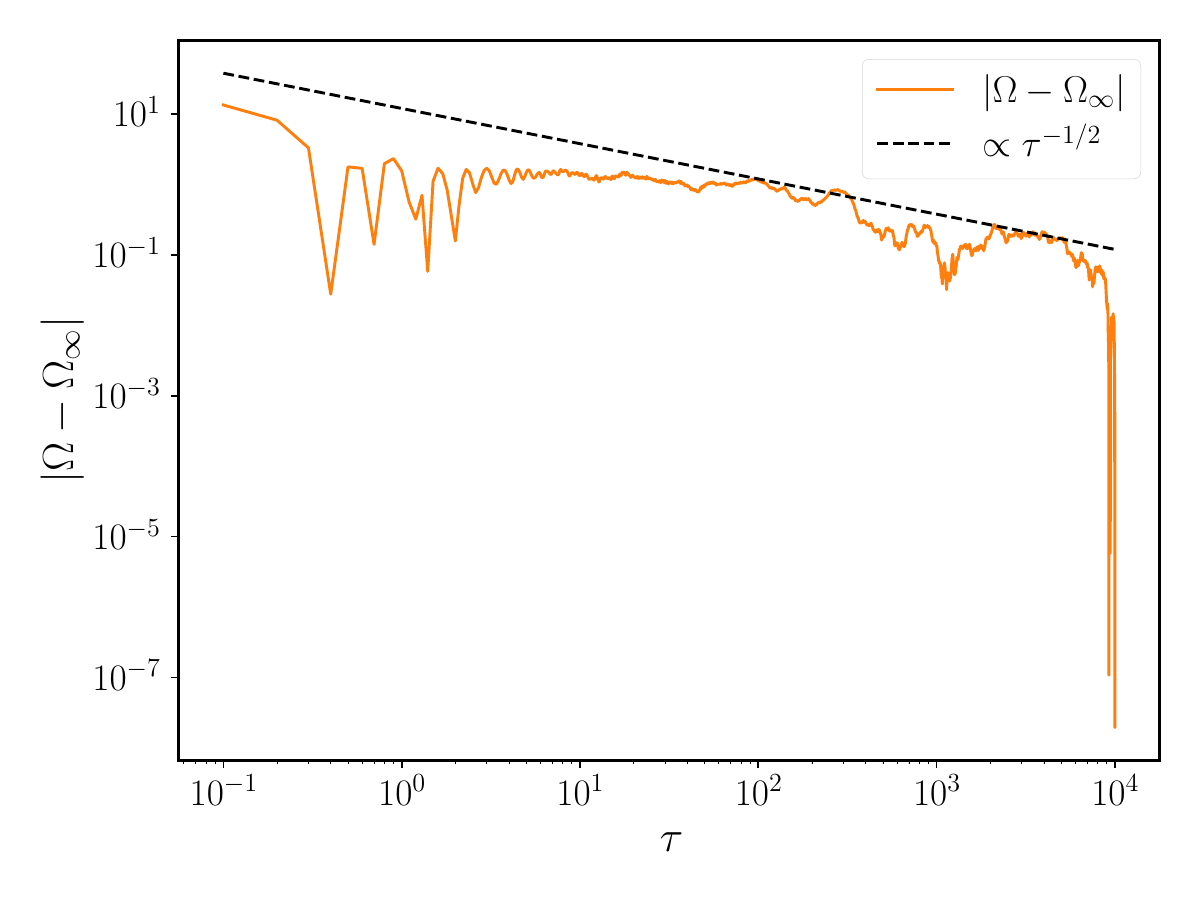}
    \caption{Analysis of $\Omega$ in the Fermi-Pasta-Ulam (FPU) system with $N = 8$. A) Time evolution of $\Omega$ for a regular (blue) and a chaotic (orange) initial condition in the FPU system with $k = 2$, $\alpha = 3$ and $\beta = 4$ for $\tau = 10^4$ time units. B) Evolution of $|\Omega - \Omega_{\infty}|$ for a regular trajectory, in perfect agreement with the prediction of Theorem~\ref{thm:regular_omega}. C) Analogous results for a chaotic initial condition, also in agreement with Theorem~\ref{thm:chaotic_omega}. The regular initial condition is defined by $q_i = 0.1$ and $p_i = 0$ for $i = 1,\ldots,8$, while the chaotic initial condition is given by $\mathbf{q} = (1.00097, 0, 1.04003, 0, -1.04003, 0, -1.00097, 0)$ and $\mathbf{p} = (0.57060, 0, -0.57060, 0, 0.57060, 0, -0.57060, 0)$. }
    \label{fig:FPU_figs}
\end{figure}

Given that the theoretical derivation of $\Omega$ does not assume any specific functional form of the Hamiltonian, the method is independent of both the system under consideration and its dimension. This makes it particularly well suited for the analysis of higher-dimensional Hamiltonian systems, where traditional approaches can become computationally expensive. To illustrate the advantages of $\Omega$ as a chaos indicator, we apply it to the Fermi--Pasta--Ulam system, which is an $N$ DoF Hamiltonian given by \cite{fermi1955}: 
\begin{equation} \label{eq:ham_FPU}
    H(\mathbf{q}, \mathbf{p}, k, \alpha, \beta) = \sum_{i = 1}^{N} \frac{p_i^2}{2} + \frac{k}{2} \sum_{i = 1}^{N} (q_{i+1} - q_i)^2 + \frac{\alpha}{3} \sum_{i = 1}^{N} (q_{i+1} - q_i)^3 + \frac{\beta}{4} \sum_{i = 1}^{N} (q_{i+1} - q_i)^4 \, ,
\end{equation}  
where $k,\alpha$ and $\beta$ are the system parameters, and $q_0$ and $q_{N+1}$ are held fixed. Here $\alpha$ denotes the cubic interaction parameter, distinct from the fitted block-fluctuation parameter. We set  $k = 2$, $\alpha = 3$ and $\beta = 4$ and integrate the two initial conditions specified in Fig.~\ref{fig:FPU_figs} up to an integration time of $\tau = 10^4$ with the Dormand-Prince $8(9)$ ODE solver \cite{prince1981}, yielding a conservation of energy within $10^{-9}$, and the regular or chaotic character of the selected trajectories is determined with SALI.

In Fig.~\ref{fig:FPU_figs} A) we observe that the regular trajectory rapidly develops a nearly stationary arclength average, whereas the chaotic trajectory retains visible fluctuations for a longer time. In Fig.~\ref{fig:FPU_figs} B), the regular residual has an envelope consistent with the $\mathcal{O}(\tau^{-1})$ decay predicted by Theorem~\ref{thm:regular_omega}. Panel C) displays the more persistent fluctuations of the chaotic orbit, with a decay broadly consistent with the $\tau^{-1/2}$ trend in Theorem~\ref{thm:chaotic_omega}. This single trajectory comparison illustrates the fluctuation contrast predicted by the theory for a system with $8$ degrees of freedom, establishing the behavior of $\Omega$ as system and dimension independent. Together with the H\'enon-Heiles ensembles, these examples demonstrate that useful information about the dynamical behavior can be extracted from the time evolution of one arclength variable.

\section{Conclusions} 
\label{sec:Conclusions}

In this work, we have developed theoretical and numerical arguments showing that the time-averaged Lagrangian descriptor, $\Omega$, provides a useful single-trajectory diagnostic for distinguishing regular and chaotic dynamics in Hamiltonian systems. The indicator is obtained by integrating the equations of motion together with a single scalar equation that accumulates the phase space arclength. Consequently, its computation does not require variational equations, the evolution of tangent vectors, neighboring trajectories, or spatial derivatives of a Lagrangian descriptor field.

The theoretical analysis clarifies that the distinction between regular and chaotic dynamics is not based on convergence versus nonconvergence of $\Omega$. Under the assumptions considered here, $\Omega(\tau)$ converges to a finite non-negative limit in both regimes. For quasiperiodic trajectories evolving on Diophantine invariant tori, the deviation from the limiting value is deterministically bounded by $\mathcal{O}(\tau^{-1})$. For trajectories sampled from sufficiently mixing ergodic chaotic components, the root-mean-square fluctuations decay instead as $\tau^{-1/2}$ when the autocorrelation of the phase space speed is absolutely integrable and the corresponding asymptotic variance is positive. The different finite-time scaling of these deviations provides the theoretical basis for the proposed classification strategy.

In the numerical implementation, this distinction is exploited through two complementary procedures based on the finite-time evolution of $\Omega$: block-scaling analysis and residual-fluctuation analysis. These methods characterize the rate at which the time average stabilizes and the magnitude of the remaining fluctuations, without requiring prior knowledge of the asymptotic value $\Omega_\infty$. The numerical experiments performed for the H\'enon--Heiles and Fermi--Pasta--Ulam systems show that these quantities provide an effective separation between regular and chaotic trajectories over the phase space regions investigated. The resulting classifications are in close agreement with those obtained using SALI as a reference chaos indicator.

An important feature of the proposed approach is its simple computational structure. Once the trajectory is integrated, only a single additional scalar arclength variable is required, while the subsequent classification is based on inexpensive post-processing of the resulting $\Omega$ time series. This makes the method particularly attractive when large ensembles of trajectories must be analyzed or when the phase space dimension becomes large. The present results therefore support $\Omega$ as a complementary diagnostic to established chaos indicators, especially in applications where a scalar trajectory-based observable and a low-overhead classification procedure are desirable.

This work can be naturally extended into different directions. The method could be further tested in systems exhibiting slow transport, stickiness, or Arnold diffusion, where long transient time scales may affect the finite-time convergence and fluctuation properties of $\Omega$. Its behavior in higher-dimensional Hamiltonian systems also deserves systematic investigation, particularly with respect to the integration times required for reliable classification. In addition, the scalar features derived from the $\Omega$ time series provide a natural basis for constructing large datasets of regular and chaotic trajectories, which could be useful in machine-learning approaches to classification and phase space exploration. Finally, since Lagrangian descriptors are defined for general dynamical systems of the form $\dot{\mathbf{x}}=\mathbf{f}(\mathbf{x},t)$, extending the present analysis to dissipative and non-autonomous systems constitutes a further direction for future work.


\section*{Acknowledgements}

The authors acknowledge Prof.~Jos\'{e} Bienvenido S\'{a}ez Landete (Universidad de Alcal\'{a}) for providing the computational resources used in this work, and also for the valuable discussions which contributed to improving the results presented in the paper.

\section*{CRediT authorship contribution statement}

\textbf{Javier Jim\'enez L\'opez:} Conceptualization, Data curation, Formal analysis, Methodology, Investigation, Resources, Software, Validation, Visualization, Writing - original draft, Writing - review \& editing. \textbf{V\'{i}ctor J. Garc\'{i}a-Garrido:} Conceptualization, Formal analysis, Investigation, Project administration, Methodology, Resources, Supervision, Validation, Visualization, Writing - original draft, Writing - review \& editing.

\section*{Data availability}

The data that support the findings of this study are available from the corresponding author upon reasonable request.

\section*{Funding}

The authors received no financial support for the research, authorship, or publication of this article

\subsection*{Conflicts of interest}

The authors declare that they have no conflicts of interest.

\subsection*{Competing Interests}

The authors have no relevant financial or non-financial interests to disclose.

\appendix

\section{The Smaller Alignment Index (SALI)} \label{sec:ground-truth}

In this appendix we give a brief description of the Smaller Alignment Index (SALI) \cite{Skokos2001,Skokos2003,Skokos2004}, which is the ground truth indicator we have used to assess the trajectory classification accuracy of the time-averaged arclength $\Omega$, as well as to evaluate its computational performance.

Following the notation used in Eq.~\eqref{eq:eq_motion}, the variational equations of a system with $N$ degrees of freedom, which determine how a small perturbation (deviation vector) $\delta \mathbf{x}$ evolves over time, can be written in matrix form as:
\begin{equation}
    \dot{\delta \mathbf{x}} = J \,\nabla^{2}H  \,  \delta \mathbf{x} \, ,
\end{equation}
where the matrix $J$ is the Poisson matrix given in Eq.~\eqref{eq:Poisson_matrix} and $\nabla^{2} H$ is the Hessian matrix, which contains the second partial derivatives of the Hamiltonian.

\begin{figure}[htbp]
    \centering
    \includegraphics[scale = 0.5]{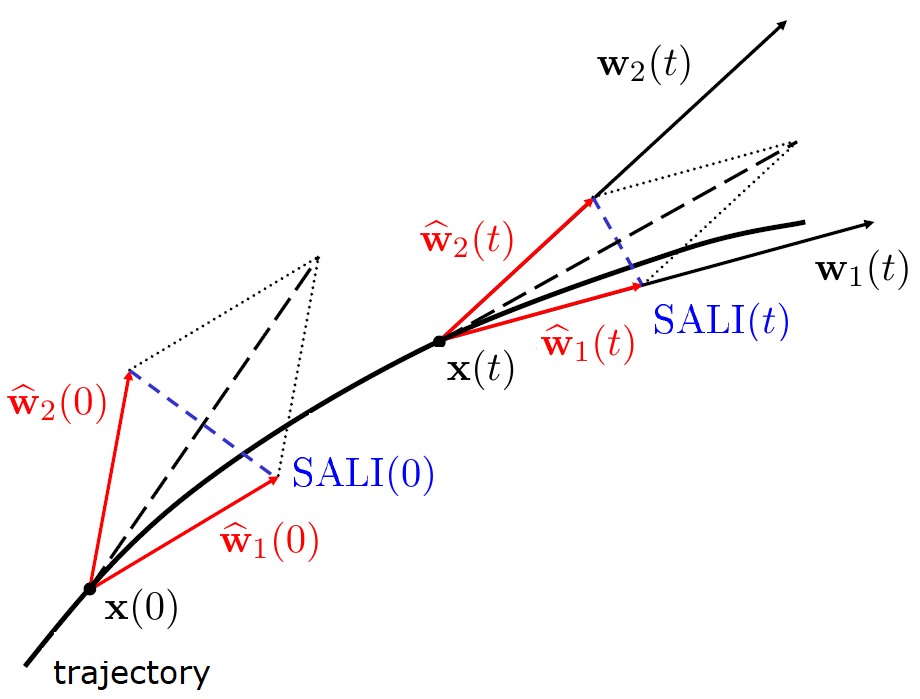}
    \caption{Evolution of a trajectory starting at $\mathbf{x}(0)$ and of two deviation vectors $\mathbf{w}_1(t)$ and $\mathbf{w}_2(t)$ that characterize the behavior of two neighboring trajectories. The SALI indicator at time $t$ corresponds to the minimum length of the two diagonals of the parallelogram generated by the normalized deviation vectors. Figure adapted from \cite{skokos2016chaos}.}
    \label{fig:SALI_diagram}
\end{figure}

In order to classify a trajectory using SALI, we simultaneously follow the evolution of the initial condition $\mathbf{x}(0)$ and of two initially linearly independent deviation vectors, $\mathbf{w}_1(0)$ and $\mathbf{w}_2(0)$. Since SALI depends only on the directions of the deviation vectors, these vectors are normalized during the integration in order to avoid numerical overflow and to retain only directional information:
\begin{equation}
    \widehat{\mathbf{w}}_i(t) = \dfrac{\mathbf{w}_i(t)}{\|\mathbf{w}_{i} (t)\|} \, , \hspace{0.3cm} i = 1,2 \, .
\end{equation}
Introducing the parallel and antiparallel alignment indices:
\begin{equation}
    d_{+}(t) = \| \widehat{\mathbf{w}}_{1}(t) + \widehat{\mathbf{w}}_{2}(t) \| \, , \hspace{1.0cm} 
    d_{-}(t) = \| \widehat{\mathbf{w}}_{1}(t) - \widehat{\mathbf{w}}_{2}(t) \| \, ,
\end{equation}
SALI is defined as:
\begin{equation} \label{eq:SALI_def}
    \mathrm{SALI}(t) = \min\{d_{+}(t), d_{-}(t)\} \, .
\end{equation}
Thus, SALI measures the minimum length, at time $t$, of the two diagonals of the parallelogram generated by the two normalized deviation vectors. A schematic illustration of this construction is shown in Fig.~\ref{fig:SALI_diagram}.

For continuous Hamiltonian systems, SALI displays the asymptotic behavior \cite{skokos2016chaos}:
\begin{equation}
    \mathrm{SALI}(t) \; \propto \; 
    \begin{cases}
        \mathrm{constant} &,\;\; \text{for a regular trajectory} \\[.1cm]
        \exp[-(\lambda_1-\lambda_2)t] \;&,\;\; \text{for a chaotic trajectory}
    \end{cases}
    \label{eq:asymp_SALI_contHam}
\end{equation}
where $\lambda_1 \geq \lambda_2$ denote the two largest Lyapunov exponents. For chaotic trajectories, both deviation vectors tend to align with the most unstable direction, causing SALI to decay exponentially to zero. In autonomous Hamiltonian systems with two degrees of freedom, this decay is governed by the largest positive Lyapunov exponent, recovering the standard asymptotic behavior commonly used in this setting. By contrast, for regular trajectories, the deviation vectors typically remain tangent to the invariant torus on which the motion evolves, and SALI fluctuates around a nonzero value. This clear asymptotic distinction makes SALI a sharp and efficient chaos indicator: in practice, the computation can be stopped once SALI falls below a prescribed numerical threshold. For classification purposes, SALI is commonly analyzed on a logarithmic scale.

\section{Analysis of misclassified initial conditions in the H\'enon-Heiles system} \label{sec:miss}

This appendix supplements the ensemble results obtained for the H\'enon-Heiles system in Sec. \ref{subsec:HH} by examining the initial conditions whose block-fluctuation exponent or coefficient of variation classification disagree with the one SALI provides at a total integration time $\tau = 10^5$. The amount of misclassified trajectories and their contribution to the accuracy are given in Table~\ref{tab:accuracy} and Figs.~\ref{fig:alpha_cm} and ~\ref{fig:CV_cm}. 

\begin{figure}[htbp]
    \centering
    A) \includegraphics[scale = 0.36]{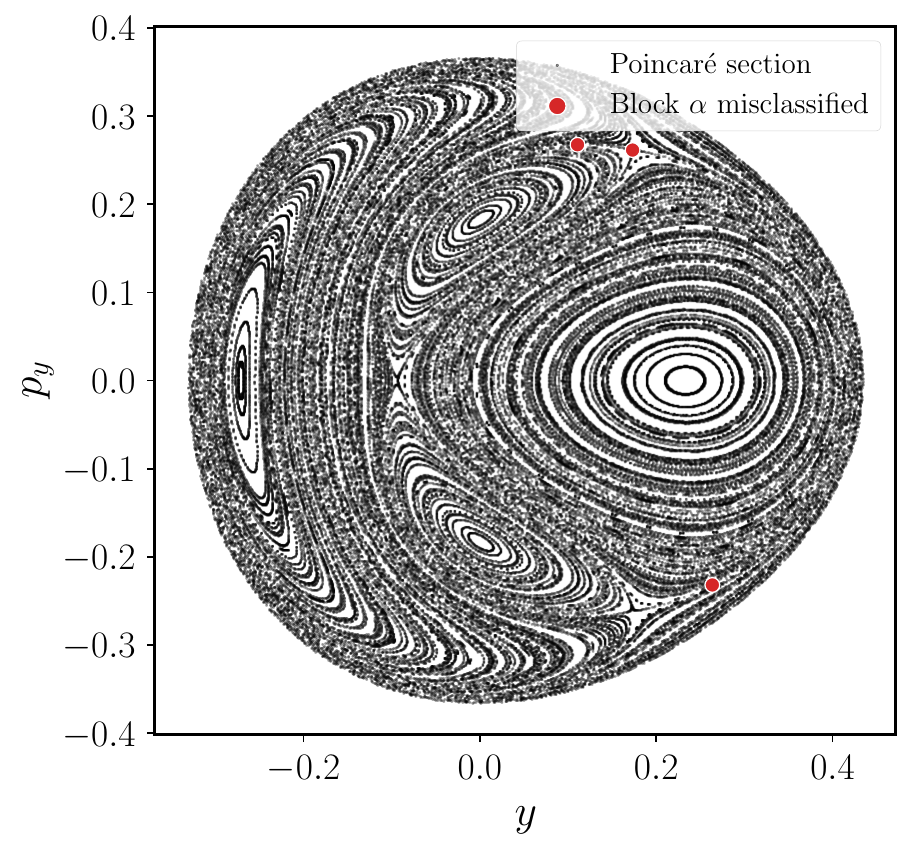} 
    B) \includegraphics[scale = 0.36]{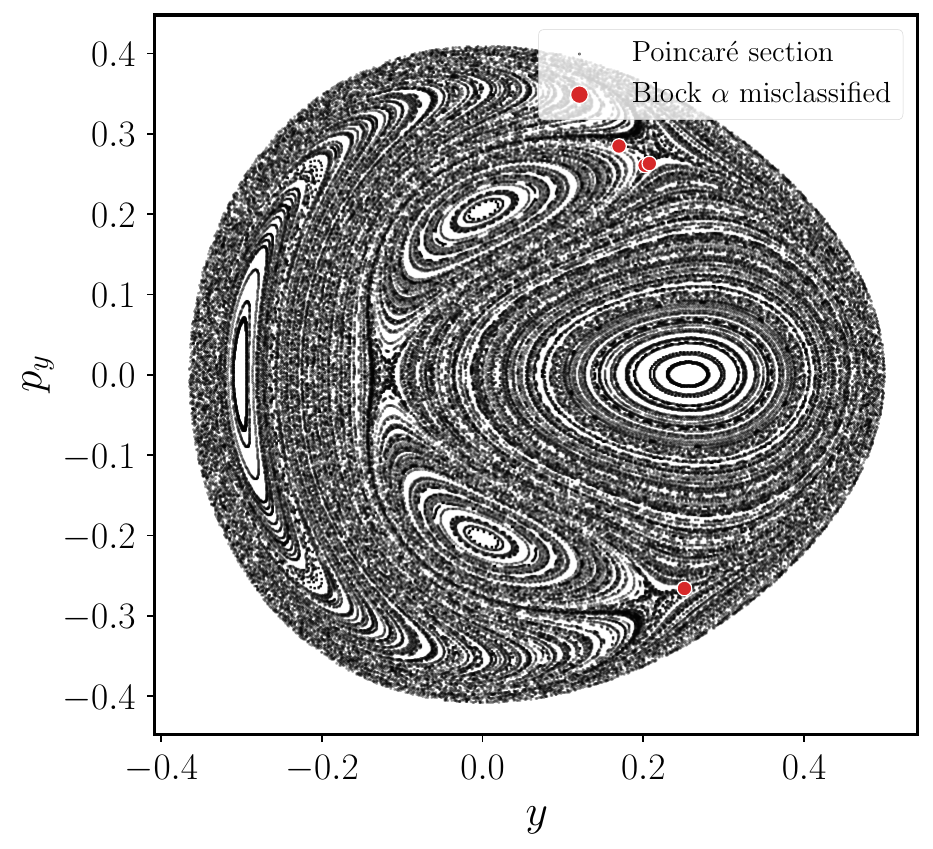} \\
    C) \includegraphics[scale = 0.36]{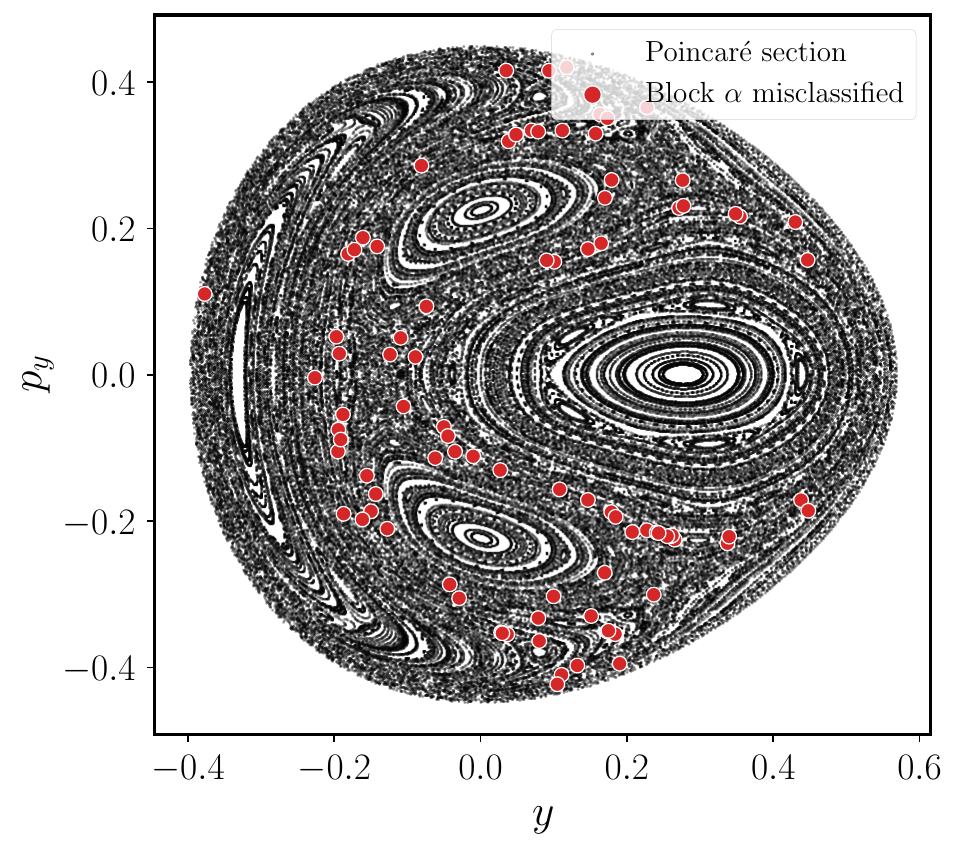} 
    D) \includegraphics[scale = 0.36]{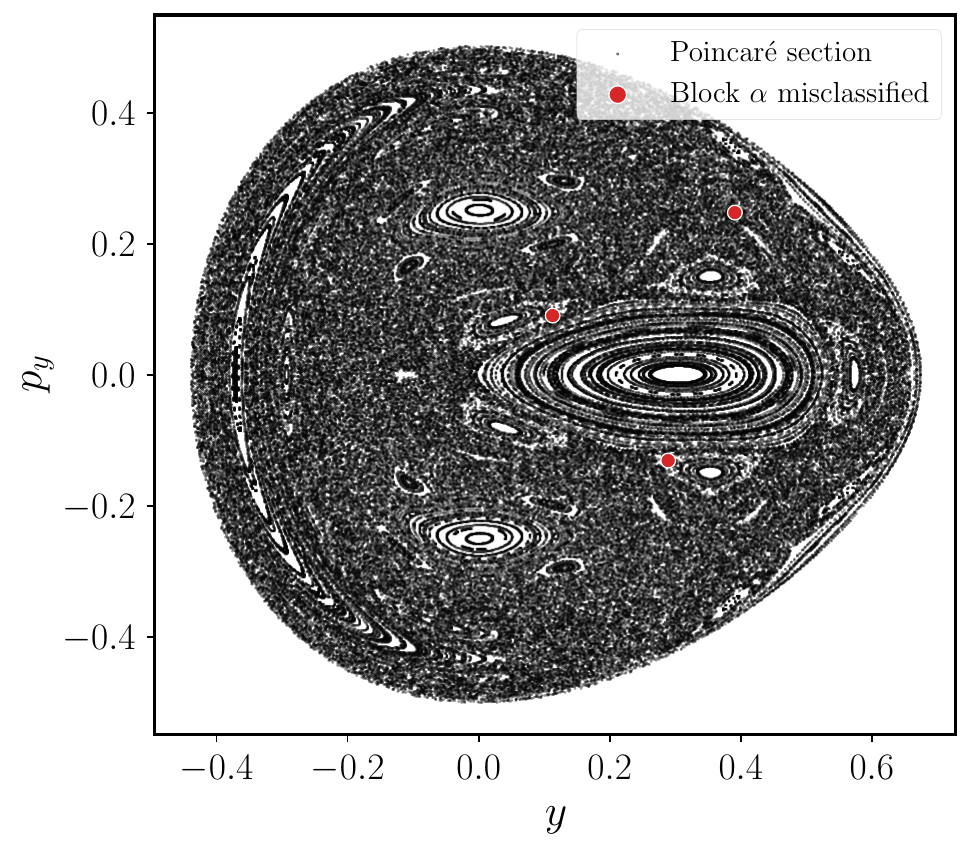}
    \caption{Poincar\'e sections for the H\'enon-Heiles system, defined for $x = 0$ and $p_x > 0$, at four different energies: A) $H = 1/15$, B) $H = 1/12$, C) $H = 1/10$ and D) $H = 1/8$. The overlaid red points identify the initial conditions incorrectly classified by the block-fluctuation exponent $\alpha$, obtained from the scaling relation $\sigma_{\mathrm{block}}(\Delta) \propto \Delta^{-\alpha}$. These misclassifications are determined by comparing the predictions of this method with the classification of SALI, considered as ground truth.}
    \label{fig:mis_P_sec_alpha}
\end{figure}

\begin{figure}[htbp]
    \centering
    A) \includegraphics[scale = 0.36]{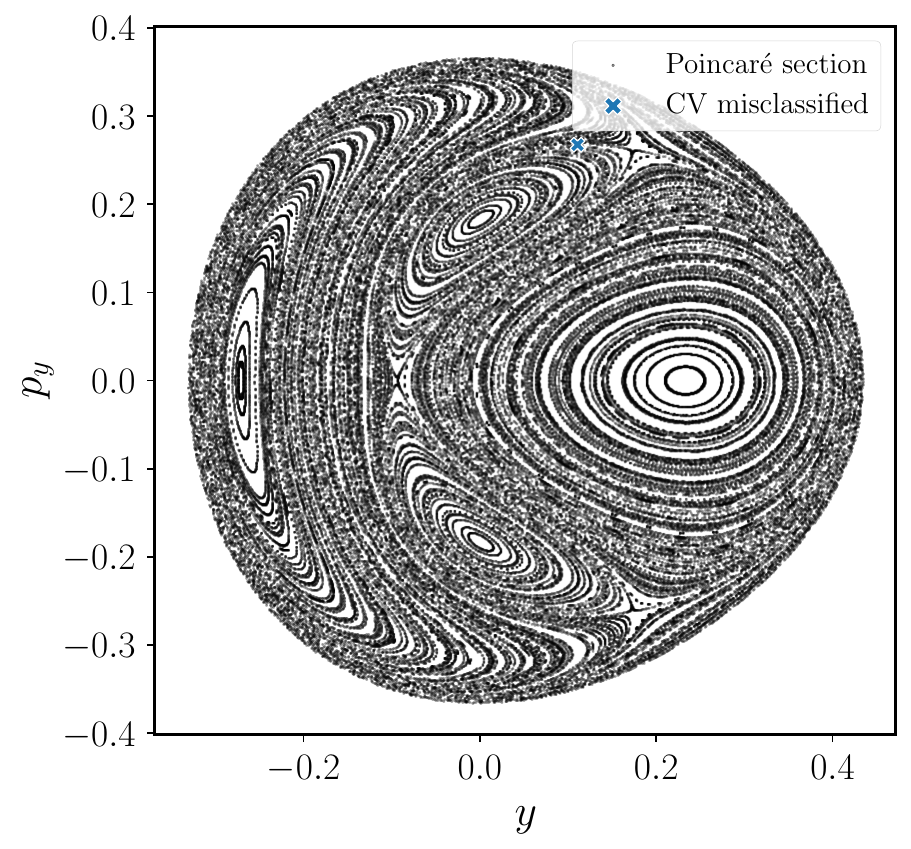} 
    B) \includegraphics[scale = 0.36]{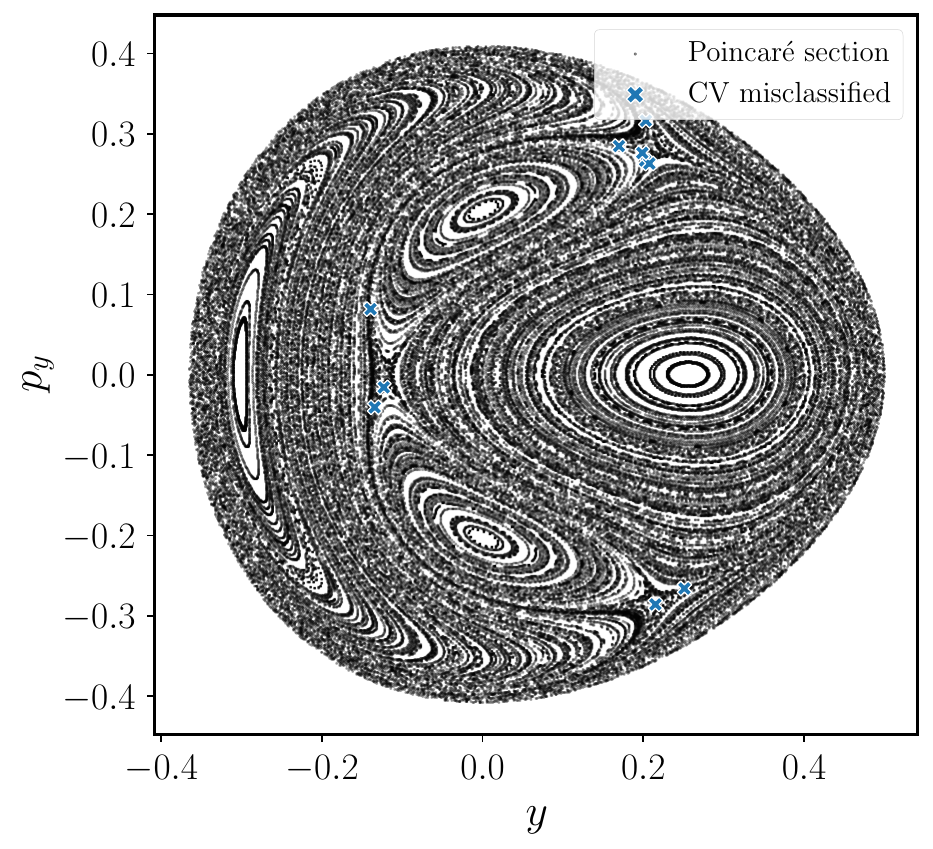} \\
    C) \includegraphics[scale = 0.36]{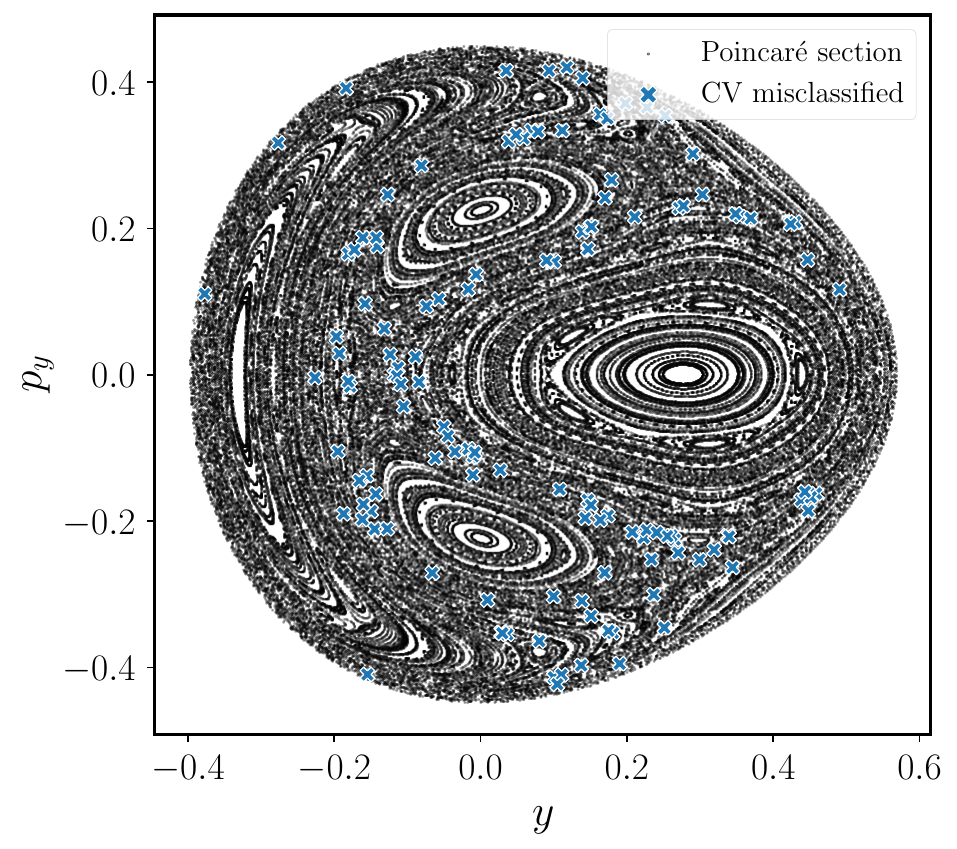} 
    D) \includegraphics[scale = 0.36]{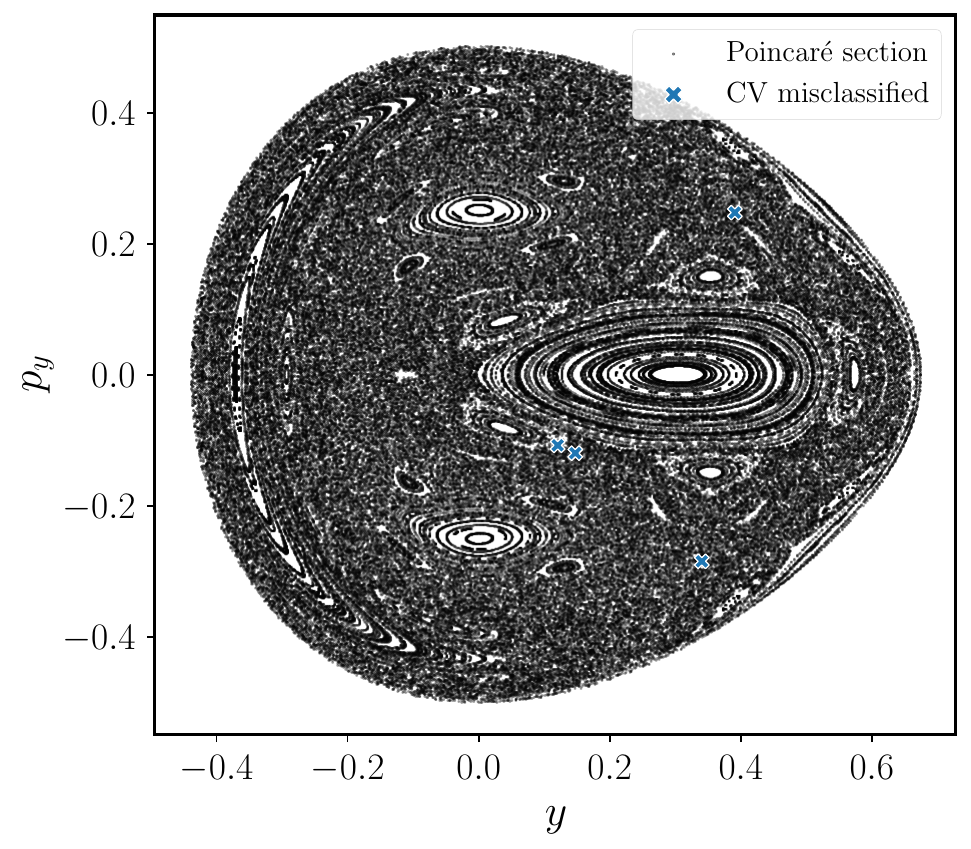}
    \caption{Poincar\'e sections for the H\'enon-Heiles system, defined for $x = 0$ and $p_x > 0$, at four different energies: A) $H = 1/15$, B) $H = 1/12$, C) $H = 1/10$ and D) $H = 1/8$. The overlaid blue points indicate the initial conditions that have been misclassified by the normalized fluctuation amplitude $\mathrm{CV} = \sigma_{\Omega} / \Omega_{\infty}$. As in Fig.\ref{fig:mis_P_sec_alpha}, the misclassifications are identified using SALI as ground truth.}
    \label{fig:mis_P_sec_cv}
\end{figure}

\begin{figure}[htbp]
    \centering
    A) \includegraphics[scale = 0.34]{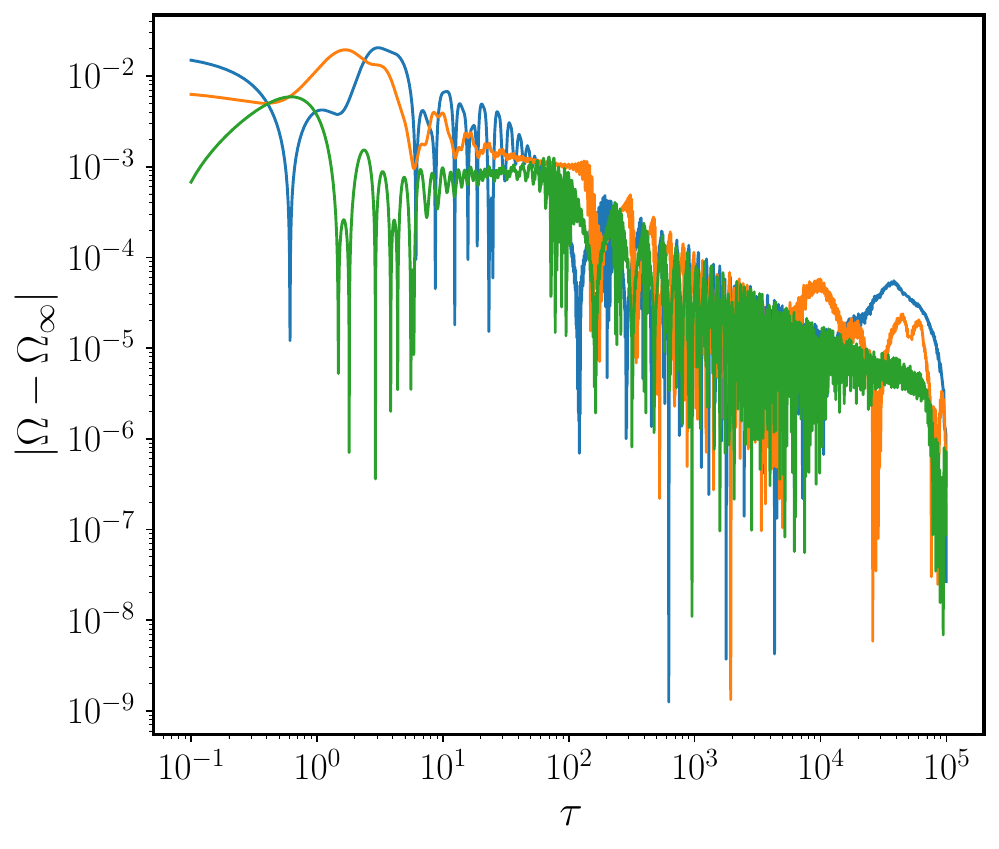} 
    E) \includegraphics[scale = 0.34]{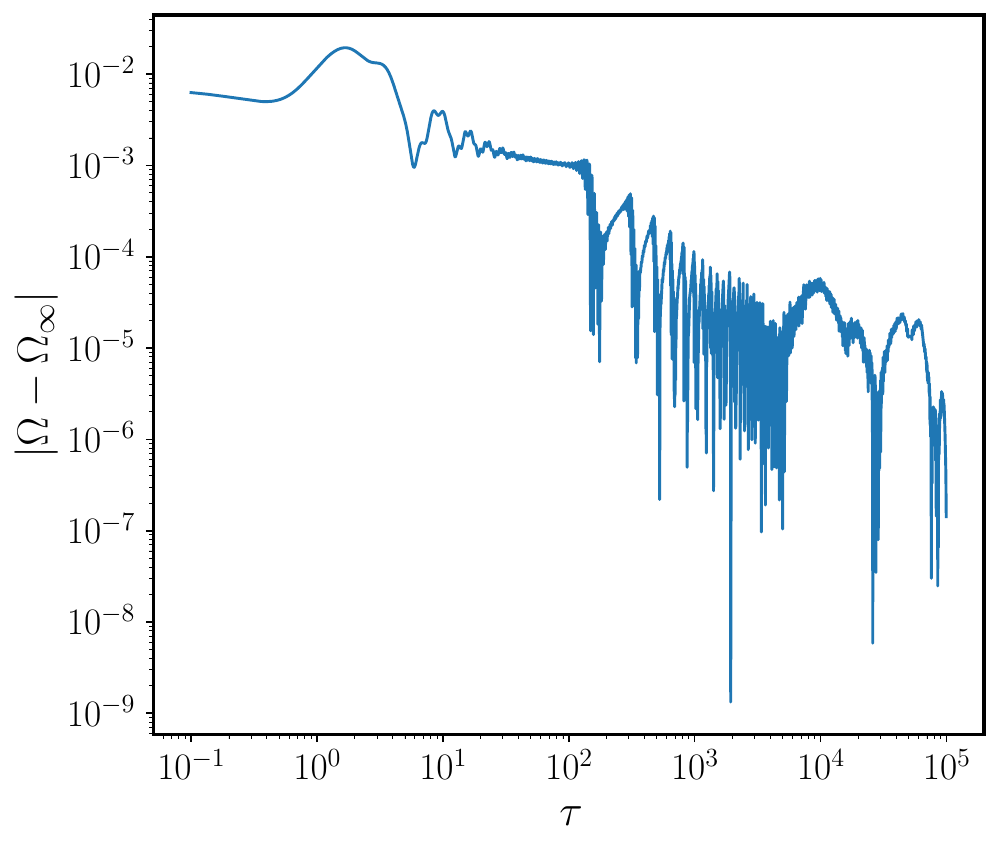} 
    \\
    B) \includegraphics[scale = 0.34]{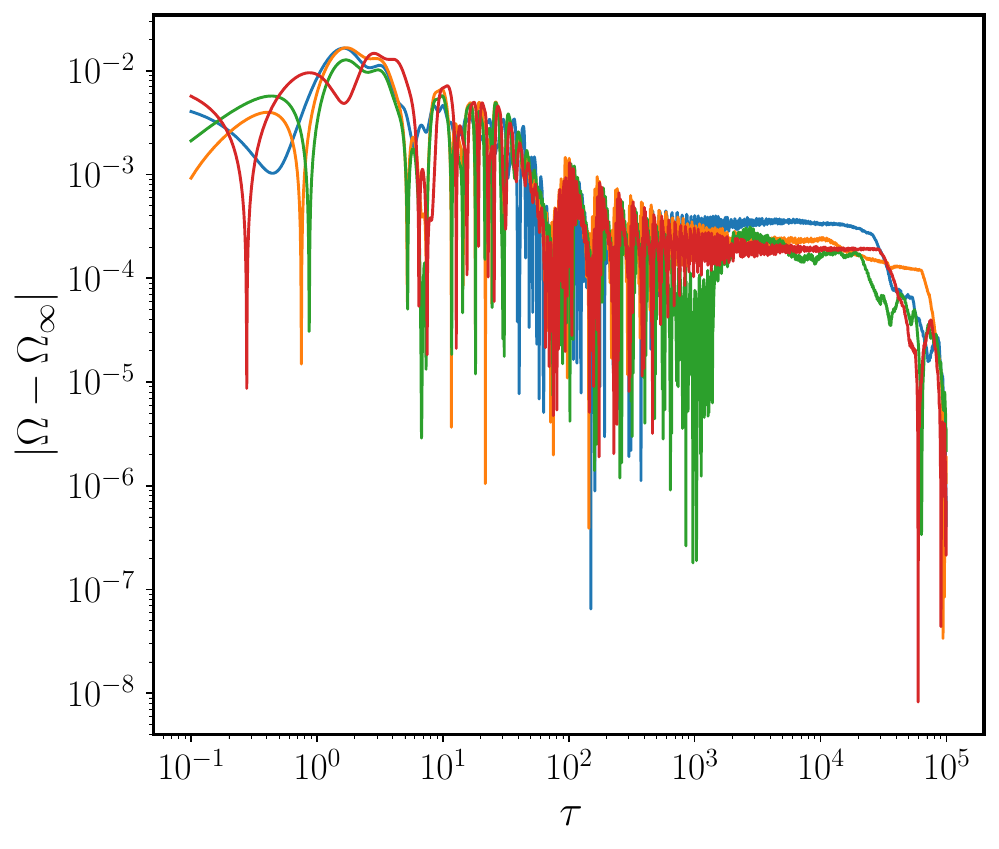} 
    F) \includegraphics[scale = 0.34]{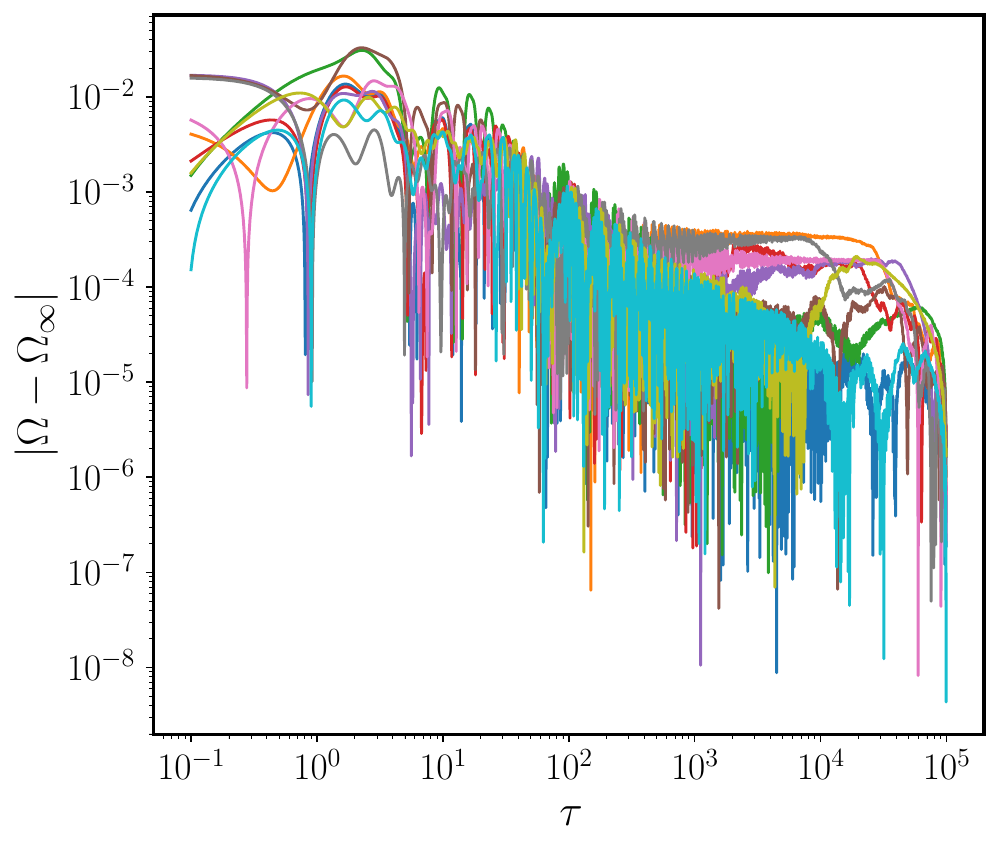}
    \\
    C) \includegraphics[scale = 0.34]{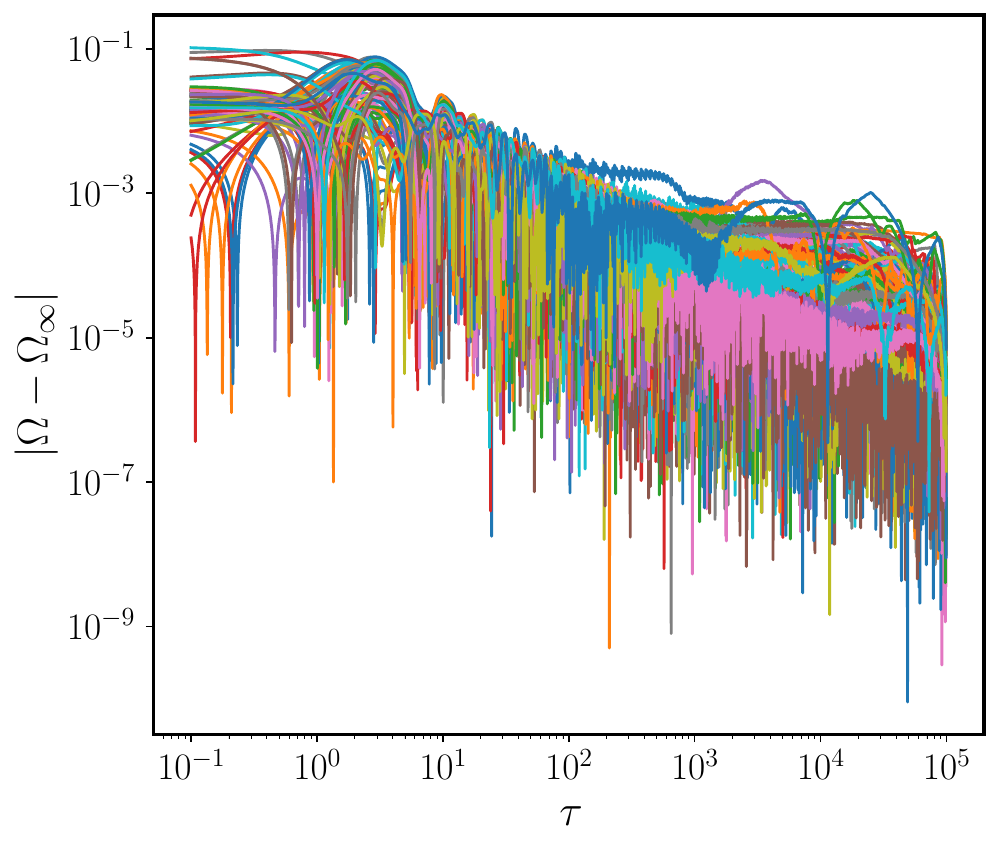} 
    G) \includegraphics[scale = 0.34]{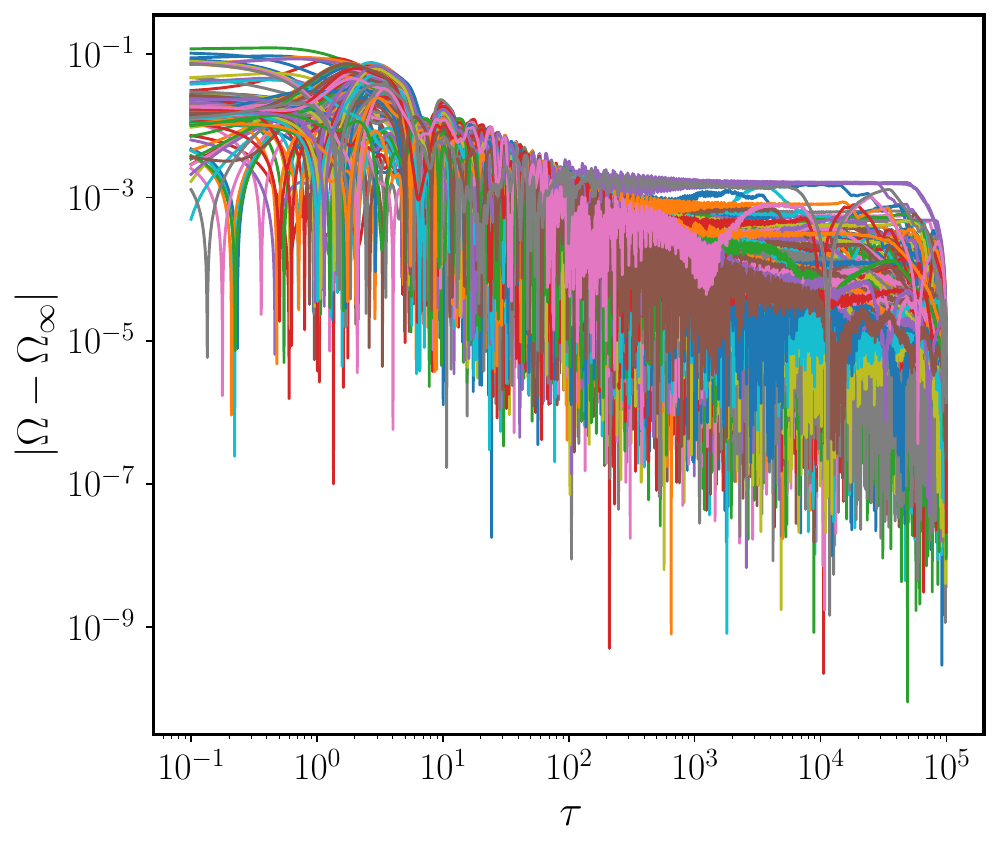} 
    \\
    D) \includegraphics[scale = 0.34]{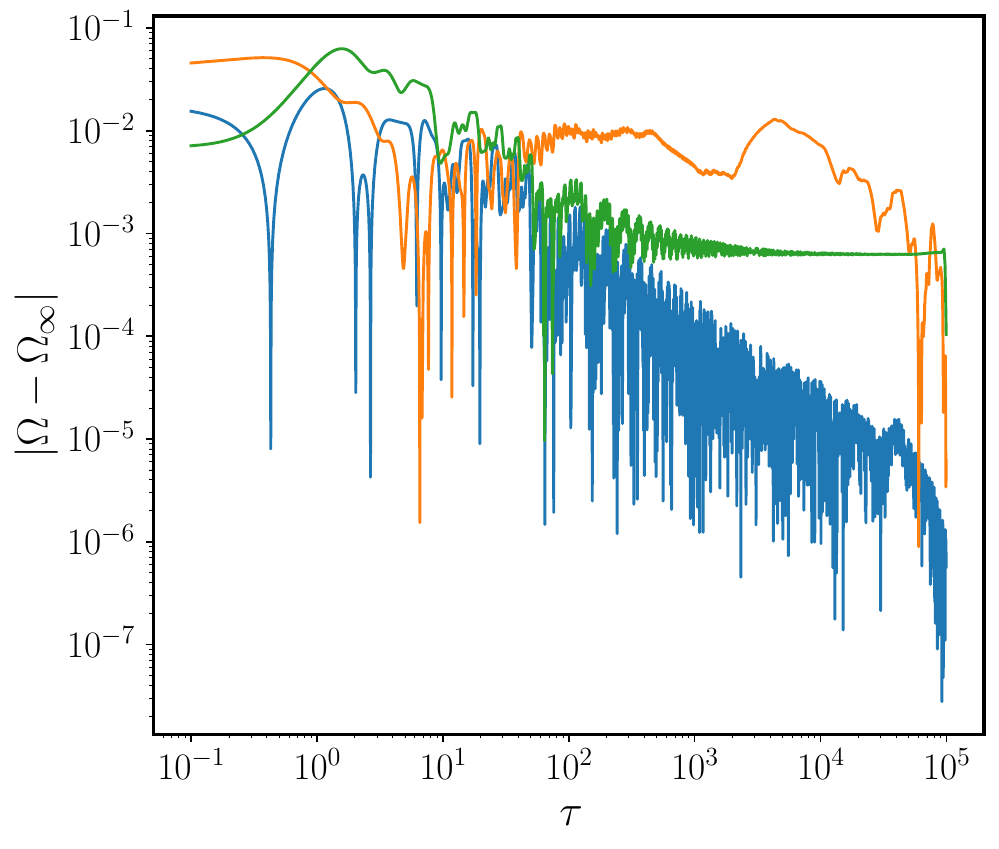}
    H) \includegraphics[scale = 0.34]{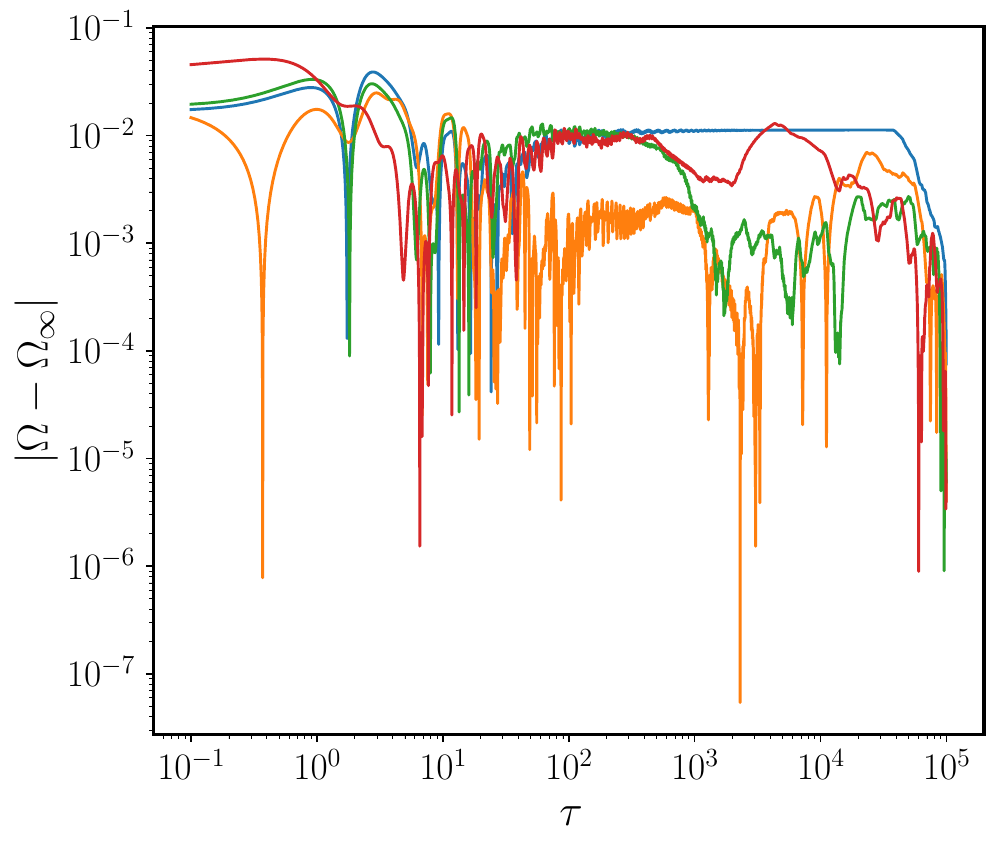}
    \caption{Time evolution of $|\Omega - \Omega_{\infty}|$ for the misclassified initial conditions by the block-fluctuation exponent (left column) and by the normalized fluctuation amplitude (right column) using the classification yield by SALI as reference. Each curve corresponds to a different misclassified initial condition of the H\'enon-Heiles system. The panels correspond to the four different energies we have considered for this system: $H = 1/15$ (A and E), $H = 1/12$ (B and F), $H = 1/10$ (C and G) and $H = 1/8$ (D and H).}
    \label{fig:Omega_mis_convergence}
\end{figure}

Figures \ref{fig:mis_P_sec_alpha} and \ref{fig:mis_P_sec_cv} show the location of the misclassified trajectories on the Poincar\'e sections defined by $x = 0$ and $p_x > 0$. The marked points include initial conditions near small island chains and island boundaries. At $H = 1/10$, they occupy a broader part of the mixed phase space. These locations suggest examining the influence of transient trapping, although proximity to an island by itself does not imply that the trajectory will undergo stickiness along its evolution. The corresponding residuals in Fig.~\ref{fig:Omega_mis_convergence} display oscillatory decay, extended plateaus and irregular changes in amplitude. These transients can affect the fitted block-fluctuation exponent and the residual amplitude differently, leading to overlap between the finite time features of the two reference classes. Stickiness is a possible contributor to this ambiguity \cite{PhysRevLett.55.2741,Zaslavsky2002}, but the curves do not establish this mechanism for every disagreement. In particular, the misclassified subsets contain both types of trajectories: those labeled by SALI as regular and by $\Omega$ as chaotic and viceversa.

\begin{figure}[htbp]
    \centering
    A) \includegraphics[scale = 0.34]{new_figures/omega_convergence_misclassified_initial_conditions_block_alpha_0p125.pdf} 
    D) \includegraphics[scale = 0.34]{new_figures/omega_convergence_misclassified_initial_conditions_cv_0p125.pdf} \\
    B) \includegraphics[scale = 0.34]{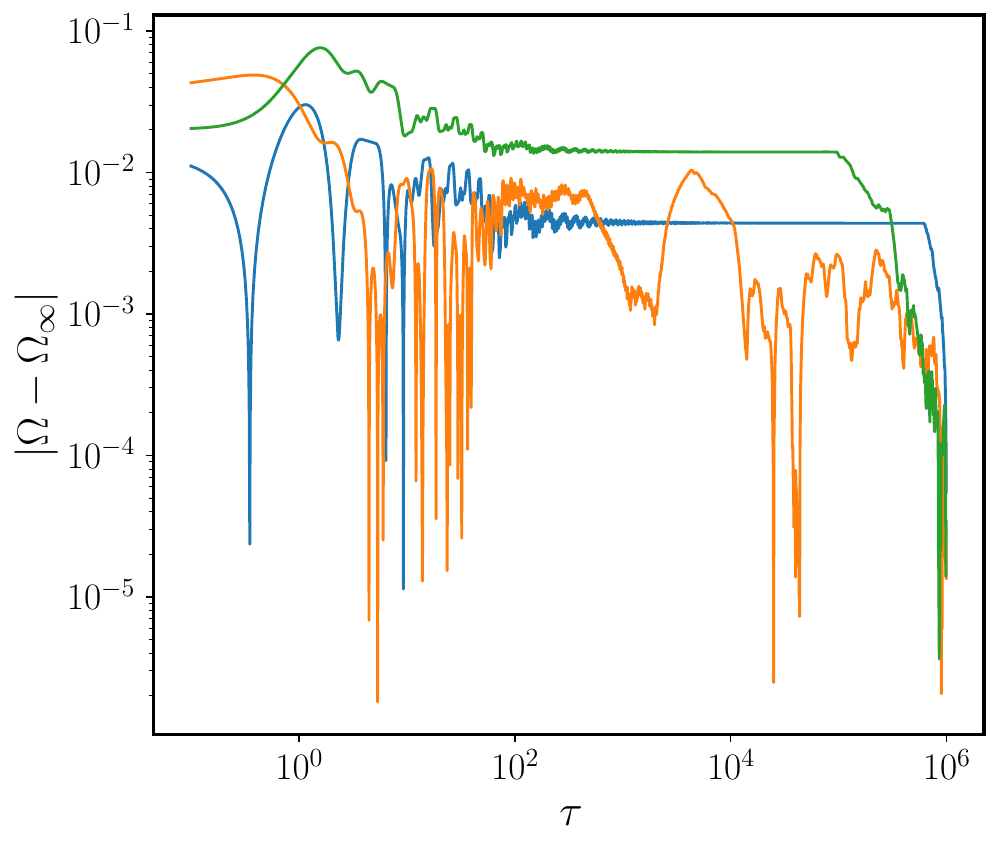} 
    E) \includegraphics[scale = 0.34]{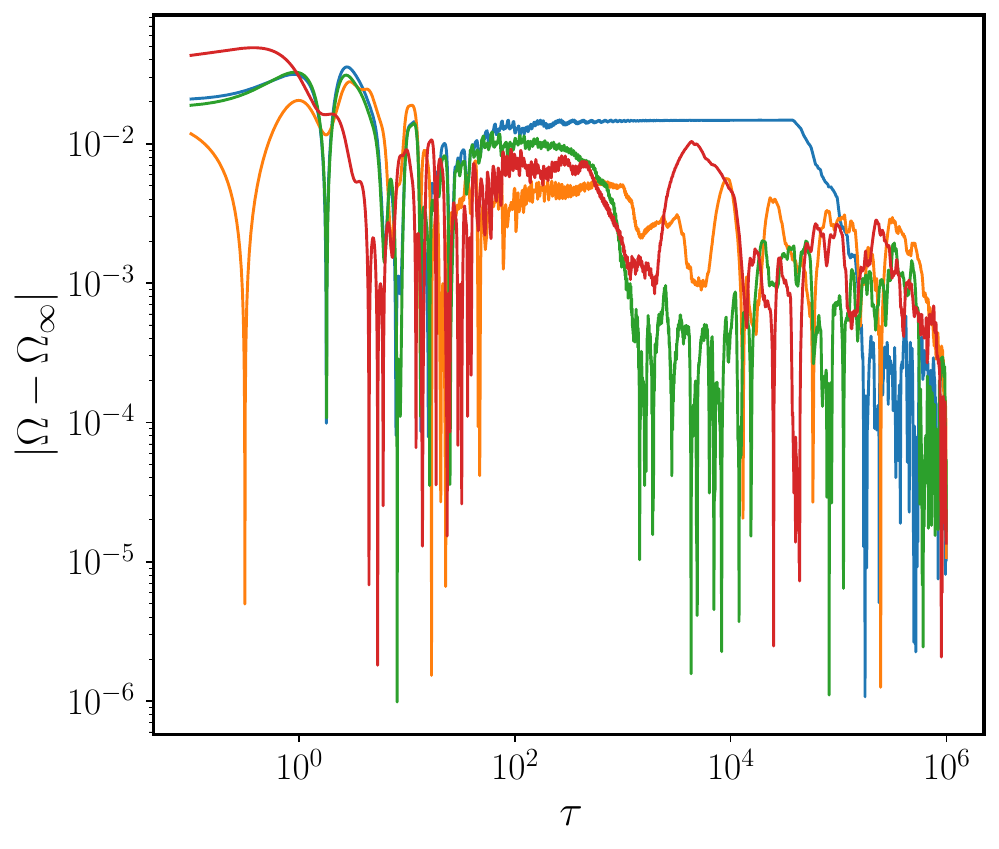} \\
    C) \includegraphics[scale = 0.34]{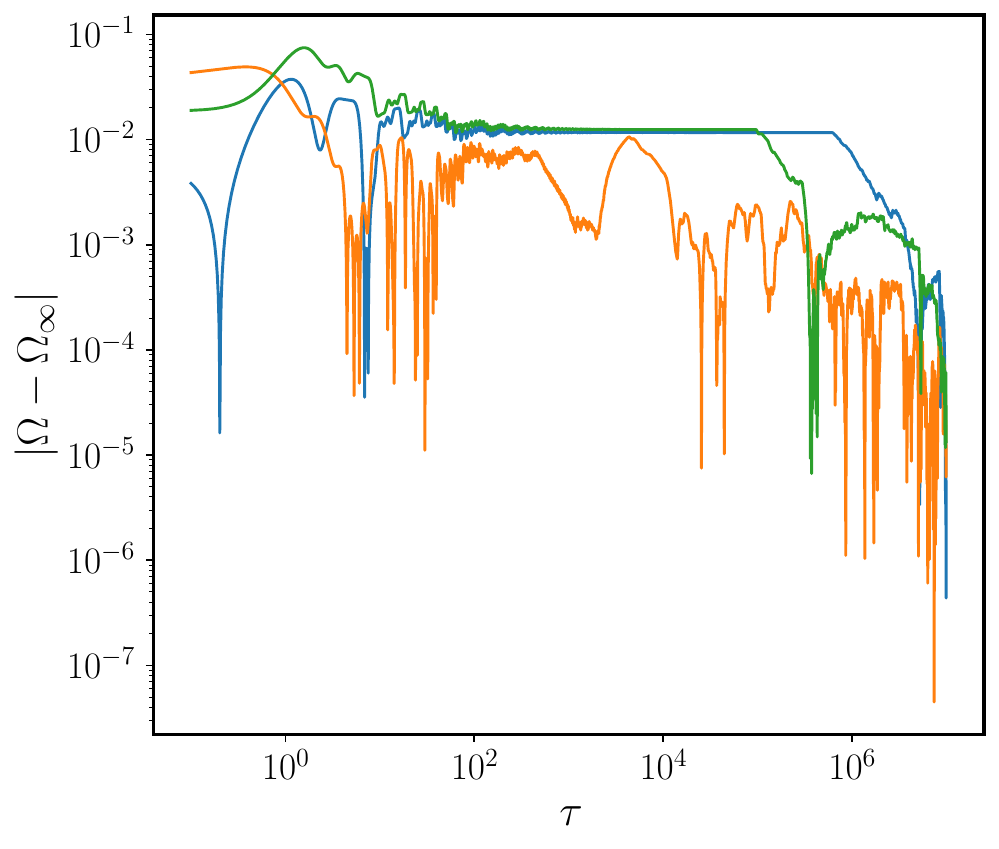}
    F) \includegraphics[scale = 0.34]{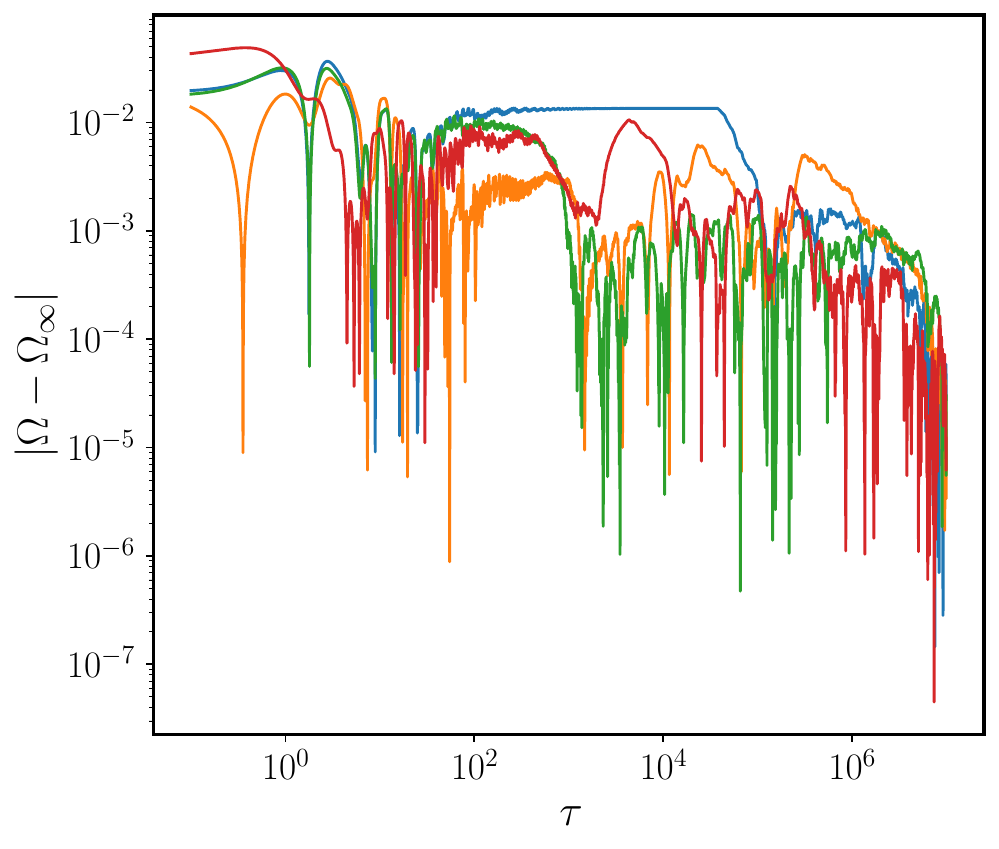}
    \caption{Time evolution of $|\Omega - \Omega_{\infty}|$ for the misclassified initial conditions by the block-fluctuation exponent (left column row) and by the normalized fluctuation amplitude (right column) using the classification yield by SALI as reference in the H\'enon-Heiles system for $H = 1/8$. The integration time is increased from $\tau = 10^5$ (A, D), to $\tau = 10^6$ (B, E) and $\tau = 10^7$ (C, F). This comparison allows us to investigate the influence of finite integration in the observed fluctuations of $\Omega$ and their role in the incorrect classification of trajectories exhibiting slow convergence.}
    \label{fig:Omega_longer_time}
\end{figure}

For $H = 1/8$, Fig.~\ref{fig:Omega_longer_time} follows the initially misclassified subsets through integrations extended to $\tau = 10^6$ and $\tau = 10^7$, where continued changes in the residuals illustrate their sensitivity to the observation interval. Therefore, the comparison helps to interpret the exceptional trajectories, but does not establish an improvement in ensemble classification accuracy at longer integration times.

\newpage
\bibliography{referencias}

\end{document}